\documentclass[amsmath,amssymb,prr,aps,showpacs,superscriptaddress,twocolumn,longbibliography]{revtex4-1}
\pdfoutput=1
\usepackage{amsmath,amsfonts,amssymb,amsthm,graphics,graphicx,epsfig,bbm}
\usepackage[colorlinks=true,citecolor=blue,linkcolor=blue,urlcolor=blue]{hyperref}
\usepackage[usenames]{color}
\usepackage{graphicx}
\usepackage{subfigure}
\usepackage{amsmath}
\usepackage{epsfig}
\usepackage{dcolumn}
\usepackage{bm}
\usepackage{color}
\usepackage{epstopdf}
\usepackage{amssymb}
\usepackage{amstext}
\usepackage{latexsym}
\usepackage{hyperref}
\usepackage{amsfonts}
\usepackage{psfrag}
\usepackage{xcolor}
\usepackage[normalem]{ulem}
\usepackage{dsfont}
\usepackage{txfonts}
\usepackage{cases}
\usepackage{ifthen}
\usepackage{algorithm}
\usepackage{algpseudocode}

\graphicspath{{./Images/},{./ImagesAppendix/}}

\newcommand{\ket}[1]{\ensuremath{\left\vert #1 \right\rangle}}
\newcommand{\bra}[1]{\ensuremath{\left\langle #1 \right\vert}}

\newcommand{\braket}[2]{\langle #1 \vert #2 \rangle}

\newcommand{\mean}[1]{\left\langle #1 \right\rangle}

\newcommand{\an}[2]{\ensuremath{\hat{#1}^{\protect\phantom{\dagger}}_{#2}}}
\newcommand{\cn}[2]{\ensuremath{\hat{#1}^\dagger_{#2}}}
\newcommand{\nn}[2]{\ensuremath{\hat{n}^{#1}_{#2}}}

\newcommand{\expU}[1]{\ensuremath{e^{#1}}}
\newcommand{\abs}[1]{\left|#1\right|}

\newcommand{\revD}[1]{#1}
\newcommand{\revE}[1]{#1}

\newcommand{\subfigimg}[3][,]{%
	\setbox1=\hbox{\includegraphics[#1]{#3}}
	\leavevmode\rlap{\usebox1}
	\rlap{\hspace*{2pt}\raisebox{\dimexpr\ht1-0.5\baselineskip}{{\bfseries \large\textsf{#2}}}}
	\phantom{\usebox1}
}

\newcommand{\idg}[1]{{\bfseries #1)}}

\begin{document}
	

\title{Machine learning engineering of \revD{quantum currents}}






\author{Tobias Haug}
\affiliation{Centre for Quantum Technologies, National University of Singapore,
3 Science Drive 2, Singapore 117543, Singapore}
\author{Rainer Dumke}
\affiliation{Centre for Quantum Technologies, National University of Singapore, 3 Science Drive 2, Singapore 117543, Singapore}
\affiliation{Division of Physics and Applied Physics, Nanyang Technological University, 21 Nanyang Link, Singapore 637371, Singapore}
\affiliation{MajuLab, CNRS-UCA-SU-NUS-NTU International Joint Research Unit, Singapore}
\author{Leong-Chuan Kwek}
\affiliation{Centre for Quantum Technologies, National University of Singapore,
	3 Science Drive 2, Singapore 117543, Singapore}
\affiliation{MajuLab, CNRS-UCA-SU-NUS-NTU International Joint Research Unit, Singapore}
\affiliation{School of Electrical and Electronic Engineering, 50 Nanyang Ave., Singapore 637553}
\affiliation{National Institute of Education, Nanyang Technological University,
	1 Nanyang Walk, Singapore 637616, Singapore}
\author{Christian Miniatura} 
\affiliation{MajuLab, CNRS-UCA-SU-NUS-NTU International Joint Research Unit, Singapore}
\affiliation{Centre for Quantum Technologies, National University of Singapore,
	3 Science Drive 2, Singapore 117543, Singapore}
\affiliation{Department of Physics, National University of Singapore, 2 Science Drive 3, Singapore 117542, Singapore}
\affiliation{School of Physical and Mathematical Sciences, Nanyang Technological University, 637371 Singapore, Singapore}
\affiliation{Yale-NUS College, 16 College Avenue West, Singapore 138527, Singapore}
\affiliation{Université Côte d'Azur, CNRS, Institut de Physique de Nice, 1361 route des Lucioles, 06560 Valbonne, France}
\author{Luigi Amico}
\affiliation{Quantum  Research  Centre,  Technology  Innovation  Institute,  Abu  Dhabi,  UAE}
\affiliation{Centre for Quantum Technologies, National University of Singapore,
	3 Science Drive 2, Singapore 117543, Singapore}
\affiliation{MajuLab, CNRS-UCA-SU-NUS-NTU International Joint Research Unit, Singapore}
\affiliation{Dipartimento di Fisica e Astronomia, Via S. Sofia 64, 95127 Catania, Italy}
\affiliation{CNR-MATIS-IMM \&   INFN-Sezione di Catania, Via S. Sofia 64, 95127 Catania, Italy}
\affiliation{LANEF {\it 'Chaire d'excellence'}, Universit\`e Grenoble-Alpes \& CNRS, F-38000 Grenoble, France}

\date{\today}

\begin{abstract}
The design, accurate preparation and manipulation of quantum states in quantum circuits are essential operational tasks at the heart of quantum technologies. Nowadays, circuits can be designed with physical parameters that  can be controlled with unprecedented accuracy and flexibility. However, the generation of well-controlled current states is still a nagging bottleneck, especially when different circuit elements are integrated together. In this work, we show how machine learning can effectively address this challenge and outperform the current existing methods.
\revE{To this end, we exploit deep reinforcement learning to prepare prescribed quantum current states  in circuits composed of lumped elements.} 
To highlight our method, we show how to engineer bosonic persistent currents as they are relevant in different quantum technologies as cold atoms and superconducting circuits.
\revE{We demonstrate the use of deep reinforcement learning to re-discover established protocols, as well as solve configurations that are difficult to treat with other methods.}
With our approach, quantum current states characterised by a single winding number or entangled currents of multiple winding numbers can be prepared \revE{in a robust manner}, superseding the existing protocols. 
\end{abstract}

\maketitle


\begin{figure*}[htbp]
	\centering
	\subfigimg[width=0.85\textwidth]{}{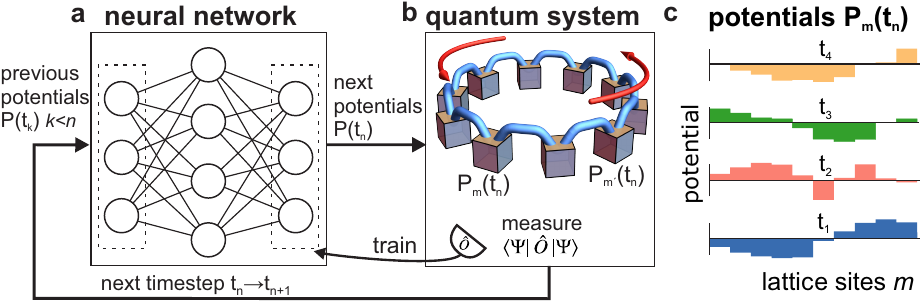}
	\caption{\idg{a,b} Deep reinforcement learning to optimize generation of  quantum states of currents. The quantum system is a ring-shaped circuit comprising $L$ lumped parameters schematized as lattice sites. The wavefunction resides at site $m$ with a local potential $P_m(t)$ that can be changed in time. The full control protocol (FCP) adjusts the potential at all sites individually in discrete timesteps $t_n$ to generate a current efficiently. To optimize the protocol, the neural network takes the potentials $P(t< t_n)$ at earlier timesteps and returns the potential $P_m(t_{n})$ to be applied to the quantum system in the next timestep. A measure $\bra{\Psi}\hat{O}\ket{\Psi}$ given by observable $\hat{O}$ is measured and used to train the neural network. This process is repeated until convergence. For further details see appendix section F.  \idg{c} Example driving potential found by the optimization algorithm.}
	\label{NN}
\end{figure*}

With the advent of quantum technologies, new forms of quantum circuits have emerged.
The architecture and  circuit performance depends on the specific physical implementation and the type of 'quantum fluid'  operating in the quantum network.  
\revE{Prominent candidates are atomtronic circuits involving neutral matterwaves
of cold atoms in optically generated structures with micrometric
resolution~\cite{Amico_NJP}. Other examples range from electronic and superconducting
circuits~\cite{barends2015digital} based on charged matter-wave on nanolithography
to photonic circuits employing photons in fiberoptics~\cite{politi2008silica}.}
These systems allow for precise control over the circuit properties such as interactions or particle statistics (fermions/bosons). In addition, it is possible with the latest achievements in the field, \revD{particularly in Atomtronics}, to dynamically adjust the spatial features of the circuit locally, {while avoiding cross-talk effects~} 
\cite{muldoon2012control,Boshier_painting,rubinsztein2016roadmap}. Finally,  quantum circuits with increasingly complex architecture and  hybrid systems, in which different technologies are  interfaced, are at a mature stage of technology readiness  level~\cite{kurizki2015quantum,aspuru2012photonic,houck2012chip,bloch2012quantum,gauthier2019quantitative}. \revD{Although important for the very definition of the quantum circuits, the generation and control of current states remain a difficult task to achieve.
Particularly for cold atoms quantum technology, 
matter wave currents have been imparted so far only  in continuous atomic rings \cite{wright2013driving}. In fact,  currents in even a simple circuit made of a lattice ring has never been achieved. Such problem  represents a  well known bottleneck in the field,  particularly urgent for the progress of Atomtronics in which neutral matter currents  are needed to flow in complex networks \cite{haug2019andreev,haug2019aharonov}. }
 
\revD{In the present work, we demonstrate that the problem can be overcome by applying machine learning.  In this way, we can engineer fast and high fidelity current states in circuits with lumped parameters.}

Machine learning with deep reinforcement learning (RL) has been recently recognized as a powerful tool to engineer dynamics in quantum systems~\cite{Carleo602,bukov2018reinforcement,niu2019universal,day2019glassy,zhang2019does,xu2019generalizable,RevModPhys.91.045002}. 
Here, we guide quantum systems by  reconfiguring deep RL protocols that have trained artificial intelligence agents to master complex decision processes~\cite{schulman2017proximal}.
We  demonstrate this approach to prepare  quantum current  states describing the flow of coherent matter-wave in closed circuits: persistent currents~\cite{eckern1992persistent}. 
Persistent currents are a direct manifestation of  the phenomenon of  quantum coherence and are therefore of central  interest in fundamental aspects of  many-body physics like superfluidity, superconductivity  and mesoscopic physics~\cite{akkermans2007mesoscopic}. 
\revE{At the same time, such concepts play a vital role for important emerging applications such as rotation sensors based on guided matter waves in atomtronic circuits~\cite{Ryu2013} which employ similar ideas to their superconducting counterpart the Superconducting Quantum Interference Devices (SQUID's)~\cite{tinkham2004introduction}.}
Persistent currents have been the object of  intense studies in different  contexts of quantum technologies like cold atoms~\cite{wright2013driving,pandey2019hypersonic,guo2020supersonic}, superconducting circuits~\cite{roushan2017chiral}, optical cavities~\cite{metelmann2015nonreciprocal},  opto-mechanical cavities~\cite{fang2017generalized} and tailored reservoirs~\cite{keck2018persistent}.

While  charged or neutral  matter-wave persistent currents  have been obtained in simplified situations, protocols for more general settings are still missing. 
We consider quantum systems that can be mapped onto ring-shaped circuits comprising local units that we call sites.  These systems  encompass  most of the general features and challenging aspects  for the generation of quantum currents in closed circuits.  We propose to create current states by locally driving the circuit parameters. If the system can be  driven by changing  few control parameters, state engineering can be carried out through optimal control theory~\cite{jirari2009optimal}.  For increasingly large number  of parameters, however,  the circuit driving cannot be handled with standard means. In this work, we employ deep RL  to implement  current state engineering  by driving  each 
lattice site of the ring circuit  independently.  With our approach, we demonstrate that persistent currents with specific winding number can be imparted,  on a timescale that is much shorter than other known protocols. Additionally, we can create entangled current states of up to three winding numbers, for which there is no known protocol. Our protocols can be readily trained and applied on experiments. 
\begin{figure*}[htbp]
	\centering
	\subfigimg[width=0.3\textwidth]{a}{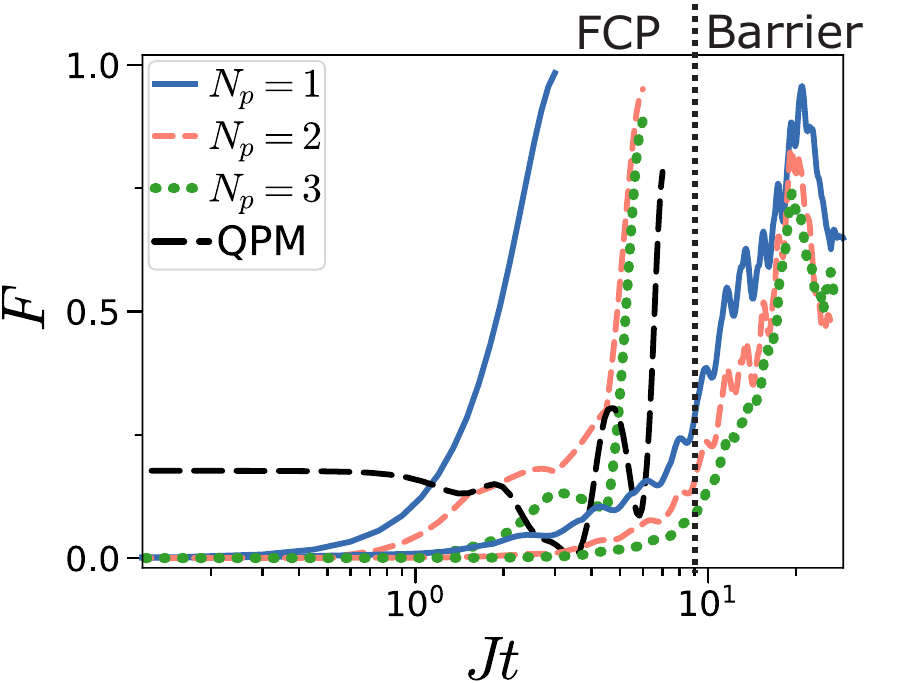}
	\subfigimg[width=0.29\textwidth]{b}{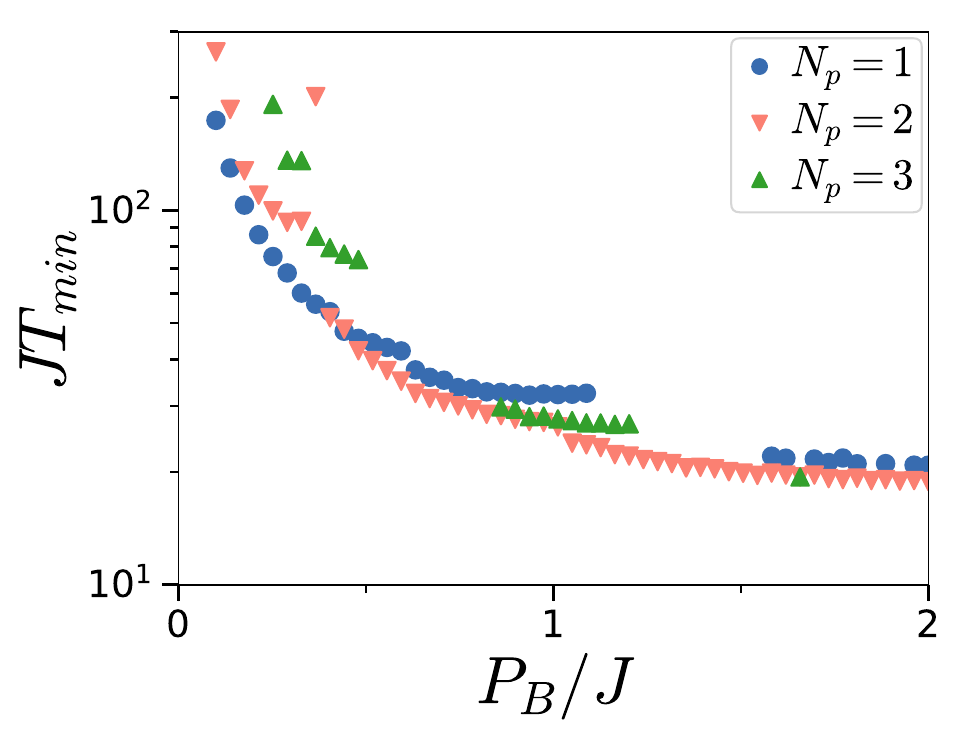}
	\subfigimg[width=0.265\textwidth]{c}{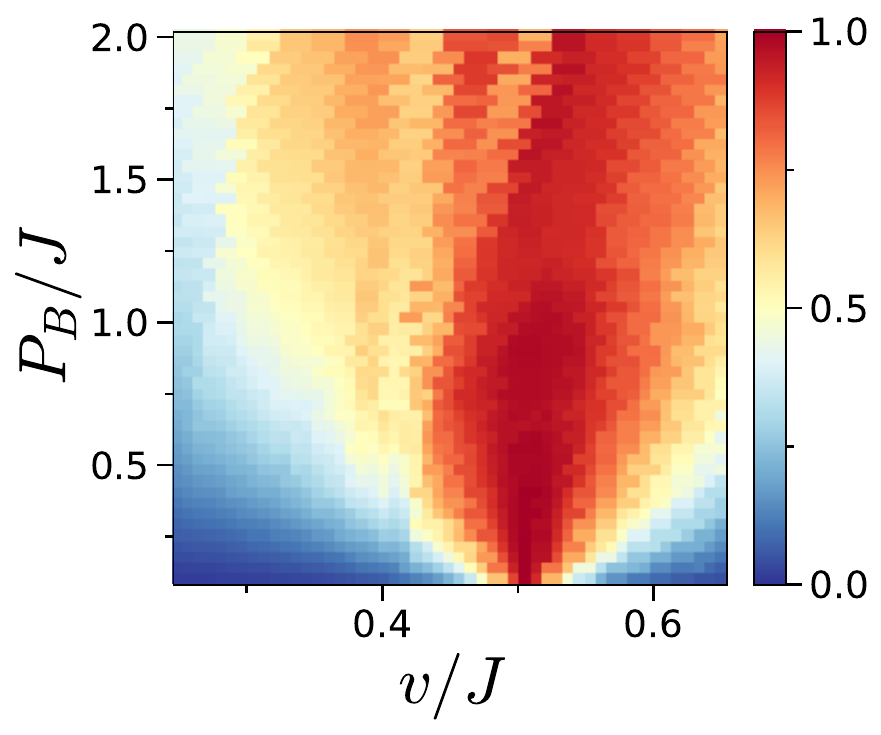}
	\caption{
		\idg{a} Generating currents by driving the potential of a ring circuit. We plot the time evolution of the fidelity ${F=\abs{\braket{\Psi(T)}{\Psi_\text{target}}}^2}$ obtained with two different protocols, comparing different numbers of particles (Bose-Hubbard model Eq. \ref{HamiltonSource} with $L=12$ sites, $U=J$,  target state  $|\Psi_\text{target}\rangle=\ket{k=1}^{\otimes N_\text{p}}$). The first protocol (Barrier) stirs the wavefunction by moving a single-site barrier ($P_\text{B}=2J$ for $N_\text{P}=1,2$, $P_\text{B}=1.6J$ for $N_\text{P}=3$ ) at a constant speed $v=0.53J$ (see curves on the right of the vertical dotted line), with Bose-Hubbard model ($L=12$ ring sites, interaction $U=J$). The second protocol (full control protocol: FCP) is fully controlling  the potential at every lattice site in time (see curves on the left of the vertical dotted line) with $N_\text{T}=6$ control steps: Solid blue curves $N_\text{p}=1$; Dashed red line: $N_\text{p}=2$; Dotted green curves: $N_\text{p}=3$. Limit of large particle number with quantum phase model (QPM) Eq.\ref{Eq:QP} with $L=7$, $U=3J_\text{E}$, and target state $|\Psi_\text{target}\rangle=|\Psi_\text{QP}(\Phi_1)\rangle$ (see appendix section A): Long-dashed black curve.
		\idg{b} Minimal time $T_\text{min}$ required to create rotational states above a threshold fidelity ($F_\text{min}=0.95$ for $N_\text{p}=1$, \revD{for more elusive higher particle number states $F_\text{min}=0.85$}) for different values of barrier amplitude $P_\text{B}$. 
		\idg{c} Maximal fidelity achieved when rotating barrier with amplitude $P_\text{B}$ and speed $v$ for $N_\text{p}=1$ particles. We find that the best rotation speed of barrier is at ${v\approx0.5J}$. 
	}
	\label{BarrierML}
\end{figure*}

{\em Local drive of Bose-Hubbard ring circuits.}
As sketched in Fig.~\ref{NN}b, our model system is a ring circuit comprising of $L$ sites, a natural architecture to consider to generate persistent currents, described with the Bose-Hubbard Hamiltonian. $N_\text{p}$ interacting  bosonic particles are filling the ring lattice and can hop between nearest-neighbor sites $j$ and ${j+1}$ with an amplitude $J$, and interact on-site with each other with strength $U$. The ring lattice can be locally driven by externally varying in time each on-site potential $P_j(t)$.
\begin{equation}\label{HamiltonSource}
\mathcal{H}_\text{BH}=\sum_{j=1}^L\Big[ -J \, ( \cn{a}{j}\an{a}{j+1} + \cn{a}{j+1}\an{a}{j})+\ P_j(t) \, \nn{}{j}+ \frac{U}{2}\nn{}{j}(\nn{}{j}-1)\Big].
\end{equation}
Here $\an{a}{j}$, $\cn{a}{j}$ and ${\nn{}{j}=\cn{a}{j}\an{a}{j}}$ are the usual bosonic creation, annihilation and number operators on site $j$, satisfying the commutation relation ${[\an{a}{i},\cn{a}{j}] = \delta_{ij}}$ and periodic boundary conditions $\cn{a}{L+1}=\cn{a}{1}$. 


In the limit of a large average number of particles per site ${N_s=N_\text{p}/L\gg1}$,
the Bose-Hubbard Hamiltonian effectively reduces to the so-called quantum phase model (QPM)
\begin{equation} 
\mathcal{H}_\text{QP}= \sum_{j=1}^L \Big[ -2 J_\text{E} \, \cos (\hat{\phi}_j-\hat{\phi}_{j+1} ) + P_j(t) \, \hat{Q}_j + \frac{U}{2} \hat{Q}^2_j \Big], 
\label{Eq:QP}
\end{equation}
where 
$J_\text{E}=J N_\text{s}$, $\hat{Q}_j = \nn{}{j}-N_\text{s}$ is the on-site particle number fluctuations and  $\phi_j$ the phase operators~\cite{garcia2004variational,fazio2001quantum}. The operators satisfy the commutation relations ${[\hat{\phi}_i,\hat{Q}_j] = \mathrm{i}\hbar \delta_{ij}}$.
Hamiltonians (\ref{HamiltonSource}) and (\ref{Eq:QP}) describe a wide class of different physical quantum systems ranging from $1d$ arrays of Josephson junctions and qubits~\cite{ye2019propagation} to atomtronic circuits.


{\em Quantum current states.} 
In a coherent quantum circuit the current states in the ring are quantized as the phase along a closed path can only change by integer multiples of $2\pi$.  We describe these winding numbers $k$ by defining the single-particle winding state $\ket{k} \doteq \cn{b}{k}\ket{\text{vac}}$, where $\ket{\text{vac}}$ denotes the vacuum state and $\cn{b}{k}=\frac{1}{\sqrt{L}}\sum_n \expU{i2\pi k n/L}\cn{a}{n}$ the quasi-momentum creation operator (details in appendix section A). In a ring system, the quasi-momentum corresponds to the winding number \revE{and therefore the quantized current in the ring}.
We choose  $\Omega=\{k_1,k_2,\dots,k_{N_\text{C}}\}$ as a set of $N_\text{C}$ winding numbers that we want to prepare in an entangled superposition state. 
The generation of such quantum current states is one of the defining goals of quantum technology and are notoriously difficult to generate. 
We consider states in the form $\ket{\Psi_{EC}}=\frac{1}{\sqrt{N_\text{C}}}\sum_{k \in \Omega}\expU{i\phi_k}\ket{k}^{\otimes N_\text{p}}$ consisting of $N_\text{C}$ winding numbers {with arbitrary phase $\phi_k$}. Important  examples of entangled current states that we will specifically consider in the present work are  the NOON-state ($N_\text{C}=2$) and the W-state ($N_\text{C}=3$).
We characterize the ability of our protocols to generate these states with the fidelity ${F=\abs{\braket{\Psi(T)}{\Psi_\text{EC}}}^2}$. 
\revD{We also refer  to a  certification measure that is related to observables  in cold atoms settings (time-of-flight measurements)~\cite{bloch2008many,haug2018readout}} and reflects the behaviour of the  fidelity: $W_{\Psi} \propto\prod_{k \in \Omega} \bra{\Psi}\nn{}{k}\nn{}{k}\ket{\Psi}$ (see appendix section B).
{
}

We introduce the full control protocol (FCP) to generate currents efficiently. \revD{In such protocol, the  potential at each lattice site is driven  freely within a range ${\abs{P_j}<P_\text{max}}$.} The  total \revD{driving time} $T$ is discretized into $N_\text{T}$ timesteps of equal length $\Delta t=T/N_\text{T}$. 
Within each timestep, the system evolves under constant parameters and we assume that the potential parameters change instantaneously between two timesteps.
We use deep RL with proximal policy optimization using actor-critic method and implementation in Tensorflow to optimize the protocols~\cite{mnih2015human,schulman2017proximal,tensorflow2015-whitepaper} (Sketch of algorithm in Fig.~\ref{NN}, details in appendix section F).
Here, we note that  our scheme relies on a model-free optimization algorithm: The learning algorithm does not make any assumptions about the specific system or knows about quantum mechanics. 
As possible application, the algorithm could be supplied with experimental data (in our case the potential parameters and measurement observables) to optimize the experiment directly. 

\revE{We compare two fundamental approaches to generate quantum current states. First, we investigate the established method of stirring with a barrier, \revD{but here} applied to a discrete ring lattice. Then, we investigate the FCP protocol.}

{\em Stirring a localized barrier potential.} Transforming the non-rotating ground state to a specific rotating state requires perturbing the state in a manner that explicitly breaks time-reversal symmetry.  This has been implemented in cold atoms settings~\cite{wright2013driving,Ramanathan2011} {for continuous systems, however has not been investigated for discrete settings such as a ring lattice.}   Here, we move a potential barrier initially localized at one site of amplitude $P_\text{B}$ to the next site at a fixed frequency $v$. We target generating current states with one winding number, starting from the ground state of the system. We find that a high fidelity is reached by driving the ring for a time $T\approx 20/J$ for different particle numbers $N_\text{p}$ (rightmost curves in Fig.~\ref{BarrierML}a).  Matching calculations for continuous systems, we find that  the first rotational state is created best by stirring with about a speed of half the desired atom velocity $v\approx0.5/J$ (see Fig.~\ref{BarrierML}d, \cite{schenke2011nonadiabatic}). 
\revE{In a rotational invariant system,  the energies  for consecutive winding numbers $k_1,k_2$ are degenerate when the barrier is driven at the mean speed $(v_2-v_1)/2$. A localized  barrier splits the degeneracy and introduces non-adiabatic transitions (Landau-Zener) between the two states. A similar mechanism works also for a ring condensate interrupted by three barriers~\cite{aghamalyan2016atomtronic}.}

The actual implementation of the barrier protocol implies a trade-off between achieving high fidelity $F$ and short protocol time $T$:  With increasing $P_\text{B}$ the maximal achievable fidelity decreases, however it is reached in a shorter time $T$ (see Fig.~\ref{BarrierML}c). 
To shorten the protocol run time without sacrificing fidelity, more complex protocols are required. 

\begin{figure}[htbp]
	\centering
	\subfigimg[width=0.24\textwidth]{a}{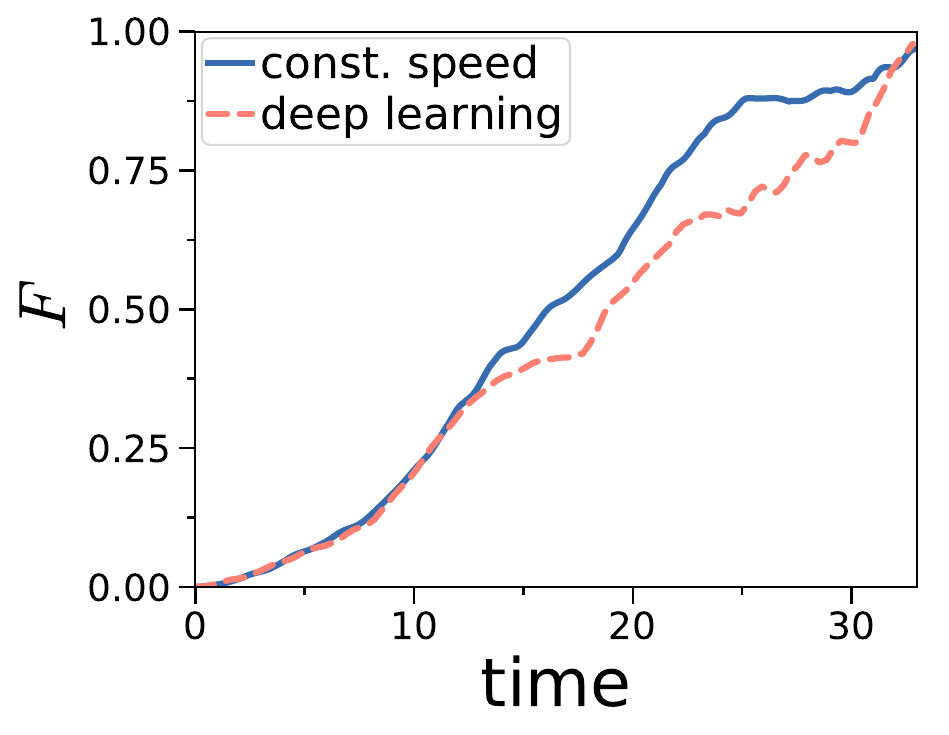}\hfill
	\subfigimg[width=0.24\textwidth]{b}{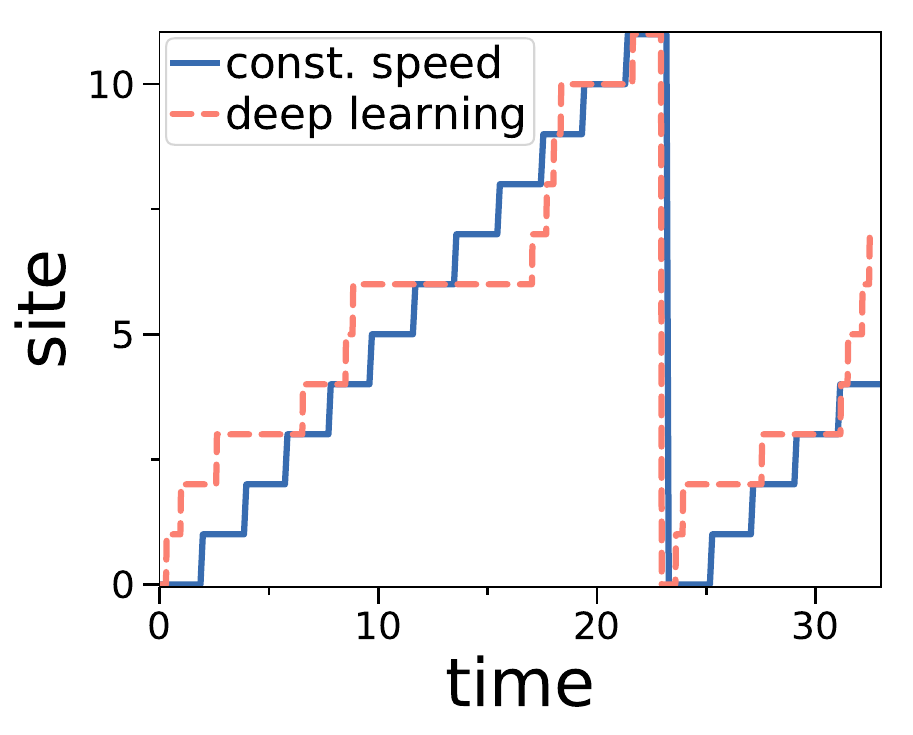}
	\caption{\revE{Current generation by moving a single delta-like potential (with constant amplitude $P=J$). We compare two protocols: Moving potential at constant speed $v=0.52J$ (solid line) or optimize with deep RL (dashed line). Over $n_t=100$ timesteps, the RL agent can either move the barrier by one site, or keep it at the current position, trained  over 50000 epochs. \idg{a} Fidelity of best found protocol to create phase winding state $\Omega=1$ for ${N_p=1}$. \idg{b} Barrier position in time.} }
	\label{BarrierComp}
\end{figure}

\revE{As a benchmark, we apply the deep learning algorithm on a setting with simple instructions  (see Fig.~\ref{BarrierComp}).  The ring is initialized with a single barrier with strength $P=J$. The neural network can at each time-step choose to either move the barrier forward by one site, or keep the barrier at the current position. \revD{This way, we see} that  the deep RL  is able to find a  protocol with nearly constant velocity (dashed line Fig.\ref{BarrierComp}b), that is very similar to the analytic solution (solid line Fig.\ref{BarrierComp}b). This demonstrate the ability of the learning algorithm to re-construct the known solution by starting from a \revD{general control scheme.}}

{\em Local control of the circuit}.
\revE{To achieve a higher control over the dynamics of the quantum system, we now apply the FCP in which  each lattice site is driven individually by varying the potential in a continuous fashion between ${-J<P_j<J}$.}
In Fig.~\ref{BarrierML}a, we compare stirring (right-hand side) and FCP (left-hand side). 
The FCP reaches a better fidelity than the barrier driving protocol, requiring only half the time or less. An example protocol that optimizes the state generation is shown in Fig.~\ref{NN}c.
\revD{It is very remarkable} that the FCP works also in the limit of many particles by employing the QPM Eq.(\ref{Eq:QP}) in the regime of intermediate interaction (see Fig.~\ref{BarrierML}a,b). 
\revE{In experimental settings, the driving potential is subject to fluctuations due to environmental noise and experimental imperfections. We find that the FCP scheme is robust to noise in the driving potentials to up to 20\% (see appendix section D).}

\begin{figure}[htbp]
	\centering
	\subfigimg[width=0.24\textwidth]{a}{EntFidelityAll1DComposedPPOInputRRMCED20TKCdtPPON1L12M0E2S0R1i0A2t0_1n6e0_002v0_5l0_0002U0O0M1m-1c6a2u200d2E120000G0_99w1b500o11O1P2u1I3.pdf}\hfill
	\subfigimg[width=0.24\textwidth]{b}{EntFidelityAll1DComposedPPOInputRRMCED23NGntimestepsPPON1L12M0E2S0R1i0A3t10_0n1e0_002v0_5l0_0002U0_0O0M1m-1c6a2u200d2E120000G0_99w1b500o1P2u1I3.pdf}\\
	\subfigimg[width=0.24\textwidth]{c}{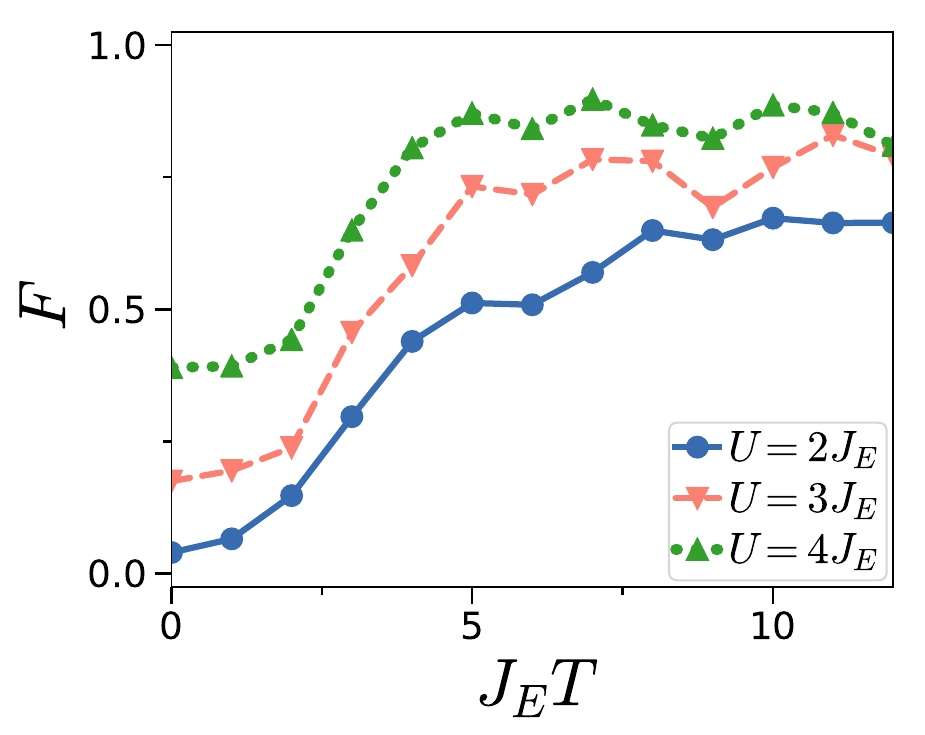}\hfill
	\subfigimg[width=0.24\textwidth]{d}{rewardAll1DComposedPPOInputRAAdtPPON1L12M0E2S0R3i0A0t0_017n6e0_0005v0_5l0_0005U0O1M1m0c6a6u100d2E50000G0_99w1b500P1u1A2I3.pdf}
	\caption{ 
	\idg{a-c} Full control with continuous control: Potential at every site can take arbitrary values between ${-J<P_j<J}$. Generation of entangled superpositions of current states of type $\ket{\text{EC}}=\frac{1}{\sqrt{N_\text{C}}}\sum_{k \in \Omega}\ket{k}^{\otimes N_\text{p}}$ of a set of winding numbers ${\Omega=\{k_1,k_2,\dots\,k_{N_\text{C}}\}}$ using deep RL.
		\idg{a} fidelity as a function of protocol time $T$ for $N_\text{p}=2$ particles, $N_\text{T}=6$ timesteps and ${U=J}$
		\idg{b} fidelity for varying timesteps $N_\text{T}$ ($N_\text{p}=2$, ${U=J}$, $\Omega=\{0,1\}$). 
		\idg{c} fidelity for the limit of large particle number (QPM) with FCP protocol for $N_\text{T} = 6$, $L=7$, $\abs{P_m(t)} \le J_E$ and limiting number fluctuations at $\Delta \hat{Q}_m \leq 2$ and target state $|\Psi_\text{target}\rangle=|\Psi_\text{QP}(\Phi_1)\rangle$. The ground state of the QPM has a broad winding number distribution and thus there is a finite initial fidelity at $T=0$. 
		\idg{d} Full control protocol with discretized amplitudes: Local potential can take either the value $P_j(t)=0$ or $P_j(t)=J$.  Fidelity to create phase winding state $\Omega=1$ for ${N_p=1}$, for varying protocol time $T$ and number timesteps $N_\text{T}$. }
	\label{ParamGraphs}
\end{figure}

To go  beyond quantum current states composed of a single winding number, we employ FCP to engineer entangled superposition of winding numbers. This way, we demonstrate the preparation of entangled superposition states of different currents for up to three winding numbers for which we are not aware of any protocol for their generation (see Fig.~\ref{ParamGraphs}a). The fidelity improves over protocol time $T$ and reaches eventually a plateau.
A minimal number of protocol timesteps are required to reach sufficient fidelity. For interacting systems, we achieve best results for $N_\text{T}\ge4$ (Fig.~\ref{ParamGraphs}b). \revE{For the QPM, higher fidelity can be achieved with increasing interaction $U$, however the initial fidelity is higher as well as the momentum distribution of the ground state becomes broader (Fig.~\ref{ParamGraphs}c).}  \revD{Further data and comparison with the alternative optimization method GRAPE in appendix section C and D \cite{johansson2013qutip,machnes2011comparing}.}
\revD{Finally, we investigate protocols with discretized control amplitudes in Fig.~\ref{ParamGraphs}d. The driving amplitude at every site can take only the values $P_j=0$ or $P_j=J$ and have similar fidelity as the continuous case. They give rise to simplistic driving protocols (see appendix section E).}

{\em Discussion.}
In this work we demonstrated how to use machine learning for the efficient generation of currents in closed quantum circuits. 
The essential features of this problem are captured by a ring consisting of lumped elements, which we schematize as lattice sites and can be modeled using Bose-Hubbard and quantum phase model, paradigmatic frameworks for the physics of cold atoms, superconducting circuits and photonic waveguides. 

\revD{We introduced the Full Control Protocol (FCP), where all sites of the lattice are driven individually. This protocol paves the way for enhanced  control over the many-body system. Inspired by recent advances in deep RL, we use neural networks to learn these complex driving protocols. In contrast to other optimisation methods like GRAPE, deep RL is  agnostic of the underlying physical system as a kind of hybrid quantum-classical optimizer~\cite{wigley2016fast,otterbach2017unsupervised,kokail2019self}. Thus, it can directly improve experiments, optimizing the fidelity with an experimental observable measure $W_\Psi$ as input to the learning algorithm, while achieving comparable results as with GRAPE (see appendix section B).
We envision that first the time-consuming training of the protocol is performed on the experiment, after which the protocol can be readily applied whenever a specific current state is needed.}
	


\revD{We benchmarked our algorithm with the stirring protocol, the standard protocol used for cold atom technology \cite{wright2013driving}.} 
\revD{Remarkably, deep learning is able to re-discover the established protocol.}
With the FCP, we show how to generate current states consisting of a single winding number more than twice as fast and with higher fidelity compared to the standard stirring scheme used so far by the cold atom community - Fig.~\ref{BarrierML}a. 
\revD{Furthermore, we show how to produce entangled current states such as NOON states as well as W-type states involving three winding numbers, for which no protocol has been known so far (see Fig.~\ref{ParamGraphs}). \revD{This  scheme is robust to noise in the driving parameter to up to 20\% (see appendix section D).}
We find that the complexity of the driving protocol (protocol time and number of time steps) depends on the number of particles: Non-interacting systems can generate currents much faster and with simpler protocols compared to the interacting many-body system. Both small and large number of particles regimes (through Bose-Hubbard and quantum phase dynamics respectively) were explored.}

\revD{Our deep learning method can be also applied to problems with discretized driving amplitudes, where gradient based methods like GRAPE are difficult to apply \cite{bukov2018reinforcement,zhang2019does} (see Fig.~\ref{ParamGraphs}d, appendix section E). These protocols take quite a simplistic form, which we conjecture could be related to general classes of driving protocols for current generation.
Based on the physical mechanism behind the optimal stirring protocols  for a single barrier and three barriers (see the Landau-Zener argument above), it is tempting to conclude that the FCP generates the desired current state by looking at the self-avoiding crossing  in the energy landscape and by optimizing the transitions amplitudes between states with different winding numbers.}

\revD{Our findings  are of direct relevance in different contexts of quantum technology.  Especially for cold-atom quantum technology, our approach clearly shows a new path to solve  important challenges in realizing atomtronic circuits.  In particular,} we note that  our approach does not rely on  artificial gauge fields, which most other methods require to generate currents. This feature leads to a simplification of the experimental apparatus. Similarly, the FCP  could be exploited  in superconducting circuits to achieve controlled electronic currents. Our approach can be extended  to other quantum many-body systems, for current states  in more complex circuit geometries and hybrid quantum networks, as well as for quantum-enhanced sensing~\cite{ragole2016interacting}.

\bigskip

\begin{acknowledgments}
 We thank D. Basko, V. Bastidas, M. Holzmann, D. Hutchinson,  A. Minguzzi, and B. Munro for discussions. The Grenoble LANEF framework (ANR-10-LABX-51-01) is acknowledged for its support with mutualized infrastructure. The computational work for this article was partially performed on resources of the National Supercomputing Centre, Singapore (https://www.nscc.sg).
\end{acknowledgments}

\bibliography{library}

\begin{thebibliography}{53}%
\makeatletter
\providecommand \@ifxundefined [1]{%
 \@ifx{#1\undefined}
}%
\providecommand \@ifnum [1]{%
 \ifnum #1\expandafter \@firstoftwo
 \else \expandafter \@secondoftwo
 \fi
}%
\providecommand \@ifx [1]{%
 \ifx #1\expandafter \@firstoftwo
 \else \expandafter \@secondoftwo
 \fi
}%
\providecommand \natexlab [1]{#1}%
\providecommand \enquote  [1]{``#1''}%
\providecommand \bibnamefont  [1]{#1}%
\providecommand \bibfnamefont [1]{#1}%
\providecommand \citenamefont [1]{#1}%
\providecommand \href@noop [0]{\@secondoftwo}%
\providecommand \href [0]{\begingroup \@sanitize@url \@href}%
\providecommand \@href[1]{\@@startlink{#1}\@@href}%
\providecommand \@@href[1]{\endgroup#1\@@endlink}%
\providecommand \@sanitize@url [0]{\catcode `\\12\catcode `\$12\catcode
  `\&12\catcode `\#12\catcode `\^12\catcode `\_12\catcode `\%12\relax}%
\providecommand \@@startlink[1]{}%
\providecommand \@@endlink[0]{}%
\providecommand \url  [0]{\begingroup\@sanitize@url \@url }%
\providecommand \@url [1]{\endgroup\@href {#1}{\urlprefix }}%
\providecommand \urlprefix  [0]{URL }%
\providecommand \Eprint [0]{\href }%
\providecommand \doibase [0]{http://dx.doi.org/}%
\providecommand \selectlanguage [0]{\@gobble}%
\providecommand \bibinfo  [0]{\@secondoftwo}%
\providecommand \bibfield  [0]{\@secondoftwo}%
\providecommand \translation [1]{[#1]}%
\providecommand \BibitemOpen [0]{}%
\providecommand \bibitemStop [0]{}%
\providecommand \bibitemNoStop [0]{.\EOS\space}%
\providecommand \EOS [0]{\spacefactor3000\relax}%
\providecommand \BibitemShut  [1]{\csname bibitem#1\endcsname}%
\let\auto@bib@innerbib\@empty
\bibitem [{\citenamefont {Amico}\ \emph {et~al.}(2017)\citenamefont {Amico},
  \citenamefont {Birkl}, \citenamefont {Boshier},\ and\ \citenamefont
  {Kwek}}]{Amico_NJP}%
  \BibitemOpen
  \bibfield  {author} {\bibinfo {author} {\bibfnamefont {Luigi}\ \bibnamefont
  {Amico}}, \bibinfo {author} {\bibfnamefont {Gerhard}\ \bibnamefont {Birkl}},
  \bibinfo {author} {\bibfnamefont {Malcolm}\ \bibnamefont {Boshier}}, \ and\
  \bibinfo {author} {\bibfnamefont {Leong-Chuan}\ \bibnamefont {Kwek}},\
  }\bibfield  {title} {\enquote {\bibinfo {title} {Focus on atomtronics-enabled
  quantum technologies},}\ }\href
  {http://stacks.iop.org/1367-2630/19/i=2/a=020201} {\bibfield  {journal}
  {\bibinfo  {journal} {New J. Phys.}\ }\textbf {\bibinfo {volume} {19}},\
  \bibinfo {pages} {020201} (\bibinfo {year} {2017})}\BibitemShut {NoStop}%
\bibitem [{\citenamefont {Barends}\ \emph {et~al.}(2015)\citenamefont
  {Barends}, \citenamefont {Lamata}, \citenamefont {Kelly}, \citenamefont
  {Garc{\'\i}a-{\'A}lvarez}, \citenamefont {Fowler}, \citenamefont {Megrant},
  \citenamefont {Jeffrey}, \citenamefont {White}, \citenamefont {Sank},
  \citenamefont {Mutus} \emph {et~al.}}]{barends2015digital}%
  \BibitemOpen
  \bibfield  {author} {\bibinfo {author} {\bibfnamefont {R}~\bibnamefont
  {Barends}}, \bibinfo {author} {\bibfnamefont {L}~\bibnamefont {Lamata}},
  \bibinfo {author} {\bibfnamefont {J}~\bibnamefont {Kelly}}, \bibinfo {author}
  {\bibfnamefont {L}~\bibnamefont {Garc{\'\i}a-{\'A}lvarez}}, \bibinfo {author}
  {\bibfnamefont {AG}~\bibnamefont {Fowler}}, \bibinfo {author} {\bibfnamefont
  {A}~\bibnamefont {Megrant}}, \bibinfo {author} {\bibfnamefont
  {E}~\bibnamefont {Jeffrey}}, \bibinfo {author} {\bibfnamefont
  {TC}~\bibnamefont {White}}, \bibinfo {author} {\bibfnamefont {D}~\bibnamefont
  {Sank}}, \bibinfo {author} {\bibfnamefont {JY}~\bibnamefont {Mutus}},  \emph
  {et~al.},\ }\bibfield  {title} {\enquote {\bibinfo {title} {Digital quantum
  simulation of fermionic models with a superconducting circuit},}\ }\href@noop
  {} {\bibfield  {journal} {\bibinfo  {journal} {Nat. Comm.}\ }\textbf
  {\bibinfo {volume} {6}},\ \bibinfo {pages} {7654} (\bibinfo {year}
  {2015})}\BibitemShut {NoStop}%
\bibitem [{\citenamefont {Politi}\ \emph {et~al.}(2008)\citenamefont {Politi},
  \citenamefont {Cryan}, \citenamefont {Rarity}, \citenamefont {Yu},\ and\
  \citenamefont {O'brien}}]{politi2008silica}%
  \BibitemOpen
  \bibfield  {author} {\bibinfo {author} {\bibfnamefont {Alberto}\ \bibnamefont
  {Politi}}, \bibinfo {author} {\bibfnamefont {Martin~J}\ \bibnamefont
  {Cryan}}, \bibinfo {author} {\bibfnamefont {John~G}\ \bibnamefont {Rarity}},
  \bibinfo {author} {\bibfnamefont {Siyuan}\ \bibnamefont {Yu}}, \ and\
  \bibinfo {author} {\bibfnamefont {Jeremy~L}\ \bibnamefont {O'brien}},\
  }\bibfield  {title} {\enquote {\bibinfo {title} {Silica-on-silicon waveguide
  quantum circuits},}\ }\href@noop {} {\bibfield  {journal} {\bibinfo
  {journal} {Science}\ }\textbf {\bibinfo {volume} {320}},\ \bibinfo {pages}
  {646--649} (\bibinfo {year} {2008})}\BibitemShut {NoStop}%
\bibitem [{\citenamefont {Muldoon}\ \emph {et~al.}(2012)\citenamefont
  {Muldoon}, \citenamefont {Brandt}, \citenamefont {Dong}, \citenamefont
  {Stuart}, \citenamefont {Brainis}, \citenamefont {Himsworth},\ and\
  \citenamefont {Kuhn}}]{muldoon2012control}%
  \BibitemOpen
  \bibfield  {author} {\bibinfo {author} {\bibfnamefont {Cecilia}\ \bibnamefont
  {Muldoon}}, \bibinfo {author} {\bibfnamefont {Lukas}\ \bibnamefont {Brandt}},
  \bibinfo {author} {\bibfnamefont {Jian}\ \bibnamefont {Dong}}, \bibinfo
  {author} {\bibfnamefont {Dustin}\ \bibnamefont {Stuart}}, \bibinfo {author}
  {\bibfnamefont {Edouard}\ \bibnamefont {Brainis}}, \bibinfo {author}
  {\bibfnamefont {Matthew}\ \bibnamefont {Himsworth}}, \ and\ \bibinfo {author}
  {\bibfnamefont {Axel}\ \bibnamefont {Kuhn}},\ }\bibfield  {title} {\enquote
  {\bibinfo {title} {Control and manipulation of cold atoms in optical
  tweezers},}\ }\href@noop {} {\bibfield  {journal} {\bibinfo  {journal} {New
  J. Phys.}\ }\textbf {\bibinfo {volume} {14}},\ \bibinfo {pages} {073051}
  (\bibinfo {year} {2012})}\BibitemShut {NoStop}%
\bibitem [{\citenamefont {Henderson}\ \emph {et~al.}(2009)\citenamefont
  {Henderson}, \citenamefont {Ryu}, \citenamefont {MacCormick},\ and\
  \citenamefont {Boshier}}]{Boshier_painting}%
  \BibitemOpen
  \bibfield  {author} {\bibinfo {author} {\bibfnamefont {K}~\bibnamefont
  {Henderson}}, \bibinfo {author} {\bibfnamefont {C}~\bibnamefont {Ryu}},
  \bibinfo {author} {\bibfnamefont {C}~\bibnamefont {MacCormick}}, \ and\
  \bibinfo {author} {\bibfnamefont {M~G}\ \bibnamefont {Boshier}},\ }\bibfield
  {title} {\enquote {\bibinfo {title} {Experimental demonstration of painting
  arbitrary and dynamic potentials for bose-einstein condensates},}\ }\href
  {http://stacks.iop.org/1367-2630/11/i=4/a=043030} {\bibfield  {journal}
  {\bibinfo  {journal} {New J. Phys.}\ }\textbf {\bibinfo {volume} {11}},\
  \bibinfo {pages} {043030} (\bibinfo {year} {2009})}\BibitemShut {NoStop}%
\bibitem [{\citenamefont {Rubinsztein-Dunlop}\ \emph
  {et~al.}(2016)\citenamefont {Rubinsztein-Dunlop}, \citenamefont {Forbes},
  \citenamefont {Berry}, \citenamefont {Dennis}, \citenamefont {Andrews},
  \citenamefont {Mansuripur}, \citenamefont {Denz}, \citenamefont {Alpmann},
  \citenamefont {Banzer}, \citenamefont {Bauer} \emph
  {et~al.}}]{rubinsztein2016roadmap}%
  \BibitemOpen
  \bibfield  {author} {\bibinfo {author} {\bibfnamefont {Halina}\ \bibnamefont
  {Rubinsztein-Dunlop}}, \bibinfo {author} {\bibfnamefont {Andrew}\
  \bibnamefont {Forbes}}, \bibinfo {author} {\bibfnamefont {Michael~V}\
  \bibnamefont {Berry}}, \bibinfo {author} {\bibfnamefont {Mark~R}\
  \bibnamefont {Dennis}}, \bibinfo {author} {\bibfnamefont {David~L}\
  \bibnamefont {Andrews}}, \bibinfo {author} {\bibfnamefont {Masud}\
  \bibnamefont {Mansuripur}}, \bibinfo {author} {\bibfnamefont {Cornelia}\
  \bibnamefont {Denz}}, \bibinfo {author} {\bibfnamefont {Christina}\
  \bibnamefont {Alpmann}}, \bibinfo {author} {\bibfnamefont {Peter}\
  \bibnamefont {Banzer}}, \bibinfo {author} {\bibfnamefont {Thomas}\
  \bibnamefont {Bauer}},  \emph {et~al.},\ }\bibfield  {title} {\enquote
  {\bibinfo {title} {Roadmap on structured light},}\ }\href@noop {} {\bibfield
  {journal} {\bibinfo  {journal} {Journal of Optics}\ }\textbf {\bibinfo
  {volume} {19}},\ \bibinfo {pages} {013001} (\bibinfo {year}
  {2016})}\BibitemShut {NoStop}%
\bibitem [{\citenamefont {Kurizki}\ \emph {et~al.}(2015)\citenamefont
  {Kurizki}, \citenamefont {Bertet}, \citenamefont {Kubo}, \citenamefont
  {M{\o}lmer}, \citenamefont {Petrosyan}, \citenamefont {Rabl},\ and\
  \citenamefont {Schmiedmayer}}]{kurizki2015quantum}%
  \BibitemOpen
  \bibfield  {author} {\bibinfo {author} {\bibfnamefont {Gershon}\ \bibnamefont
  {Kurizki}}, \bibinfo {author} {\bibfnamefont {Patrice}\ \bibnamefont
  {Bertet}}, \bibinfo {author} {\bibfnamefont {Yuimaru}\ \bibnamefont {Kubo}},
  \bibinfo {author} {\bibfnamefont {Klaus}\ \bibnamefont {M{\o}lmer}}, \bibinfo
  {author} {\bibfnamefont {David}\ \bibnamefont {Petrosyan}}, \bibinfo {author}
  {\bibfnamefont {Peter}\ \bibnamefont {Rabl}}, \ and\ \bibinfo {author}
  {\bibfnamefont {J{\"o}rg}\ \bibnamefont {Schmiedmayer}},\ }\bibfield  {title}
  {\enquote {\bibinfo {title} {Quantum technologies with hybrid systems},}\
  }\href@noop {} {\bibfield  {journal} {\bibinfo  {journal} {Proc. Natl. Acad.
  Sci. U.S.A.}\ }\textbf {\bibinfo {volume} {112}},\ \bibinfo {pages}
  {3866--3873} (\bibinfo {year} {2015})}\BibitemShut {NoStop}%
\bibitem [{\citenamefont {Aspuru-Guzik}\ and\ \citenamefont
  {Walther}(2012)}]{aspuru2012photonic}%
  \BibitemOpen
  \bibfield  {author} {\bibinfo {author} {\bibfnamefont {Al{\'a}n}\
  \bibnamefont {Aspuru-Guzik}}\ and\ \bibinfo {author} {\bibfnamefont {Philip}\
  \bibnamefont {Walther}},\ }\bibfield  {title} {\enquote {\bibinfo {title}
  {Photonic quantum simulators},}\ }\href@noop {} {\bibfield  {journal}
  {\bibinfo  {journal} {Nat. Phys.}\ }\textbf {\bibinfo {volume} {8}},\
  \bibinfo {pages} {285} (\bibinfo {year} {2012})}\BibitemShut {NoStop}%
\bibitem [{\citenamefont {Houck}\ \emph {et~al.}(2012)\citenamefont {Houck},
  \citenamefont {T{\"u}reci},\ and\ \citenamefont {Koch}}]{houck2012chip}%
  \BibitemOpen
  \bibfield  {author} {\bibinfo {author} {\bibfnamefont {Andrew~A}\
  \bibnamefont {Houck}}, \bibinfo {author} {\bibfnamefont {Hakan~E}\
  \bibnamefont {T{\"u}reci}}, \ and\ \bibinfo {author} {\bibfnamefont {Jens}\
  \bibnamefont {Koch}},\ }\bibfield  {title} {\enquote {\bibinfo {title}
  {On-chip quantum simulation with superconducting circuits},}\ }\href@noop {}
  {\bibfield  {journal} {\bibinfo  {journal} {Nat. Phys.}\ }\textbf {\bibinfo
  {volume} {8}},\ \bibinfo {pages} {292} (\bibinfo {year} {2012})}\BibitemShut
  {NoStop}%
\bibitem [{\citenamefont {Bloch}\ \emph {et~al.}(2012)\citenamefont {Bloch},
  \citenamefont {Dalibard},\ and\ \citenamefont
  {Nascimbene}}]{bloch2012quantum}%
  \BibitemOpen
  \bibfield  {author} {\bibinfo {author} {\bibfnamefont {Immanuel}\
  \bibnamefont {Bloch}}, \bibinfo {author} {\bibfnamefont {Jean}\ \bibnamefont
  {Dalibard}}, \ and\ \bibinfo {author} {\bibfnamefont {Sylvain}\ \bibnamefont
  {Nascimbene}},\ }\bibfield  {title} {\enquote {\bibinfo {title} {Quantum
  simulations with ultracold quantum gases},}\ }\href@noop {} {\bibfield
  {journal} {\bibinfo  {journal} {Nat. Phys.}\ }\textbf {\bibinfo {volume}
  {8}},\ \bibinfo {pages} {267} (\bibinfo {year} {2012})}\BibitemShut {NoStop}%
\bibitem [{\citenamefont {Gauthier}\ \emph {et~al.}(2019)\citenamefont
  {Gauthier}, \citenamefont {Szigeti}, \citenamefont {Reeves}, \citenamefont
  {Baker}, \citenamefont {Bell}, \citenamefont {Rubinsztein-Dunlop},
  \citenamefont {Davis},\ and\ \citenamefont
  {Neely}}]{gauthier2019quantitative}%
  \BibitemOpen
  \bibfield  {author} {\bibinfo {author} {\bibfnamefont {Guillaume}\
  \bibnamefont {Gauthier}}, \bibinfo {author} {\bibfnamefont {Stuart~S.}\
  \bibnamefont {Szigeti}}, \bibinfo {author} {\bibfnamefont {Matthew~T.}\
  \bibnamefont {Reeves}}, \bibinfo {author} {\bibfnamefont {Mark}\ \bibnamefont
  {Baker}}, \bibinfo {author} {\bibfnamefont {Thomas~A.}\ \bibnamefont {Bell}},
  \bibinfo {author} {\bibfnamefont {Halina}\ \bibnamefont
  {Rubinsztein-Dunlop}}, \bibinfo {author} {\bibfnamefont {Matthew~J.}\
  \bibnamefont {Davis}}, \ and\ \bibinfo {author} {\bibfnamefont {Tyler~W.}\
  \bibnamefont {Neely}},\ }\bibfield  {title} {\enquote {\bibinfo {title}
  {Quantitative acoustic models for superfluid circuits},}\ }\href {\doibase
  10.1103/PhysRevLett.123.260402} {\bibfield  {journal} {\bibinfo  {journal}
  {Phys. Rev. Lett.}\ }\textbf {\bibinfo {volume} {123}},\ \bibinfo {pages}
  {260402} (\bibinfo {year} {2019})}\BibitemShut {NoStop}%
\bibitem [{\citenamefont {Wright}\ \emph {et~al.}(2013)\citenamefont {Wright},
  \citenamefont {Blakestad}, \citenamefont {Lobb}, \citenamefont {Phillips},\
  and\ \citenamefont {Campbell}}]{wright2013driving}%
  \BibitemOpen
  \bibfield  {author} {\bibinfo {author} {\bibfnamefont {K~C}\ \bibnamefont
  {Wright}}, \bibinfo {author} {\bibfnamefont {R~B}\ \bibnamefont {Blakestad}},
  \bibinfo {author} {\bibfnamefont {Christopher~J}\ \bibnamefont {Lobb}},
  \bibinfo {author} {\bibfnamefont {William~D}\ \bibnamefont {Phillips}}, \
  and\ \bibinfo {author} {\bibfnamefont {Gretchen~K}\ \bibnamefont
  {Campbell}},\ }\bibfield  {title} {\enquote {\bibinfo {title} {Driving phase
  slips in a superfluid atom circuit with a rotating weak link},}\ }\href@noop
  {} {\bibfield  {journal} {\bibinfo  {journal} {Phys. Rev. Lett.}\ }\textbf
  {\bibinfo {volume} {110}},\ \bibinfo {pages} {025302} (\bibinfo {year}
  {2013})}\BibitemShut {NoStop}%
\bibitem [{\citenamefont {Haug}\ \emph
  {et~al.}(2019{\natexlab{a}})\citenamefont {Haug}, \citenamefont {Dumke},
  \citenamefont {Kwek},\ and\ \citenamefont {Amico}}]{haug2019andreev}%
  \BibitemOpen
  \bibfield  {author} {\bibinfo {author} {\bibfnamefont {Tobias}\ \bibnamefont
  {Haug}}, \bibinfo {author} {\bibfnamefont {Rainer}\ \bibnamefont {Dumke}},
  \bibinfo {author} {\bibfnamefont {Leong-Chuan}\ \bibnamefont {Kwek}}, \ and\
  \bibinfo {author} {\bibfnamefont {Luigi}\ \bibnamefont {Amico}},\ }\bibfield
  {title} {\enquote {\bibinfo {title} {Andreev-reflection and aharonov--bohm
  dynamics in atomtronic circuits},}\ }\href {\doibase
  10.1088/2058-9565/ab2e61} {\bibfield  {journal} {\bibinfo  {journal} {Quantum
  Sci. Technol.}\ }\textbf {\bibinfo {volume} {4}},\ \bibinfo {pages} {045001}
  (\bibinfo {year} {2019}{\natexlab{a}})}\BibitemShut {NoStop}%
\bibitem [{\citenamefont {Haug}\ \emph
  {et~al.}(2019{\natexlab{b}})\citenamefont {Haug}, \citenamefont {Heimonen},
  \citenamefont {Dumke}, \citenamefont {Kwek},\ and\ \citenamefont
  {Amico}}]{haug2019aharonov}%
  \BibitemOpen
  \bibfield  {author} {\bibinfo {author} {\bibfnamefont {Tobias}\ \bibnamefont
  {Haug}}, \bibinfo {author} {\bibfnamefont {Hermanni}\ \bibnamefont
  {Heimonen}}, \bibinfo {author} {\bibfnamefont {Rainer}\ \bibnamefont
  {Dumke}}, \bibinfo {author} {\bibfnamefont {Leong-Chuan}\ \bibnamefont
  {Kwek}}, \ and\ \bibinfo {author} {\bibfnamefont {Luigi}\ \bibnamefont
  {Amico}},\ }\bibfield  {title} {\enquote {\bibinfo {title} {Aharonov-bohm
  effect in mesoscopic bose-einstein condensates},}\ }\href {\doibase
  10.1103/PhysRevA.100.041601} {\bibfield  {journal} {\bibinfo  {journal}
  {Phys. Rev. A}\ }\textbf {\bibinfo {volume} {100}},\ \bibinfo {pages}
  {041601} (\bibinfo {year} {2019}{\natexlab{b}})}\BibitemShut {NoStop}%
\bibitem [{\citenamefont {Carleo}\ and\ \citenamefont
  {Troyer}(2017)}]{Carleo602}%
  \BibitemOpen
  \bibfield  {author} {\bibinfo {author} {\bibfnamefont {Giuseppe}\
  \bibnamefont {Carleo}}\ and\ \bibinfo {author} {\bibfnamefont {Matthias}\
  \bibnamefont {Troyer}},\ }\bibfield  {title} {\enquote {\bibinfo {title}
  {Solving the quantum many-body problem with artificial neural networks},}\
  }\href {\doibase 10.1126/science.aag2302} {\bibfield  {journal} {\bibinfo
  {journal} {Science}\ }\textbf {\bibinfo {volume} {355}},\ \bibinfo {pages}
  {602--606} (\bibinfo {year} {2017})},\ \Eprint
  {http://arxiv.org/abs/https://science.sciencemag.org/content/355/6325/602.full.pdf}
  {https://science.sciencemag.org/content/355/6325/602.full.pdf} \BibitemShut
  {NoStop}%
\bibitem [{\citenamefont {Bukov}\ \emph {et~al.}(2018)\citenamefont {Bukov},
  \citenamefont {Day}, \citenamefont {Sels}, \citenamefont {Weinberg},
  \citenamefont {Polkovnikov},\ and\ \citenamefont
  {Mehta}}]{bukov2018reinforcement}%
  \BibitemOpen
  \bibfield  {author} {\bibinfo {author} {\bibfnamefont {Marin}\ \bibnamefont
  {Bukov}}, \bibinfo {author} {\bibfnamefont {Alexandre~GR}\ \bibnamefont
  {Day}}, \bibinfo {author} {\bibfnamefont {Dries}\ \bibnamefont {Sels}},
  \bibinfo {author} {\bibfnamefont {Phillip}\ \bibnamefont {Weinberg}},
  \bibinfo {author} {\bibfnamefont {Anatoli}\ \bibnamefont {Polkovnikov}}, \
  and\ \bibinfo {author} {\bibfnamefont {Pankaj}\ \bibnamefont {Mehta}},\
  }\bibfield  {title} {\enquote {\bibinfo {title} {Reinforcement learning in
  different phases of quantum control},}\ }\href@noop {} {\bibfield  {journal}
  {\bibinfo  {journal} {Phys. Rev. X}\ }\textbf {\bibinfo {volume} {8}},\
  \bibinfo {pages} {031086} (\bibinfo {year} {2018})}\BibitemShut {NoStop}%
\bibitem [{\citenamefont {Niu}\ \emph {et~al.}(2019)\citenamefont {Niu},
  \citenamefont {Boixo}, \citenamefont {Smelyanskiy},\ and\ \citenamefont
  {Neven}}]{niu2019universal}%
  \BibitemOpen
  \bibfield  {author} {\bibinfo {author} {\bibfnamefont {Murphy~Yuezhen}\
  \bibnamefont {Niu}}, \bibinfo {author} {\bibfnamefont {Sergio}\ \bibnamefont
  {Boixo}}, \bibinfo {author} {\bibfnamefont {Vadim~N}\ \bibnamefont
  {Smelyanskiy}}, \ and\ \bibinfo {author} {\bibfnamefont {Hartmut}\
  \bibnamefont {Neven}},\ }\bibfield  {title} {\enquote {\bibinfo {title}
  {Universal quantum control through deep reinforcement learning},}\
  }\href@noop {} {\bibfield  {journal} {\bibinfo  {journal} {npj Quantum Inf.}\
  }\textbf {\bibinfo {volume} {5}},\ \bibinfo {pages} {33} (\bibinfo {year}
  {2019})}\BibitemShut {NoStop}%
\bibitem [{\citenamefont {Day}\ \emph {et~al.}(2019)\citenamefont {Day},
  \citenamefont {Bukov}, \citenamefont {Weinberg}, \citenamefont {Mehta},\ and\
  \citenamefont {Sels}}]{day2019glassy}%
  \BibitemOpen
  \bibfield  {author} {\bibinfo {author} {\bibfnamefont {Alexandre~GR}\
  \bibnamefont {Day}}, \bibinfo {author} {\bibfnamefont {Marin}\ \bibnamefont
  {Bukov}}, \bibinfo {author} {\bibfnamefont {Phillip}\ \bibnamefont
  {Weinberg}}, \bibinfo {author} {\bibfnamefont {Pankaj}\ \bibnamefont
  {Mehta}}, \ and\ \bibinfo {author} {\bibfnamefont {Dries}\ \bibnamefont
  {Sels}},\ }\bibfield  {title} {\enquote {\bibinfo {title} {Glassy phase of
  optimal quantum control},}\ }\href@noop {} {\bibfield  {journal} {\bibinfo
  {journal} {Phys. Rev. Lett.}\ }\textbf {\bibinfo {volume} {122}},\ \bibinfo
  {pages} {020601} (\bibinfo {year} {2019})}\BibitemShut {NoStop}%
\bibitem [{\citenamefont {Zhang}\ \emph {et~al.}(2019)\citenamefont {Zhang},
  \citenamefont {Wei}, \citenamefont {Asad}, \citenamefont {Yang},\ and\
  \citenamefont {Wang}}]{zhang2019does}%
  \BibitemOpen
  \bibfield  {author} {\bibinfo {author} {\bibfnamefont {Xiao-Ming}\
  \bibnamefont {Zhang}}, \bibinfo {author} {\bibfnamefont {Zezhu}\ \bibnamefont
  {Wei}}, \bibinfo {author} {\bibfnamefont {Raza}\ \bibnamefont {Asad}},
  \bibinfo {author} {\bibfnamefont {Xu-Chen}\ \bibnamefont {Yang}}, \ and\
  \bibinfo {author} {\bibfnamefont {Xin}\ \bibnamefont {Wang}},\ }\bibfield
  {title} {\enquote {\bibinfo {title} {When does reinforcement learning stand
  out in quantum control? a comparative study on state preparation},}\
  }\href@noop {} {\bibfield  {journal} {\bibinfo  {journal} {npj Quantum Inf.}\
  }\textbf {\bibinfo {volume} {5}},\ \bibinfo {pages} {1--7} (\bibinfo {year}
  {2019})}\BibitemShut {NoStop}%
\bibitem [{\citenamefont {Xu}\ \emph {et~al.}(2019)\citenamefont {Xu},
  \citenamefont {Li}, \citenamefont {Liu}, \citenamefont {Wang}, \citenamefont
  {Yuan},\ and\ \citenamefont {Wang}}]{xu2019generalizable}%
  \BibitemOpen
  \bibfield  {author} {\bibinfo {author} {\bibfnamefont {Han}\ \bibnamefont
  {Xu}}, \bibinfo {author} {\bibfnamefont {Junning}\ \bibnamefont {Li}},
  \bibinfo {author} {\bibfnamefont {Liqiang}\ \bibnamefont {Liu}}, \bibinfo
  {author} {\bibfnamefont {Yu}~\bibnamefont {Wang}}, \bibinfo {author}
  {\bibfnamefont {Haidong}\ \bibnamefont {Yuan}}, \ and\ \bibinfo {author}
  {\bibfnamefont {Xin}\ \bibnamefont {Wang}},\ }\bibfield  {title} {\enquote
  {\bibinfo {title} {Generalizable control for quantum parameter estimation
  through reinforcement learning},}\ }\href@noop {} {\bibfield  {journal}
  {\bibinfo  {journal} {Npj Quantum Inf.}\ }\textbf {\bibinfo {volume} {5}},\
  \bibinfo {pages} {1--8} (\bibinfo {year} {2019})}\BibitemShut {NoStop}%
\bibitem [{\citenamefont {Carleo}\ \emph {et~al.}(2019)\citenamefont {Carleo},
  \citenamefont {Cirac}, \citenamefont {Cranmer}, \citenamefont {Daudet},
  \citenamefont {Schuld}, \citenamefont {Tishby}, \citenamefont
  {Vogt-Maranto},\ and\ \citenamefont {Zdeborov\'a}}]{RevModPhys.91.045002}%
  \BibitemOpen
  \bibfield  {author} {\bibinfo {author} {\bibfnamefont {Giuseppe}\
  \bibnamefont {Carleo}}, \bibinfo {author} {\bibfnamefont {Ignacio}\
  \bibnamefont {Cirac}}, \bibinfo {author} {\bibfnamefont {Kyle}\ \bibnamefont
  {Cranmer}}, \bibinfo {author} {\bibfnamefont {Laurent}\ \bibnamefont
  {Daudet}}, \bibinfo {author} {\bibfnamefont {Maria}\ \bibnamefont {Schuld}},
  \bibinfo {author} {\bibfnamefont {Naftali}\ \bibnamefont {Tishby}}, \bibinfo
  {author} {\bibfnamefont {Leslie}\ \bibnamefont {Vogt-Maranto}}, \ and\
  \bibinfo {author} {\bibfnamefont {Lenka}\ \bibnamefont {Zdeborov\'a}},\
  }\bibfield  {title} {\enquote {\bibinfo {title} {Machine learning and the
  physical sciences},}\ }\href {\doibase 10.1103/RevModPhys.91.045002}
  {\bibfield  {journal} {\bibinfo  {journal} {Rev. Mod. Phys.}\ }\textbf
  {\bibinfo {volume} {91}},\ \bibinfo {pages} {045002} (\bibinfo {year}
  {2019})}\BibitemShut {NoStop}%
\bibitem [{\citenamefont {Schulman}\ \emph {et~al.}(2017)\citenamefont
  {Schulman}, \citenamefont {Wolski}, \citenamefont {Dhariwal}, \citenamefont
  {Radford},\ and\ \citenamefont {Klimov}}]{schulman2017proximal}%
  \BibitemOpen
  \bibfield  {author} {\bibinfo {author} {\bibfnamefont {John}\ \bibnamefont
  {Schulman}}, \bibinfo {author} {\bibfnamefont {Filip}\ \bibnamefont
  {Wolski}}, \bibinfo {author} {\bibfnamefont {Prafulla}\ \bibnamefont
  {Dhariwal}}, \bibinfo {author} {\bibfnamefont {Alec}\ \bibnamefont
  {Radford}}, \ and\ \bibinfo {author} {\bibfnamefont {Oleg}\ \bibnamefont
  {Klimov}},\ }\bibfield  {title} {\enquote {\bibinfo {title} {Proximal policy
  optimization algorithms},}\ }\href@noop {} {\bibfield  {journal} {\bibinfo
  {journal} {arXiv:1707.06347}\ } (\bibinfo {year} {2017})}\BibitemShut
  {NoStop}%
\bibitem [{\citenamefont {Eckern}\ and\ \citenamefont
  {Schmid}(1992)}]{eckern1992persistent}%
  \BibitemOpen
  \bibfield  {author} {\bibinfo {author} {\bibfnamefont {Ulrich}\ \bibnamefont
  {Eckern}}\ and\ \bibinfo {author} {\bibfnamefont {Albert}\ \bibnamefont
  {Schmid}},\ }\bibfield  {title} {\enquote {\bibinfo {title} {Persistent
  currents of single mesoscopic rings},}\ }\href@noop {} {\bibfield  {journal}
  {\bibinfo  {journal} {EPL (Europhysics Letters)}\ }\textbf {\bibinfo {volume}
  {18}},\ \bibinfo {pages} {457} (\bibinfo {year} {1992})}\BibitemShut
  {NoStop}%
\bibitem [{\citenamefont {Akkermans}\ and\ \citenamefont
  {Montambaux}(2007)}]{akkermans2007mesoscopic}%
  \BibitemOpen
  \bibfield  {author} {\bibinfo {author} {\bibfnamefont {Eric}\ \bibnamefont
  {Akkermans}}\ and\ \bibinfo {author} {\bibfnamefont {Gilles}\ \bibnamefont
  {Montambaux}},\ }\href@noop {} {\emph {\bibinfo {title} {Mesoscopic physics
  of electrons and photons}}}\ (\bibinfo  {publisher} {Cambridge university
  press},\ \bibinfo {year} {2007})\BibitemShut {NoStop}%
\bibitem [{\citenamefont {Ryu}\ \emph {et~al.}(2013)\citenamefont {Ryu},
  \citenamefont {Blackburn}, \citenamefont {Blinova},\ and\ \citenamefont
  {Boshier}}]{Ryu2013}%
  \BibitemOpen
  \bibfield  {author} {\bibinfo {author} {\bibfnamefont {C.}~\bibnamefont
  {Ryu}}, \bibinfo {author} {\bibfnamefont {P.~W.}\ \bibnamefont {Blackburn}},
  \bibinfo {author} {\bibfnamefont {A.~A.}\ \bibnamefont {Blinova}}, \ and\
  \bibinfo {author} {\bibfnamefont {M.~G.}\ \bibnamefont {Boshier}},\
  }\bibfield  {title} {\enquote {\bibinfo {title} {Experimental realization of
  josephson junctions for an atom squid},}\ }\href {\doibase
  10.1103/PhysRevLett.111.205301} {\bibfield  {journal} {\bibinfo  {journal}
  {Phys. Rev. Lett.}\ }\textbf {\bibinfo {volume} {111}},\ \bibinfo {pages}
  {205301} (\bibinfo {year} {2013})}\BibitemShut {NoStop}%
\bibitem [{\citenamefont {Tinkham}(2004)}]{tinkham2004introduction}%
  \BibitemOpen
  \bibfield  {author} {\bibinfo {author} {\bibfnamefont {Michael}\ \bibnamefont
  {Tinkham}},\ }\href@noop {} {\emph {\bibinfo {title} {Introduction to
  superconductivity}}}\ (\bibinfo  {publisher} {Courier Corporation},\ \bibinfo
  {year} {2004})\BibitemShut {NoStop}%
\bibitem [{\citenamefont {Pandey}\ \emph {et~al.}(2019)\citenamefont {Pandey},
  \citenamefont {Mas}, \citenamefont {Drougakis}, \citenamefont {Thekkeppatt},
  \citenamefont {Bolpasi}, \citenamefont {Vasilakis}, \citenamefont {Poulios},\
  and\ \citenamefont {von Klitzing}}]{pandey2019hypersonic}%
  \BibitemOpen
  \bibfield  {author} {\bibinfo {author} {\bibfnamefont {Saurabh}\ \bibnamefont
  {Pandey}}, \bibinfo {author} {\bibfnamefont {Hector}\ \bibnamefont {Mas}},
  \bibinfo {author} {\bibfnamefont {Giannis}\ \bibnamefont {Drougakis}},
  \bibinfo {author} {\bibfnamefont {Premjith}\ \bibnamefont {Thekkeppatt}},
  \bibinfo {author} {\bibfnamefont {Vasiliki}\ \bibnamefont {Bolpasi}},
  \bibinfo {author} {\bibfnamefont {Georgios}\ \bibnamefont {Vasilakis}},
  \bibinfo {author} {\bibfnamefont {Konstantinos}\ \bibnamefont {Poulios}}, \
  and\ \bibinfo {author} {\bibfnamefont {Wolf}\ \bibnamefont {von Klitzing}},\
  }\bibfield  {title} {\enquote {\bibinfo {title} {Hypersonic bose--einstein
  condensates in accelerator rings},}\ }\href@noop {} {\bibfield  {journal}
  {\bibinfo  {journal} {Nature}\ }\textbf {\bibinfo {volume} {570}},\ \bibinfo
  {pages} {205–209} (\bibinfo {year} {2019})}\BibitemShut {NoStop}%
\bibitem [{\citenamefont {Roushan}\ \emph {et~al.}(2017)\citenamefont
  {Roushan}, \citenamefont {Neill}, \citenamefont {Megrant}, \citenamefont
  {Chen}, \citenamefont {Babbush}, \citenamefont {Barends}, \citenamefont
  {Campbell}, \citenamefont {Chen}, \citenamefont {Chiaro}, \citenamefont
  {Dunsworth} \emph {et~al.}}]{roushan2017chiral}%
  \BibitemOpen
  \bibfield  {author} {\bibinfo {author} {\bibfnamefont {P}~\bibnamefont
  {Roushan}}, \bibinfo {author} {\bibfnamefont {C}~\bibnamefont {Neill}},
  \bibinfo {author} {\bibfnamefont {A}~\bibnamefont {Megrant}}, \bibinfo
  {author} {\bibfnamefont {Y}~\bibnamefont {Chen}}, \bibinfo {author}
  {\bibfnamefont {R}~\bibnamefont {Babbush}}, \bibinfo {author} {\bibfnamefont
  {R}~\bibnamefont {Barends}}, \bibinfo {author} {\bibfnamefont
  {B}~\bibnamefont {Campbell}}, \bibinfo {author} {\bibfnamefont
  {Z}~\bibnamefont {Chen}}, \bibinfo {author} {\bibfnamefont {B}~\bibnamefont
  {Chiaro}}, \bibinfo {author} {\bibfnamefont {A}~\bibnamefont {Dunsworth}},
  \emph {et~al.},\ }\bibfield  {title} {\enquote {\bibinfo {title} {Chiral
  ground-state currents of interacting photons in a synthetic magnetic
  field},}\ }\href@noop {} {\bibfield  {journal} {\bibinfo  {journal} {Nat.
  Phys.}\ }\textbf {\bibinfo {volume} {13}},\ \bibinfo {pages} {146--151}
  (\bibinfo {year} {2017})}\BibitemShut {NoStop}%
\bibitem [{\citenamefont {Metelmann}\ and\ \citenamefont
  {Clerk}(2015)}]{metelmann2015nonreciprocal}%
  \BibitemOpen
  \bibfield  {author} {\bibinfo {author} {\bibfnamefont {Anja}\ \bibnamefont
  {Metelmann}}\ and\ \bibinfo {author} {\bibfnamefont {Aashish~A}\ \bibnamefont
  {Clerk}},\ }\bibfield  {title} {\enquote {\bibinfo {title} {Nonreciprocal
  photon transmission and amplification via reservoir engineering},}\
  }\href@noop {} {\bibfield  {journal} {\bibinfo  {journal} {Phys. Rev. X}\
  }\textbf {\bibinfo {volume} {5}},\ \bibinfo {pages} {021025} (\bibinfo {year}
  {2015})}\BibitemShut {NoStop}%
\bibitem [{\citenamefont {Fang}\ \emph {et~al.}(2017)\citenamefont {Fang},
  \citenamefont {Luo}, \citenamefont {Metelmann}, \citenamefont {Matheny},
  \citenamefont {Marquardt}, \citenamefont {Clerk},\ and\ \citenamefont
  {Painter}}]{fang2017generalized}%
  \BibitemOpen
  \bibfield  {author} {\bibinfo {author} {\bibfnamefont {Kejie}\ \bibnamefont
  {Fang}}, \bibinfo {author} {\bibfnamefont {Jie}\ \bibnamefont {Luo}},
  \bibinfo {author} {\bibfnamefont {Anja}\ \bibnamefont {Metelmann}}, \bibinfo
  {author} {\bibfnamefont {Matthew~H}\ \bibnamefont {Matheny}}, \bibinfo
  {author} {\bibfnamefont {Florian}\ \bibnamefont {Marquardt}}, \bibinfo
  {author} {\bibfnamefont {Aashish~A}\ \bibnamefont {Clerk}}, \ and\ \bibinfo
  {author} {\bibfnamefont {Oskar}\ \bibnamefont {Painter}},\ }\bibfield
  {title} {\enquote {\bibinfo {title} {Generalized non-reciprocity in an
  optomechanical circuit via synthetic magnetism and reservoir engineering},}\
  }\href@noop {} {\bibfield  {journal} {\bibinfo  {journal} {Nat. Phys.}\
  }\textbf {\bibinfo {volume} {13}},\ \bibinfo {pages} {465} (\bibinfo {year}
  {2017})}\BibitemShut {NoStop}%
\bibitem [{\citenamefont {Keck}\ \emph {et~al.}(2018)\citenamefont {Keck},
  \citenamefont {Rossini},\ and\ \citenamefont {Fazio}}]{keck2018persistent}%
  \BibitemOpen
  \bibfield  {author} {\bibinfo {author} {\bibfnamefont {Maximilian}\
  \bibnamefont {Keck}}, \bibinfo {author} {\bibfnamefont {Davide}\ \bibnamefont
  {Rossini}}, \ and\ \bibinfo {author} {\bibfnamefont {Rosario}\ \bibnamefont
  {Fazio}},\ }\bibfield  {title} {\enquote {\bibinfo {title} {Persistent
  currents by reservoir engineering},}\ }\href@noop {} {\bibfield  {journal}
  {\bibinfo  {journal} {Phys. Rev. A}\ }\textbf {\bibinfo {volume} {98}},\
  \bibinfo {pages} {053812} (\bibinfo {year} {2018})}\BibitemShut {NoStop}%
\bibitem [{\citenamefont {Jirari}\ \emph {et~al.}(2009)\citenamefont {Jirari},
  \citenamefont {Hekking},\ and\ \citenamefont {Buisson}}]{jirari2009optimal}%
  \BibitemOpen
  \bibfield  {author} {\bibinfo {author} {\bibfnamefont {Hamza}\ \bibnamefont
  {Jirari}}, \bibinfo {author} {\bibfnamefont {Frank~WJ}\ \bibnamefont
  {Hekking}}, \ and\ \bibinfo {author} {\bibfnamefont {Olivier}\ \bibnamefont
  {Buisson}},\ }\bibfield  {title} {\enquote {\bibinfo {title} {Optimal control
  of superconducting n-level quantum systems},}\ }\href@noop {} {\bibfield
  {journal} {\bibinfo  {journal} {EPL}\ }\textbf {\bibinfo {volume} {87}},\
  \bibinfo {pages} {28004} (\bibinfo {year} {2009})}\BibitemShut {NoStop}%
\bibitem [{\citenamefont {Garc{\'\i}a-Ripoll}\ \emph
  {et~al.}(2004)\citenamefont {Garc{\'\i}a-Ripoll}, \citenamefont {Cirac},
  \citenamefont {Zoller}, \citenamefont {Kollath}, \citenamefont
  {Schollw{\"o}ck},\ and\ \citenamefont {von Delft}}]{garcia2004variational}%
  \BibitemOpen
  \bibfield  {author} {\bibinfo {author} {\bibfnamefont {Juan~Jose}\
  \bibnamefont {Garc{\'\i}a-Ripoll}}, \bibinfo {author} {\bibfnamefont
  {J~Ignacio}\ \bibnamefont {Cirac}}, \bibinfo {author} {\bibfnamefont
  {P}~\bibnamefont {Zoller}}, \bibinfo {author} {\bibfnamefont {C}~\bibnamefont
  {Kollath}}, \bibinfo {author} {\bibfnamefont {U}~\bibnamefont
  {Schollw{\"o}ck}}, \ and\ \bibinfo {author} {\bibfnamefont {J}~\bibnamefont
  {von Delft}},\ }\bibfield  {title} {\enquote {\bibinfo {title} {Variational
  ansatz for the superfluid mott-insulator transition in optical lattices},}\
  }\href@noop {} {\bibfield  {journal} {\bibinfo  {journal} {Opt. Express}\
  }\textbf {\bibinfo {volume} {12}},\ \bibinfo {pages} {42--54} (\bibinfo
  {year} {2004})}\BibitemShut {NoStop}%
\bibitem [{\citenamefont {Fazio}\ and\ \citenamefont {Van
  Der~Zant}(2001)}]{fazio2001quantum}%
  \BibitemOpen
  \bibfield  {author} {\bibinfo {author} {\bibfnamefont {Rosario}\ \bibnamefont
  {Fazio}}\ and\ \bibinfo {author} {\bibfnamefont {Herre}\ \bibnamefont {Van
  Der~Zant}},\ }\bibfield  {title} {\enquote {\bibinfo {title} {Quantum phase
  transitions and vortex dynamics in superconducting networks},}\ }\href@noop
  {} {\bibfield  {journal} {\bibinfo  {journal} {Physics Reports}\ }\textbf
  {\bibinfo {volume} {355}},\ \bibinfo {pages} {235--334} (\bibinfo {year}
  {2001})}\BibitemShut {NoStop}%
\bibitem [{\citenamefont {Ye}\ \emph {et~al.}(2019)\citenamefont {Ye},
  \citenamefont {Ge}, \citenamefont {Wu}, \citenamefont {Wang}, \citenamefont
  {Gong}, \citenamefont {Zhang}, \citenamefont {Zhu}, \citenamefont {Yang},
  \citenamefont {Li}, \citenamefont {Liang} \emph
  {et~al.}}]{ye2019propagation}%
  \BibitemOpen
  \bibfield  {author} {\bibinfo {author} {\bibfnamefont {Yangsen}\ \bibnamefont
  {Ye}}, \bibinfo {author} {\bibfnamefont {Zi-Yong}\ \bibnamefont {Ge}},
  \bibinfo {author} {\bibfnamefont {Yulin}\ \bibnamefont {Wu}}, \bibinfo
  {author} {\bibfnamefont {Shiyu}\ \bibnamefont {Wang}}, \bibinfo {author}
  {\bibfnamefont {Ming}\ \bibnamefont {Gong}}, \bibinfo {author} {\bibfnamefont
  {Yu-Ran}\ \bibnamefont {Zhang}}, \bibinfo {author} {\bibfnamefont {Qingling}\
  \bibnamefont {Zhu}}, \bibinfo {author} {\bibfnamefont {Rui}\ \bibnamefont
  {Yang}}, \bibinfo {author} {\bibfnamefont {Shaowei}\ \bibnamefont {Li}},
  \bibinfo {author} {\bibfnamefont {Futian}\ \bibnamefont {Liang}},  \emph
  {et~al.},\ }\bibfield  {title} {\enquote {\bibinfo {title} {Propagation and
  localization of collective excitations on a 24-qubit superconducting
  processor},}\ }\href@noop {} {\bibfield  {journal} {\bibinfo  {journal}
  {Phys. Rev. Lett.}\ }\textbf {\bibinfo {volume} {123}},\ \bibinfo {pages}
  {050502} (\bibinfo {year} {2019})}\BibitemShut {NoStop}%
\bibitem [{\citenamefont {Bloch}\ \emph {et~al.}(2008)\citenamefont {Bloch},
  \citenamefont {Dalibard},\ and\ \citenamefont {Zwerger}}]{bloch2008many}%
  \BibitemOpen
  \bibfield  {author} {\bibinfo {author} {\bibfnamefont {Immanuel}\
  \bibnamefont {Bloch}}, \bibinfo {author} {\bibfnamefont {Jean}\ \bibnamefont
  {Dalibard}}, \ and\ \bibinfo {author} {\bibfnamefont {Wilhelm}\ \bibnamefont
  {Zwerger}},\ }\bibfield  {title} {\enquote {\bibinfo {title} {Many-body
  physics with ultracold gases},}\ }\href@noop {} {\bibfield  {journal}
  {\bibinfo  {journal} {Rev. Mod. Phys.}\ }\textbf {\bibinfo {volume} {80}},\
  \bibinfo {pages} {885} (\bibinfo {year} {2008})}\BibitemShut {NoStop}%
\bibitem [{\citenamefont {Haug}\ \emph {et~al.}(2018)\citenamefont {Haug},
  \citenamefont {Tan}, \citenamefont {Theng}, \citenamefont {Dumke},
  \citenamefont {Kwek},\ and\ \citenamefont {Amico}}]{haug2018readout}%
  \BibitemOpen
  \bibfield  {author} {\bibinfo {author} {\bibfnamefont {Tobias}\ \bibnamefont
  {Haug}}, \bibinfo {author} {\bibfnamefont {Joel}\ \bibnamefont {Tan}},
  \bibinfo {author} {\bibfnamefont {Mark}\ \bibnamefont {Theng}}, \bibinfo
  {author} {\bibfnamefont {Rainer}\ \bibnamefont {Dumke}}, \bibinfo {author}
  {\bibfnamefont {Leong-Chuan}\ \bibnamefont {Kwek}}, \ and\ \bibinfo {author}
  {\bibfnamefont {Luigi}\ \bibnamefont {Amico}},\ }\bibfield  {title} {\enquote
  {\bibinfo {title} {Readout of the atomtronic quantum interference device},}\
  }\href@noop {} {\bibfield  {journal} {\bibinfo  {journal} {Phys. Rev. A}\
  }\textbf {\bibinfo {volume} {97}},\ \bibinfo {pages} {013633} (\bibinfo
  {year} {2018})}\BibitemShut {NoStop}%
\bibitem [{\citenamefont {Mnih}\ \emph {et~al.}(2015)\citenamefont {Mnih},
  \citenamefont {Kavukcuoglu}, \citenamefont {Silver}, \citenamefont {Rusu},
  \citenamefont {Veness}, \citenamefont {Bellemare}, \citenamefont {Graves},
  \citenamefont {Riedmiller}, \citenamefont {Fidjeland}, \citenamefont
  {Ostrovski} \emph {et~al.}}]{mnih2015human}%
  \BibitemOpen
  \bibfield  {author} {\bibinfo {author} {\bibfnamefont {Volodymyr}\
  \bibnamefont {Mnih}}, \bibinfo {author} {\bibfnamefont {Koray}\ \bibnamefont
  {Kavukcuoglu}}, \bibinfo {author} {\bibfnamefont {David}\ \bibnamefont
  {Silver}}, \bibinfo {author} {\bibfnamefont {Andrei~A}\ \bibnamefont {Rusu}},
  \bibinfo {author} {\bibfnamefont {Joel}\ \bibnamefont {Veness}}, \bibinfo
  {author} {\bibfnamefont {Marc~G}\ \bibnamefont {Bellemare}}, \bibinfo
  {author} {\bibfnamefont {Alex}\ \bibnamefont {Graves}}, \bibinfo {author}
  {\bibfnamefont {Martin}\ \bibnamefont {Riedmiller}}, \bibinfo {author}
  {\bibfnamefont {Andreas~K}\ \bibnamefont {Fidjeland}}, \bibinfo {author}
  {\bibfnamefont {Georg}\ \bibnamefont {Ostrovski}},  \emph {et~al.},\
  }\bibfield  {title} {\enquote {\bibinfo {title} {Human-level control through
  deep reinforcement learning},}\ }\href@noop {} {\bibfield  {journal}
  {\bibinfo  {journal} {Nature}\ }\textbf {\bibinfo {volume} {518}},\ \bibinfo
  {pages} {529} (\bibinfo {year} {2015})}\BibitemShut {NoStop}%
\bibitem [{\citenamefont {Abadi}\ \emph {et~al.}(2015)\citenamefont {Abadi},
  \citenamefont {Agarwal}, \citenamefont {Barham}, \citenamefont {Brevdo},
  \citenamefont {Chen}, \citenamefont {Citro}, \citenamefont {Corrado},
  \citenamefont {Davis}, \citenamefont {Dean}, \citenamefont {Devin},
  \citenamefont {Ghemawat}, \citenamefont {Goodfellow}, \citenamefont {Harp},
  \citenamefont {Irving}, \citenamefont {Isard}, \citenamefont {Jia},
  \citenamefont {Jozefowicz}, \citenamefont {Kaiser}, \citenamefont {Kudlur},
  \citenamefont {Levenberg}, \citenamefont {Man\'{e}}, \citenamefont {Monga},
  \citenamefont {Moore}, \citenamefont {Murray}, \citenamefont {Olah},
  \citenamefont {Schuster}, \citenamefont {Shlens}, \citenamefont {Steiner},
  \citenamefont {Sutskever}, \citenamefont {Talwar}, \citenamefont {Tucker},
  \citenamefont {Vanhoucke}, \citenamefont {Vasudevan}, \citenamefont
  {Vi\'{e}gas}, \citenamefont {Vinyals}, \citenamefont {Warden}, \citenamefont
  {Wattenberg}, \citenamefont {Wicke}, \citenamefont {Yu},\ and\ \citenamefont
  {Zheng}}]{tensorflow2015-whitepaper}%
  \BibitemOpen
  \bibfield  {author} {\bibinfo {author} {\bibfnamefont {Mart\'{\i}n}\
  \bibnamefont {Abadi}}, \bibinfo {author} {\bibfnamefont {Ashish}\
  \bibnamefont {Agarwal}}, \bibinfo {author} {\bibfnamefont {Paul}\
  \bibnamefont {Barham}}, \bibinfo {author} {\bibfnamefont {Eugene}\
  \bibnamefont {Brevdo}}, \bibinfo {author} {\bibfnamefont {Zhifeng}\
  \bibnamefont {Chen}}, \bibinfo {author} {\bibfnamefont {Craig}\ \bibnamefont
  {Citro}}, \bibinfo {author} {\bibfnamefont {Greg~S.}\ \bibnamefont
  {Corrado}}, \bibinfo {author} {\bibfnamefont {Andy}\ \bibnamefont {Davis}},
  \bibinfo {author} {\bibfnamefont {Jeffrey}\ \bibnamefont {Dean}}, \bibinfo
  {author} {\bibfnamefont {Matthieu}\ \bibnamefont {Devin}}, \bibinfo {author}
  {\bibfnamefont {Sanjay}\ \bibnamefont {Ghemawat}}, \bibinfo {author}
  {\bibfnamefont {Ian}\ \bibnamefont {Goodfellow}}, \bibinfo {author}
  {\bibfnamefont {Andrew}\ \bibnamefont {Harp}}, \bibinfo {author}
  {\bibfnamefont {Geoffrey}\ \bibnamefont {Irving}}, \bibinfo {author}
  {\bibfnamefont {Michael}\ \bibnamefont {Isard}}, \bibinfo {author}
  {\bibfnamefont {Yangqing}\ \bibnamefont {Jia}}, \bibinfo {author}
  {\bibfnamefont {Rafal}\ \bibnamefont {Jozefowicz}}, \bibinfo {author}
  {\bibfnamefont {Lukasz}\ \bibnamefont {Kaiser}}, \bibinfo {author}
  {\bibfnamefont {Manjunath}\ \bibnamefont {Kudlur}}, \bibinfo {author}
  {\bibfnamefont {Josh}\ \bibnamefont {Levenberg}}, \bibinfo {author}
  {\bibfnamefont {Dan}\ \bibnamefont {Man\'{e}}}, \bibinfo {author}
  {\bibfnamefont {Rajat}\ \bibnamefont {Monga}}, \bibinfo {author}
  {\bibfnamefont {Sherry}\ \bibnamefont {Moore}}, \bibinfo {author}
  {\bibfnamefont {Derek}\ \bibnamefont {Murray}}, \bibinfo {author}
  {\bibfnamefont {Chris}\ \bibnamefont {Olah}}, \bibinfo {author}
  {\bibfnamefont {Mike}\ \bibnamefont {Schuster}}, \bibinfo {author}
  {\bibfnamefont {Jonathon}\ \bibnamefont {Shlens}}, \bibinfo {author}
  {\bibfnamefont {Benoit}\ \bibnamefont {Steiner}}, \bibinfo {author}
  {\bibfnamefont {Ilya}\ \bibnamefont {Sutskever}}, \bibinfo {author}
  {\bibfnamefont {Kunal}\ \bibnamefont {Talwar}}, \bibinfo {author}
  {\bibfnamefont {Paul}\ \bibnamefont {Tucker}}, \bibinfo {author}
  {\bibfnamefont {Vincent}\ \bibnamefont {Vanhoucke}}, \bibinfo {author}
  {\bibfnamefont {Vijay}\ \bibnamefont {Vasudevan}}, \bibinfo {author}
  {\bibfnamefont {Fernanda}\ \bibnamefont {Vi\'{e}gas}}, \bibinfo {author}
  {\bibfnamefont {Oriol}\ \bibnamefont {Vinyals}}, \bibinfo {author}
  {\bibfnamefont {Pete}\ \bibnamefont {Warden}}, \bibinfo {author}
  {\bibfnamefont {Martin}\ \bibnamefont {Wattenberg}}, \bibinfo {author}
  {\bibfnamefont {Martin}\ \bibnamefont {Wicke}}, \bibinfo {author}
  {\bibfnamefont {Yuan}\ \bibnamefont {Yu}}, \ and\ \bibinfo {author}
  {\bibfnamefont {Xiaoqiang}\ \bibnamefont {Zheng}},\ }\href
  {http://tensorflow.org/} {\enquote {\bibinfo {title} {{TensorFlow}:
  Large-scale machine learning on heterogeneous systems},}\ } (\bibinfo {year}
  {2015}),\ \bibinfo {note} {software available from
  tensorflow.org}\BibitemShut {NoStop}%
\bibitem [{\citenamefont {Ramanathan}\ \emph {et~al.}(2011)\citenamefont
  {Ramanathan}, \citenamefont {Wright}, \citenamefont {Muniz}, \citenamefont
  {Zelan}, \citenamefont {Hill}, \citenamefont {Lobb}, \citenamefont
  {Helmerson}, \citenamefont {Phillips},\ and\ \citenamefont
  {Campbell}}]{Ramanathan2011}%
  \BibitemOpen
  \bibfield  {author} {\bibinfo {author} {\bibfnamefont {A.}~\bibnamefont
  {Ramanathan}}, \bibinfo {author} {\bibfnamefont {K.~C.}\ \bibnamefont
  {Wright}}, \bibinfo {author} {\bibfnamefont {S.~R.}\ \bibnamefont {Muniz}},
  \bibinfo {author} {\bibfnamefont {M.}~\bibnamefont {Zelan}}, \bibinfo
  {author} {\bibfnamefont {W.~T.}\ \bibnamefont {Hill}}, \bibinfo {author}
  {\bibfnamefont {C.~J.}\ \bibnamefont {Lobb}}, \bibinfo {author}
  {\bibfnamefont {K.}~\bibnamefont {Helmerson}}, \bibinfo {author}
  {\bibfnamefont {W.~D.}\ \bibnamefont {Phillips}}, \ and\ \bibinfo {author}
  {\bibfnamefont {G.~K.}\ \bibnamefont {Campbell}},\ }\bibfield  {title}
  {\enquote {\bibinfo {title} {Superflow in a toroidal bose-einstein
  condensate: An atom circuit with a tunable weak link},}\ }\href {\doibase
  10.1103/PhysRevLett.106.130401} {\bibfield  {journal} {\bibinfo  {journal}
  {Phys. Rev. Lett.}\ }\textbf {\bibinfo {volume} {106}},\ \bibinfo {pages}
  {130401} (\bibinfo {year} {2011})}\BibitemShut {NoStop}%
\bibitem [{\citenamefont {Schenke}\ \emph {et~al.}(2011)\citenamefont
  {Schenke}, \citenamefont {Minguzzi},\ and\ \citenamefont
  {Hekking}}]{schenke2011nonadiabatic}%
  \BibitemOpen
  \bibfield  {author} {\bibinfo {author} {\bibfnamefont {C}~\bibnamefont
  {Schenke}}, \bibinfo {author} {\bibfnamefont {A}~\bibnamefont {Minguzzi}}, \
  and\ \bibinfo {author} {\bibfnamefont {FWJ}\ \bibnamefont {Hekking}},\
  }\bibfield  {title} {\enquote {\bibinfo {title} {Nonadiabatic creation of
  macroscopic superpositions with strongly correlated one-dimensional bosons in
  a ring trap},}\ }\href@noop {} {\bibfield  {journal} {\bibinfo  {journal}
  {Phys. Rev. A}\ }\textbf {\bibinfo {volume} {84}},\ \bibinfo {pages} {053636}
  (\bibinfo {year} {2011})}\BibitemShut {NoStop}%
\bibitem [{\citenamefont {Aghamalyan}\ \emph {et~al.}(2016)\citenamefont
  {Aghamalyan}, \citenamefont {Nguyen}, \citenamefont {Auksztol}, \citenamefont
  {Gan}, \citenamefont {Valado}, \citenamefont {Condylis}, \citenamefont
  {Kwek}, \citenamefont {Dumke},\ and\ \citenamefont
  {Amico}}]{aghamalyan2016atomtronic}%
  \BibitemOpen
  \bibfield  {author} {\bibinfo {author} {\bibfnamefont {D}~\bibnamefont
  {Aghamalyan}}, \bibinfo {author} {\bibfnamefont {NT}~\bibnamefont {Nguyen}},
  \bibinfo {author} {\bibfnamefont {F}~\bibnamefont {Auksztol}}, \bibinfo
  {author} {\bibfnamefont {KS}~\bibnamefont {Gan}}, \bibinfo {author}
  {\bibfnamefont {M~Martinez}\ \bibnamefont {Valado}}, \bibinfo {author}
  {\bibfnamefont {PC}~\bibnamefont {Condylis}}, \bibinfo {author}
  {\bibfnamefont {Leong-Chuan}\ \bibnamefont {Kwek}}, \bibinfo {author}
  {\bibfnamefont {R}~\bibnamefont {Dumke}}, \ and\ \bibinfo {author}
  {\bibfnamefont {L}~\bibnamefont {Amico}},\ }\bibfield  {title} {\enquote
  {\bibinfo {title} {An atomtronic flux qubit: a ring lattice of bose--einstein
  condensates interrupted by three weak links},}\ }\href@noop {} {\bibfield
  {journal} {\bibinfo  {journal} {New Journal of Physics}\ }\textbf {\bibinfo
  {volume} {18}},\ \bibinfo {pages} {075013} (\bibinfo {year}
  {2016})}\BibitemShut {NoStop}%
\bibitem [{\citenamefont {Johansson}\ \emph {et~al.}(2013)\citenamefont
  {Johansson}, \citenamefont {Nation},\ and\ \citenamefont
  {Nori}}]{johansson2013qutip}%
  \BibitemOpen
  \bibfield  {author} {\bibinfo {author} {\bibfnamefont {J~Robert}\
  \bibnamefont {Johansson}}, \bibinfo {author} {\bibfnamefont {Paul~D}\
  \bibnamefont {Nation}}, \ and\ \bibinfo {author} {\bibfnamefont {Franco}\
  \bibnamefont {Nori}},\ }\bibfield  {title} {\enquote {\bibinfo {title} {Qutip
  2: A python framework for the dynamics of open quantum systems},}\
  }\href@noop {} {\bibfield  {journal} {\bibinfo  {journal} {Comput. Phys.
  Commun.}\ }\textbf {\bibinfo {volume} {184}},\ \bibinfo {pages} {1234--1240}
  (\bibinfo {year} {2013})}\BibitemShut {NoStop}%
\bibitem [{\citenamefont {Machnes}\ \emph {et~al.}(2011)\citenamefont
  {Machnes}, \citenamefont {Sander}, \citenamefont {Glaser}, \citenamefont
  {De~Fouqui{\`e}res}, \citenamefont {Gruslys}, \citenamefont {Schirmer},\ and\
  \citenamefont {Schulte-Herbr{\"u}ggen}}]{machnes2011comparing}%
  \BibitemOpen
  \bibfield  {author} {\bibinfo {author} {\bibfnamefont {Shai}\ \bibnamefont
  {Machnes}}, \bibinfo {author} {\bibfnamefont {U}~\bibnamefont {Sander}},
  \bibinfo {author} {\bibfnamefont {Steffen~J}\ \bibnamefont {Glaser}},
  \bibinfo {author} {\bibfnamefont {P}~\bibnamefont {De~Fouqui{\`e}res}},
  \bibinfo {author} {\bibfnamefont {A}~\bibnamefont {Gruslys}}, \bibinfo
  {author} {\bibfnamefont {S}~\bibnamefont {Schirmer}}, \ and\ \bibinfo
  {author} {\bibfnamefont {Thomas}\ \bibnamefont {Schulte-Herbr{\"u}ggen}},\
  }\bibfield  {title} {\enquote {\bibinfo {title} {Comparing, optimizing, and
  benchmarking quantum-control algorithms in a unifying programming
  framework},}\ }\href@noop {} {\bibfield  {journal} {\bibinfo  {journal}
  {Phys. Rev. A}\ }\textbf {\bibinfo {volume} {84}},\ \bibinfo {pages} {022305}
  (\bibinfo {year} {2011})}\BibitemShut {NoStop}%
\bibitem [{\citenamefont {Wigley}\ \emph {et~al.}(2016)\citenamefont {Wigley},
  \citenamefont {Everitt}, \citenamefont {van~den Hengel}, \citenamefont
  {Bastian}, \citenamefont {Sooriyabandara}, \citenamefont {McDonald},
  \citenamefont {Hardman}, \citenamefont {Quinlivan}, \citenamefont {Manju},
  \citenamefont {Kuhn} \emph {et~al.}}]{wigley2016fast}%
  \BibitemOpen
  \bibfield  {author} {\bibinfo {author} {\bibfnamefont {Paul~B}\ \bibnamefont
  {Wigley}}, \bibinfo {author} {\bibfnamefont {Patrick~J}\ \bibnamefont
  {Everitt}}, \bibinfo {author} {\bibfnamefont {Anton}\ \bibnamefont {van~den
  Hengel}}, \bibinfo {author} {\bibfnamefont {John~W}\ \bibnamefont {Bastian}},
  \bibinfo {author} {\bibfnamefont {Mahasen~A}\ \bibnamefont {Sooriyabandara}},
  \bibinfo {author} {\bibfnamefont {Gordon~D}\ \bibnamefont {McDonald}},
  \bibinfo {author} {\bibfnamefont {Kyle~S}\ \bibnamefont {Hardman}}, \bibinfo
  {author} {\bibfnamefont {Ciaron~D}\ \bibnamefont {Quinlivan}}, \bibinfo
  {author} {\bibfnamefont {P}~\bibnamefont {Manju}}, \bibinfo {author}
  {\bibfnamefont {Carlos~CN}\ \bibnamefont {Kuhn}},  \emph {et~al.},\
  }\bibfield  {title} {\enquote {\bibinfo {title} {Fast machine-learning online
  optimization of ultra-cold-atom experiments},}\ }\href@noop {} {\bibfield
  {journal} {\bibinfo  {journal} {Sci. Rep.}\ }\textbf {\bibinfo {volume}
  {6}},\ \bibinfo {pages} {25890} (\bibinfo {year} {2016})}\BibitemShut
  {NoStop}%
\bibitem [{\citenamefont {Otterbach}\ \emph {et~al.}(2017)\citenamefont
  {Otterbach}, \citenamefont {Manenti}, \citenamefont {Alidoust}, \citenamefont
  {Bestwick}, \citenamefont {Block}, \citenamefont {Bloom}, \citenamefont
  {Caldwell}, \citenamefont {Didier}, \citenamefont {Fried}, \citenamefont
  {Hong} \emph {et~al.}}]{otterbach2017unsupervised}%
  \BibitemOpen
  \bibfield  {author} {\bibinfo {author} {\bibfnamefont {JS}~\bibnamefont
  {Otterbach}}, \bibinfo {author} {\bibfnamefont {R}~\bibnamefont {Manenti}},
  \bibinfo {author} {\bibfnamefont {N}~\bibnamefont {Alidoust}}, \bibinfo
  {author} {\bibfnamefont {A}~\bibnamefont {Bestwick}}, \bibinfo {author}
  {\bibfnamefont {M}~\bibnamefont {Block}}, \bibinfo {author} {\bibfnamefont
  {B}~\bibnamefont {Bloom}}, \bibinfo {author} {\bibfnamefont {S}~\bibnamefont
  {Caldwell}}, \bibinfo {author} {\bibfnamefont {N}~\bibnamefont {Didier}},
  \bibinfo {author} {\bibfnamefont {E~Schuyler}\ \bibnamefont {Fried}},
  \bibinfo {author} {\bibfnamefont {S}~\bibnamefont {Hong}},  \emph {et~al.},\
  }\bibfield  {title} {\enquote {\bibinfo {title} {Unsupervised machine
  learning on a hybrid quantum computer},}\ }\href@noop {} {\bibfield
  {journal} {\bibinfo  {journal} {arXiv:1712.05771}\ } (\bibinfo {year}
  {2017})}\BibitemShut {NoStop}%
\bibitem [{\citenamefont {Kokail}\ \emph {et~al.}(2019)\citenamefont {Kokail},
  \citenamefont {Maier}, \citenamefont {van Bijnen}, \citenamefont {Brydges},
  \citenamefont {Joshi}, \citenamefont {Jurcevic}, \citenamefont {Muschik},
  \citenamefont {Silvi}, \citenamefont {Blatt}, \citenamefont {Roos} \emph
  {et~al.}}]{kokail2019self}%
  \BibitemOpen
  \bibfield  {author} {\bibinfo {author} {\bibfnamefont {C}~\bibnamefont
  {Kokail}}, \bibinfo {author} {\bibfnamefont {C}~\bibnamefont {Maier}},
  \bibinfo {author} {\bibfnamefont {R}~\bibnamefont {van Bijnen}}, \bibinfo
  {author} {\bibfnamefont {T}~\bibnamefont {Brydges}}, \bibinfo {author}
  {\bibfnamefont {MK}~\bibnamefont {Joshi}}, \bibinfo {author} {\bibfnamefont
  {P}~\bibnamefont {Jurcevic}}, \bibinfo {author} {\bibfnamefont
  {CA}~\bibnamefont {Muschik}}, \bibinfo {author} {\bibfnamefont
  {P}~\bibnamefont {Silvi}}, \bibinfo {author} {\bibfnamefont {R}~\bibnamefont
  {Blatt}}, \bibinfo {author} {\bibfnamefont {CF}~\bibnamefont {Roos}},  \emph
  {et~al.},\ }\bibfield  {title} {\enquote {\bibinfo {title} {Self-verifying
  variational quantum simulation of lattice models},}\ }\href@noop {}
  {\bibfield  {journal} {\bibinfo  {journal} {Nature}\ }\textbf {\bibinfo
  {volume} {569}},\ \bibinfo {pages} {355} (\bibinfo {year}
  {2019})}\BibitemShut {NoStop}%
\bibitem [{\citenamefont {Ragole}\ and\ \citenamefont
  {Taylor}(2016)}]{ragole2016interacting}%
  \BibitemOpen
  \bibfield  {author} {\bibinfo {author} {\bibfnamefont {Stephen}\ \bibnamefont
  {Ragole}}\ and\ \bibinfo {author} {\bibfnamefont {Jacob~M.}\ \bibnamefont
  {Taylor}},\ }\bibfield  {title} {\enquote {\bibinfo {title} {Interacting
  atomic interferometry for rotation sensing approaching the heisenberg
  limit},}\ }\href@noop {} {\bibfield  {journal} {\bibinfo  {journal} {Phys.
  Rev. Lett.}\ }\textbf {\bibinfo {volume} {117}},\ \bibinfo {pages} {203002}
  (\bibinfo {year} {2016})}\BibitemShut {NoStop}%
\bibitem [{\citenamefont {Gluza}\ \emph {et~al.}(2018)\citenamefont {Gluza},
  \citenamefont {Kliesch}, \citenamefont {Eisert},\ and\ \citenamefont
  {Aolita}}]{Gluza2018}%
  \BibitemOpen
  \bibfield  {author} {\bibinfo {author} {\bibfnamefont {M}~\bibnamefont
  {Gluza}}, \bibinfo {author} {\bibfnamefont {M}~\bibnamefont {Kliesch}},
  \bibinfo {author} {\bibfnamefont {J}~\bibnamefont {Eisert}}, \ and\ \bibinfo
  {author} {\bibfnamefont {L}~\bibnamefont {Aolita}},\ }\bibfield  {title}
  {\enquote {\bibinfo {title} {Fidelity witnesses for fermionic quantum
  simulations},}\ }\href@noop {} {\bibfield  {journal} {\bibinfo  {journal}
  {Phys. Rev. Lett.}\ }\textbf {\bibinfo {volume} {120}},\ \bibinfo {pages}
  {190501} (\bibinfo {year} {2018})}\BibitemShut {NoStop}%
\bibitem [{\citenamefont {Kingma}\ and\ \citenamefont
  {Ba}(2014)}]{kingma2014adam}%
  \BibitemOpen
  \bibfield  {author} {\bibinfo {author} {\bibfnamefont {Diederik~P}\
  \bibnamefont {Kingma}}\ and\ \bibinfo {author} {\bibfnamefont {Jimmy}\
  \bibnamefont {Ba}},\ }\bibfield  {title} {\enquote {\bibinfo {title} {Adam: A
  method for stochastic optimization},}\ }\href@noop {} {\bibfield  {journal}
  {\bibinfo  {journal} {arXiv:1412.6980}\ } (\bibinfo {year}
  {2014})}\BibitemShut {NoStop}%
\bibitem [{\citenamefont {Wang}\ \emph {et~al.}(2015)\citenamefont {Wang},
  \citenamefont {Schaul}, \citenamefont {Hessel}, \citenamefont {Van~Hasselt},
  \citenamefont {Lanctot},\ and\ \citenamefont {De~Freitas}}]{wang2015dueling}%
  \BibitemOpen
  \bibfield  {author} {\bibinfo {author} {\bibfnamefont {Ziyu}\ \bibnamefont
  {Wang}}, \bibinfo {author} {\bibfnamefont {Tom}\ \bibnamefont {Schaul}},
  \bibinfo {author} {\bibfnamefont {Matteo}\ \bibnamefont {Hessel}}, \bibinfo
  {author} {\bibfnamefont {Hado}\ \bibnamefont {Van~Hasselt}}, \bibinfo
  {author} {\bibfnamefont {Marc}\ \bibnamefont {Lanctot}}, \ and\ \bibinfo
  {author} {\bibfnamefont {Nando}\ \bibnamefont {De~Freitas}},\ }\bibfield
  {title} {\enquote {\bibinfo {title} {Dueling network architectures for deep
  reinforcement learning},}\ }\href@noop {} {\bibfield  {journal} {\bibinfo
  {journal} {arXiv:1511.06581}\ } (\bibinfo {year} {2015})}\BibitemShut
  {NoStop}%
\bibitem [{\citenamefont {Jaksch}\ \emph {et~al.}(1998)\citenamefont {Jaksch},
  \citenamefont {Bruder}, \citenamefont {Cirac}, \citenamefont {Gardiner},\
  and\ \citenamefont {Zoller}}]{jaksch1998cold}%
  \BibitemOpen
  \bibfield  {author} {\bibinfo {author} {\bibfnamefont {Dieter}\ \bibnamefont
  {Jaksch}}, \bibinfo {author} {\bibfnamefont {Christoph}\ \bibnamefont
  {Bruder}}, \bibinfo {author} {\bibfnamefont {Juan~Ignacio}\ \bibnamefont
  {Cirac}}, \bibinfo {author} {\bibfnamefont {Crispin~W}\ \bibnamefont
  {Gardiner}}, \ and\ \bibinfo {author} {\bibfnamefont {Peter}\ \bibnamefont
  {Zoller}},\ }\bibfield  {title} {\enquote {\bibinfo {title} {Cold bosonic
  atoms in optical lattices},}\ }\href@noop {} {\bibfield  {journal} {\bibinfo
  {journal} {Phys. Rev. Lett.}\ }\textbf {\bibinfo {volume} {81}},\ \bibinfo
  {pages} {3108} (\bibinfo {year} {1998})}\BibitemShut {NoStop}%
\bibitem [{\citenamefont {Gauthier}\ \emph {et~al.}(2016)\citenamefont
  {Gauthier}, \citenamefont {Lenton}, \citenamefont {Parry}, \citenamefont
  {Baker}, \citenamefont {Davis}, \citenamefont {Rubinsztein-Dunlop},\ and\
  \citenamefont {Neely}}]{gauthier2016direct}%
  \BibitemOpen
  \bibfield  {author} {\bibinfo {author} {\bibfnamefont {G}~\bibnamefont
  {Gauthier}}, \bibinfo {author} {\bibfnamefont {I}~\bibnamefont {Lenton}},
  \bibinfo {author} {\bibfnamefont {N~McKay}\ \bibnamefont {Parry}}, \bibinfo
  {author} {\bibfnamefont {M}~\bibnamefont {Baker}}, \bibinfo {author}
  {\bibfnamefont {MJ}~\bibnamefont {Davis}}, \bibinfo {author} {\bibfnamefont
  {H}~\bibnamefont {Rubinsztein-Dunlop}}, \ and\ \bibinfo {author}
  {\bibfnamefont {TW}~\bibnamefont {Neely}},\ }\bibfield  {title} {\enquote
  {\bibinfo {title} {Direct imaging of a digital-micromirror device for
  configurable microscopic optical potentials},}\ }\href@noop {} {\bibfield
  {journal} {\bibinfo  {journal} {Optica}\ }\textbf {\bibinfo {volume} {3}},\
  \bibinfo {pages} {1136--1143} (\bibinfo {year} {2016})}\BibitemShut {NoStop}%
\end{thebibliography}%

\appendix

\section*{Overview}
Here, we give a short overview of the sections of the appendix. 

In Sec.\ref{Theory}, we introduce complementary theoretical definitions for the quantum phase model and phase windings that we used in the main text.

In Sec.\ref{cert}, we compare the fidelity with the certification measure, which can be used to experimentally verify the quality of the prepared state.

In Sec.\ref{grape}, we compare our results with deep reinforcement learning with the standard method GRAPE.

In Sec.\ref{robust}, we investigate the robustness of our protocols against noise in the driving parameters

In Sec.\ref{numerics}, we present additional numerical results on the quantum phase model and the generation of superposition states using full control and barrier protocols.

In Sec.\ref{rl}, we introduce the deep reinforcement learning protocol in full detail

In Sec.\ref{runtime}, we show how the deep learning protocol improves with training time.

In Sec.\ref{statistics}, we show the statistical fluctuations of different repetitions of the deep learning protocol.

In Sec.\ref{exp}, we show experimental considerations how to implement our protocols with cold atoms and superconducting circuits.

\section{Theoretical models}\label{Theory}
\subsection{Winding number states}
To define our current states, we transform  the ring Hamiltonian Eq.1 in the main text of $L$ sites with $U=0$ and $P_j=0$ by Fourier transforming the operators into the quasi-momentum basis
\begin{equation}\label{HamiltonMomentum}
\mathcal{H}_\text{FT}=\sum_{k=0}^{L-1}-2J\cos\left(\frac{2\pi k}{L}\right)\nn{}{k}\;,
\end{equation}
where $\nn{}{k}=\cn{b}{k}\an{b}{k}$, with  $\cn{b}{k}=\frac{1}{\sqrt{L}}\sum_n \expU{i2\pi k n/L}\cn{a}{n}$. 
As the wavefunction around the ring is continuous, the wavefunction must be the same after going once around the ring. Thus, we demand $\exp\left(i\frac{2\pi k}{L} n \right)=\exp\left(i\frac{2\pi k}{L} (n+L) \right)$, which is only fulfilled if $k$ is an integer number, which describes how often the phase of the wavefunction winds by $2\pi$ around the ring.
The state with winding number $k$ for a single particle is defined as  $\cn{b}{k}\ket{\text{vac}}=\ket{k}$, where $\ket{\text{vac}}$ denotes the vacuum state.  Many-body states are generated as tensor products of particles, e.g. a state with $N_\text{p}$ particles with winding number $k$ is given by $\ket{\Psi_k}=\ket{k}^{\otimes N_\text{p}}$. For this state, the expectation value of the number of particles with winding number $k$ is given by $\langle\nn{}{k}\rangle=\bra{\Psi_k}\nn{}{k}\ket{\Psi_k}=N_\text{p}$.

\subsection{Quantum phase model fidelity}
To define the current state for the quantum phase model (QPM), we use the QPM with an applied artificial magnetic field $\Phi_\text{M}$ 
\begin{equation} 
\mathcal{H}_\text{QP}(\Phi_\text{M})= \sum_{j=1}^L \Big[ -2JN_s \, \cos (\hat{\phi}_j-\hat{\phi}_{j+1}-\Phi_\text{M} ) + P_j(t) \, \hat{Q}_j + \frac{U}{2} \hat{Q}^2_j \Big].
\end{equation}
To define the target state that carries a current, we refer to the ground state with a winding number distribution  that is centered around a specific winding number, depending on $\Phi_\text{M}$. For $\Phi_\text{k}=\frac{2\pi k}{L}$, the ground state $\ket{\Psi_\text{QP}(\Phi_{k})}$ winding number distribution is centered around the winding number $k$. We define the fidelity as $F=\abs{\braket{\Psi}{\Psi_\text{QP}(\Phi_{k})}}^2$.

\section{Fidelity and certification measure}\label{cert}
In the main text, we discuss entangled current states using the fidelity $F$. However, fidelity is not an observable that can be easily measured in experiments. We proposed a certification measure $W_\Psi$ that is an experimental observable (\cite{Gluza2018})
\begin{equation}\label{Eq:Fidelity}
W_{\Psi} =\frac{N_\text{C}^{N_\text{C}}}{N_\text{p}^{2N_\text{C}}}\prod_{k \in \Omega} \bra{\Psi}\nn{}{k}\nn{}{k}\ket{\Psi}.
\end{equation}
This measure behaves similarly to the standard fidelity for the entangled quantum current states $F=\abs{\braket{\Psi}{\Psi_{EC}}}^2$. For the target state $\ket{\Psi_{EC}}$ it reaches maximal value ${W_{\Psi}=1}$  and gives similar results for the state design.  In contrast to the fidelity $F$ though, we note that $W_{\Psi}$ is related to particle densities  and therefore it is an observable. In a cold atom setting, for example, $W_{\Psi}$ can be accessed by measuring the number of particles in a specific momentum mode, which can be achieved by time-of-flight measurements~\cite{bloch2008many,haug2018readout} (see below).
In Fig.~\ref{ParamGraphsFidelity}, we show results on the fidelity for the same parameters as in the main text in Fig.~4. We find that the fidelity behaves similar to the certification measure. The certification measure is always zero for the initial state (e.g. seen for datapoints with $T=0$), however the fidelity can be non-zero. This is because the initial state has in some cases a finite overlap with the target state. In contrast, the certification measure is constructed such that this initial overlap does not affect it.
\begin{figure*}[htbp]
	\centering
	\subfigimg[width=0.24\textwidth]{a1}{EntFidelityAll1DComposedPPOInputRRMCED20TKCdtPPON1L12M0E2S0R1i0A2t0_1n6e0_002v0_5l0_0002U0O0M1m-1c6a2u200d2E120000G0_99w1b500o11O1P2u1I3.pdf}
	\subfigimg[width=0.24\textwidth]{a2}{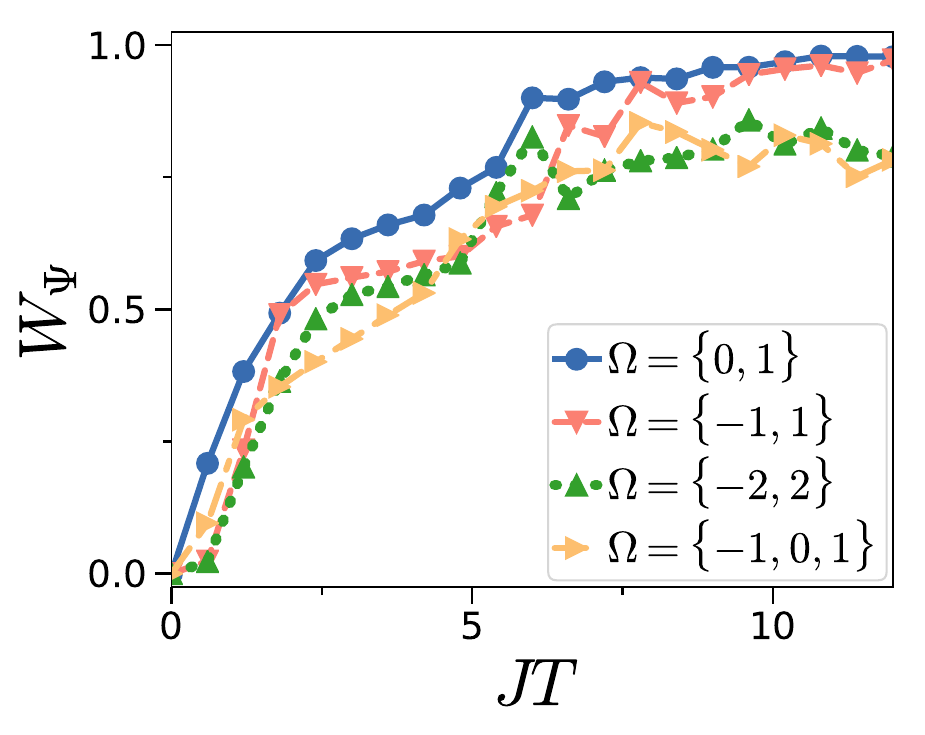}
	\subfigimg[width=0.24\textwidth]{b1}{EntFidelityAll1DComposedPPOInputRRMCED23NGntimestepsPPON1L12M0E2S0R1i0A3t10_0n1e0_002v0_5l0_0002U0_0O0M1m-1c6a2u200d2E120000G0_99w1b500o1P2u1I3.pdf}
	\subfigimg[width=0.24\textwidth]{b2}{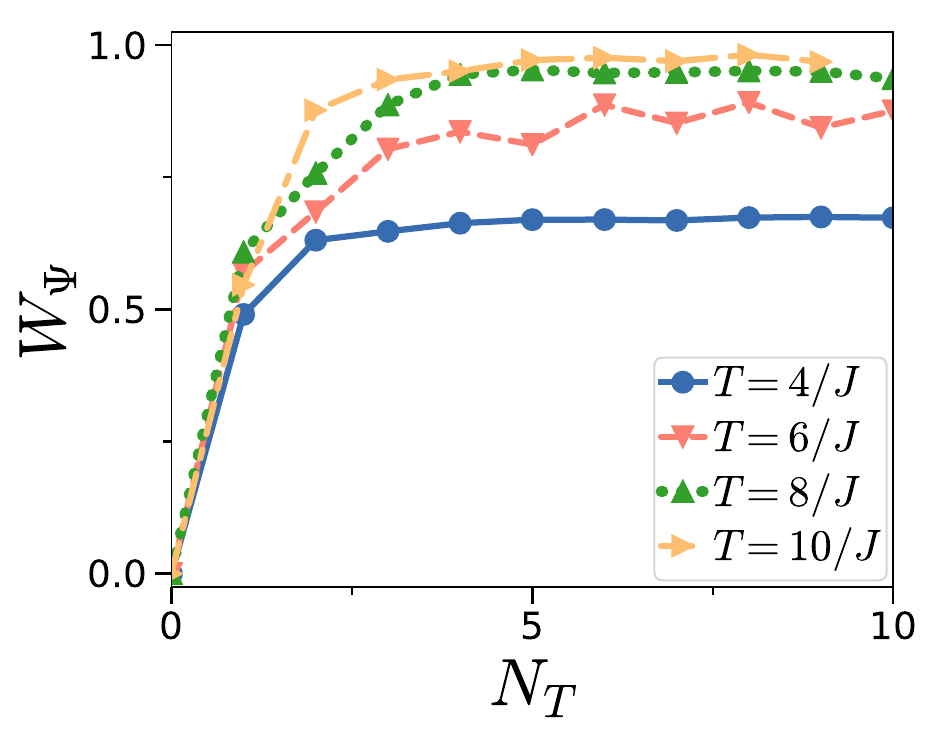}\\
	\subfigimg[width=0.24\textwidth]{c1}{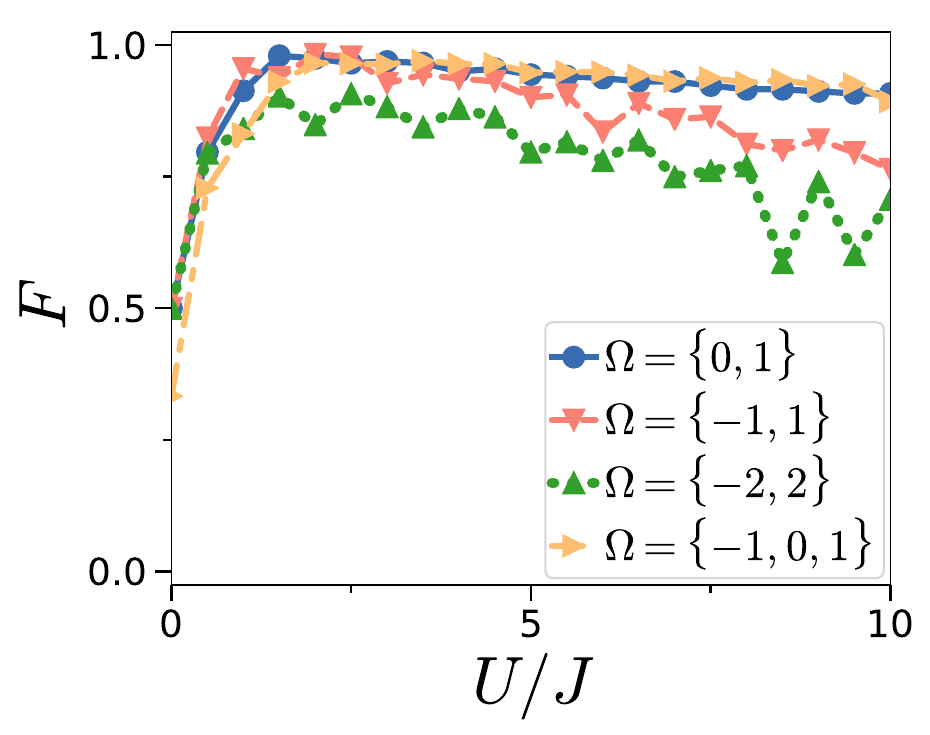}
	\subfigimg[width=0.24\textwidth]{c2}{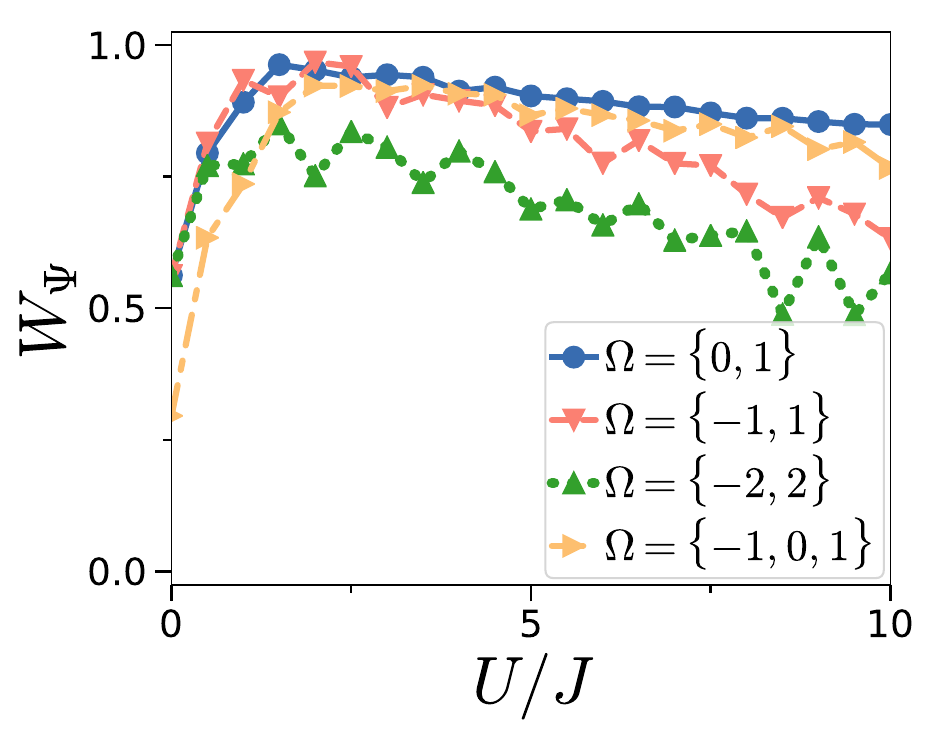}
	\subfigimg[width=0.24\textwidth]{d1}{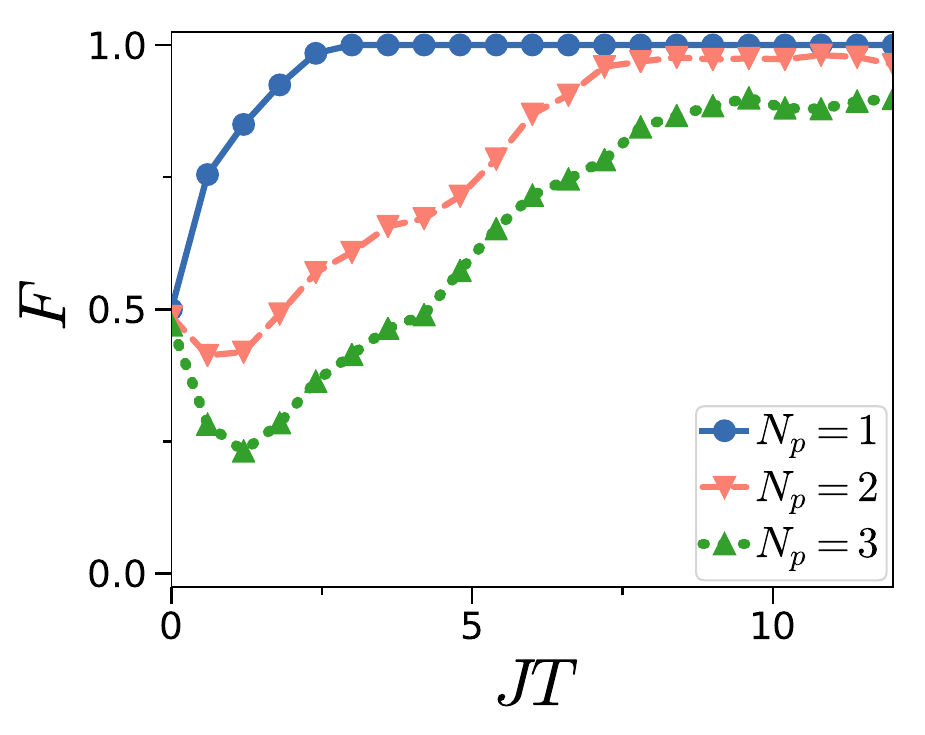}
	\subfigimg[width=0.24\textwidth]{d2}{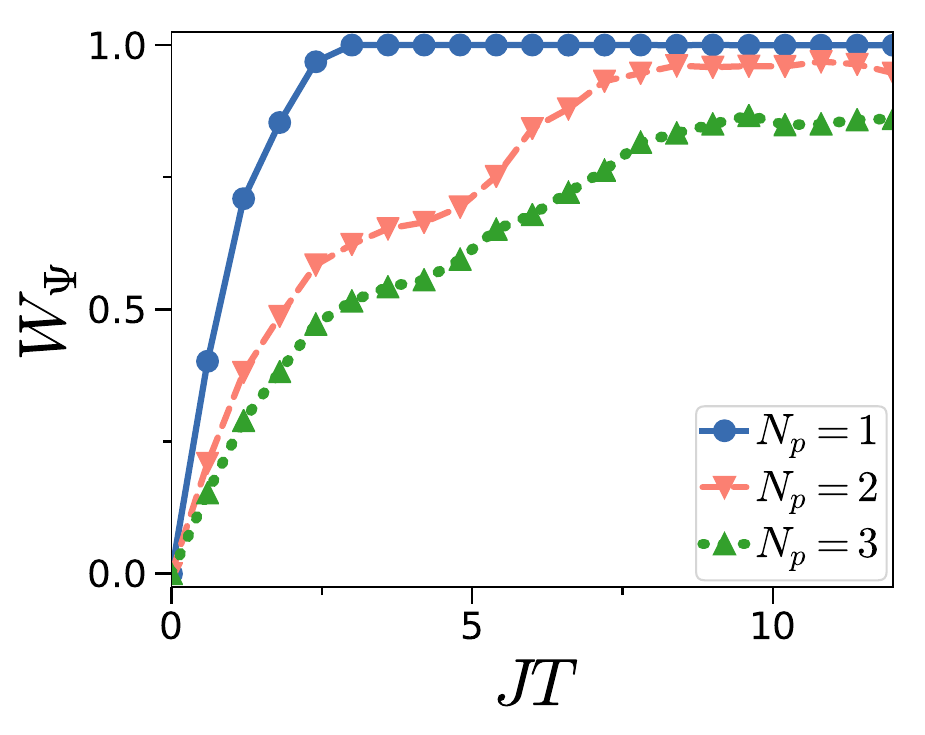}
	\caption{Comparison of fidelity  with entangled state $F=\abs{\braket{\Psi}{\text{ES}}}^2$ (for figure index 1) and certification measure $W_\Psi$ (for index 2) for ${L=12}$ sites for various parameters. Same parameters as in Fig.~4 of main text. We optimize for equal weight entangled states of $N_\text{C}$ winding number $k$ of type $\ket{\text{ES}}=\frac{1}{\sqrt{N_\text{C}}}\sum_{k \in \Omega}\ket{k}^{\otimes N_\text{p}}$ of a set of winding number $\Omega=\{k_1,k_2,\dots\,k_{N_\text{C}}\}$.  \idg{a} Varying time $T$ to generate different entangled states for $N_\text{p}=2$ particles, $N_\text{T}=6$ timesteps and ${U=J}$
		\idg{b} For varying timesteps $N_\text{T}$ to reach state $\Omega=\{0,1\}$ for $N_\text{p}=2$ particles.
		\idg{c} Interaction $U$ dependence for different types of states for $N_\text{T}=6$ timesteps and protocol time $T=9/J$.
		\idg{d} total protocol time $T$ for $N_\text{T}=4$ timesteps to generate entangled superposition state of winding number $\Omega=\{0,1\}$.}
	\label{ParamGraphsFidelity}
\end{figure*}

To characterize the entangled current states in an experimental setting, we defined the certification measure Eq.\ref{Eq:Fidelity}, which is a product of expectation values of observables. Experimentally, one is required to measure the square of the particle-number operator $\mean{\nn{2}{k}}$ of the winding number mode $k$. For cold atoms condensates, this measure can be determined from time-of-flight measurements, where the prepared state is expanded in free space~\cite{bloch2008many}. After free expansion, the density of the atoms develops a hole in the center, which is proportional to the winding number $k$ \cite{wright2013driving}. $\mean{\nn{2}{k}}$ can be calculated by repeating the time-of-flight measurement several times and calculating the fluctuations in the winding numbers.  

For superconducting circuits, the expectation value of the square of the particle number of a specific winding number $k$ can be derived from the expectation value of fourth order correlators between different qubits. We find
\begin{equation*}
\nn{2}{k}=\frac{1}{L^2}\sum_{n,m,r,s}\expU{i2\pi k (n+r-m-s)/L}\cn{a}{n}\an{a}{m}\cn{a}{r}\an{a}{s}\, ,
\end{equation*}
where the correlators can be derived by Fourier transforming the annihilation and creation operators of the operator.

\section{Comparison to GRAPE}\label{grape}
\begin{figure}[htbp]
	\centering
	\subfigimg[width=0.24\textwidth]{a}{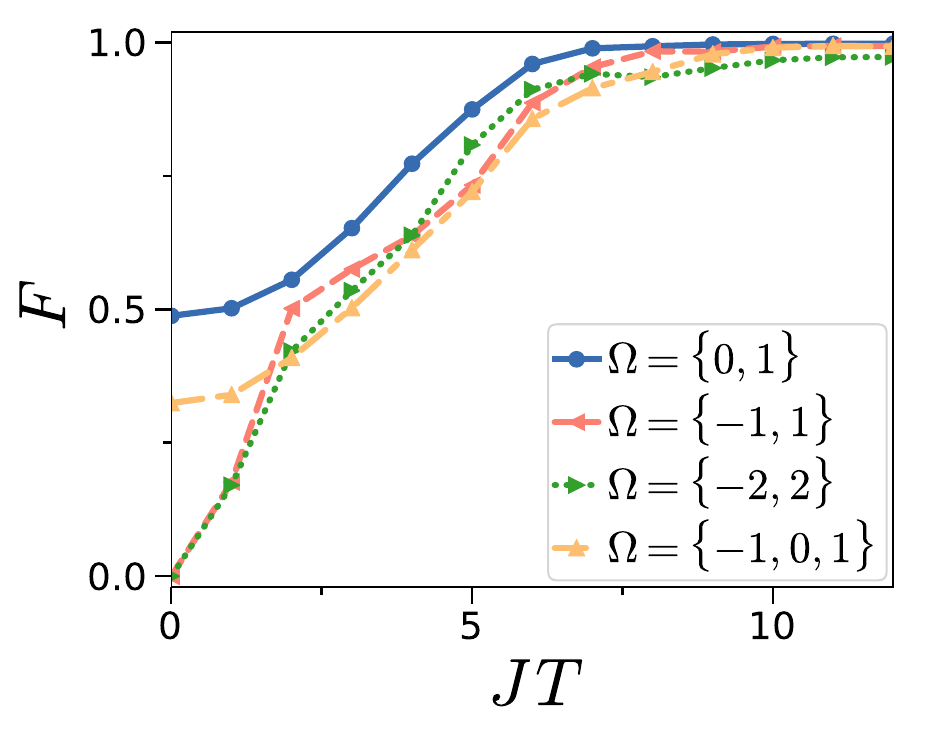}\hfill
	\subfigimg[width=0.24\textwidth]{b}{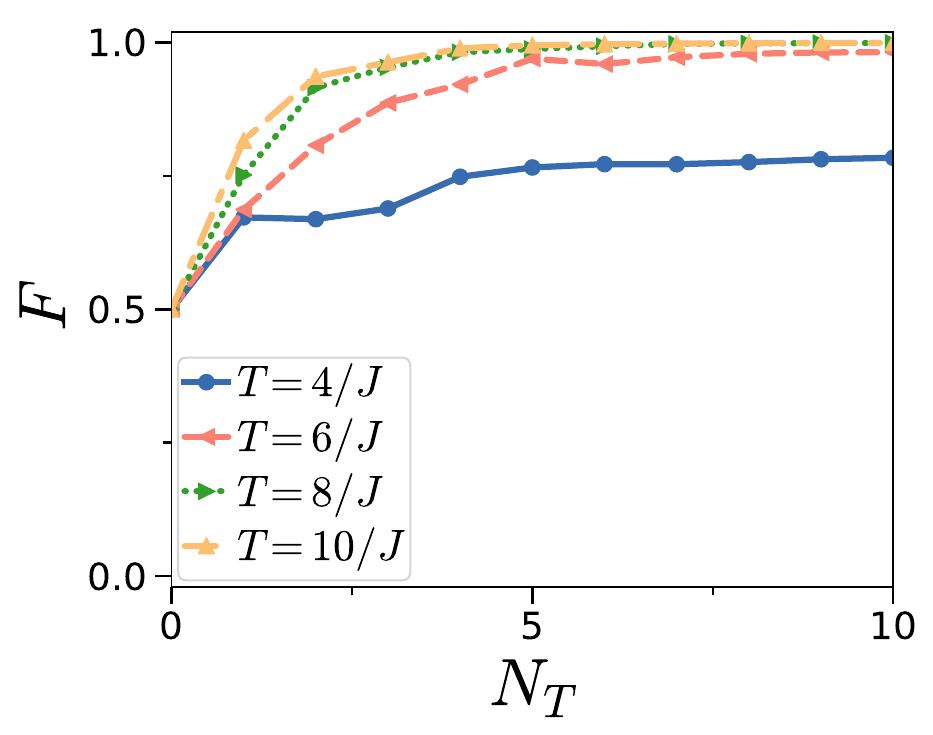}\\
	\subfigimg[width=0.24\textwidth]{c}{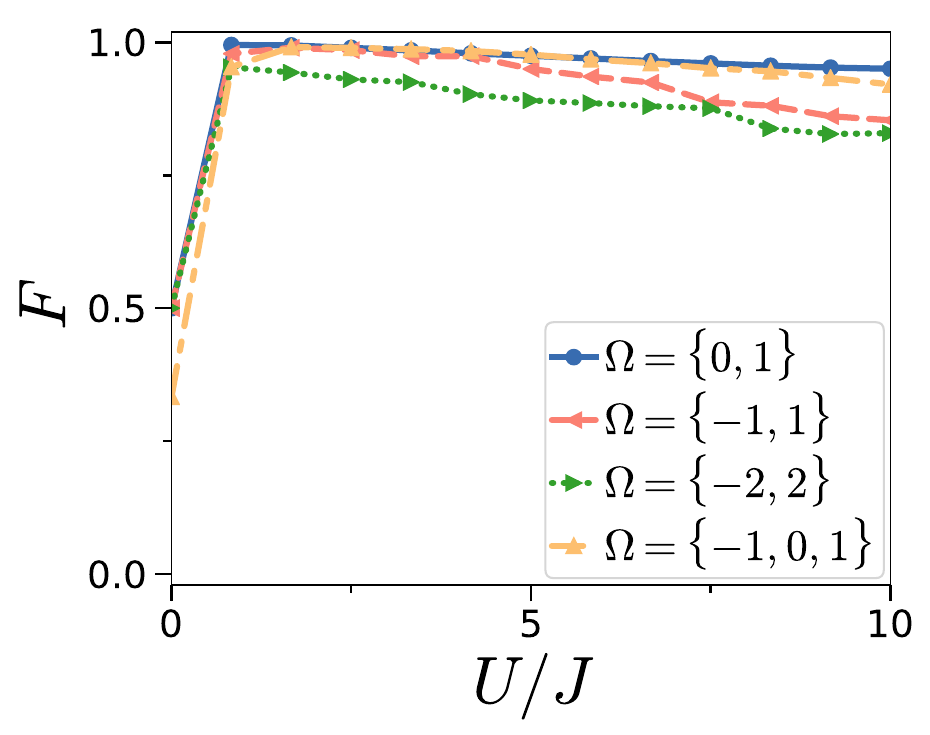}\hfill
	\subfigimg[width=0.24\textwidth]{d}{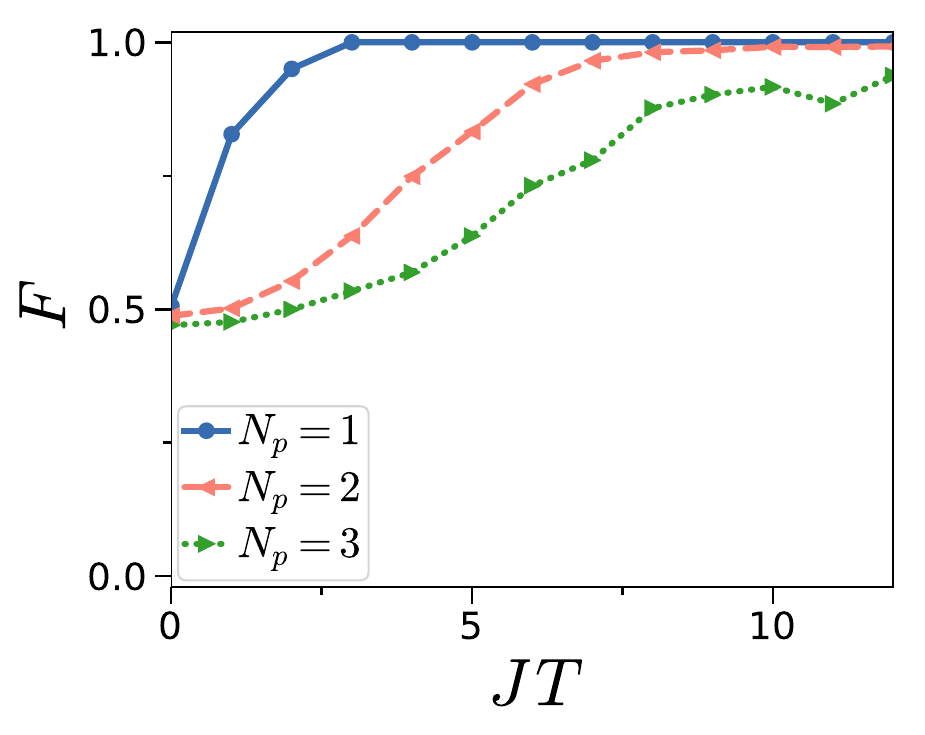}
	\caption{ 
		Generation of entangled superpositions of current states of type $\ket{\text{EC}}=\frac{1}{\sqrt{N_\text{C}}}\sum_{k \in \Omega}\ket{k}^{\otimes N_\text{p}}$ of a set of winding numbers ${\Omega=\{k_1,k_2,\dots\,k_{N_\text{C}}\}}$ using GRAPE.
		\idg{a} fidelity as a function of protocol time $T$ for $N_\text{p}=2$ particles, $N_\text{T}=6$ timesteps and ${U=J}$
		\idg{b} fidelity for varying timesteps $N_\text{T}$ ($N_\text{p}=2$, ${U=J}$, $\Omega=\{0,1\}$). 
		\idg{c} fidelity as a function of interaction $U$ for different types of states ($N_\text{p}=2$, $N_\text{T}=6$, $T=9/J$).
		\idg{d} fidelity as a function of protocol time $T$ for three different particle numbers $N_\text{P}$ ($N_\text{T}=4$, ${U=J}$, $\Omega=\{0,1\}$). All data for ${L=12}$ sites. GRAPE algorithm is run multiple times to avoid solutions that became stuck in local extrema.}
	\label{ParamGraphsGRAPE}
\end{figure}

A standard approach to optimise quantum dynamics is GRAPE \cite{johansson2013qutip,machnes2011comparing}. It relies on calculating the gradients of the control unitaries, then updating the driving parameters with gradient descent. As such, it requires access to the full model parameters as well as the underlying unitaries, in contrast to deep RL which suffices with observables only and is model free. We compare GRAPE in Fig.\ref{ParamGraphsGRAPE} with the data generated with deep RL in Fig.\ref{ParamGraphsFidelity}. GRAPE achieves same or a bit higher fidelity in most cases.  However, this is not surprising since it has access to all the information about the quantum system itself. GRAPE relies on gradients, and as such cannot optimise problems with discretized driving parameters as considered in Fig.4d in the main text.

\section{Robustness to noise in driving}\label{robust}
\begin{figure}[htbp]
	\centering
	\subfigimg[width=0.24\textwidth]{a}{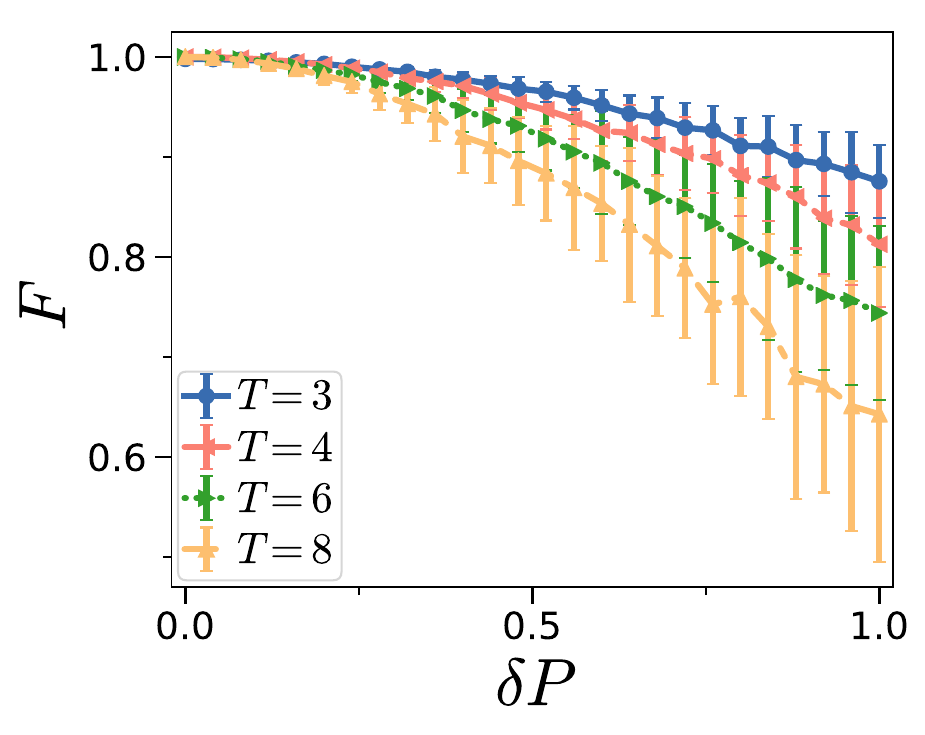}\hfill
	\subfigimg[width=0.24\textwidth]{b}{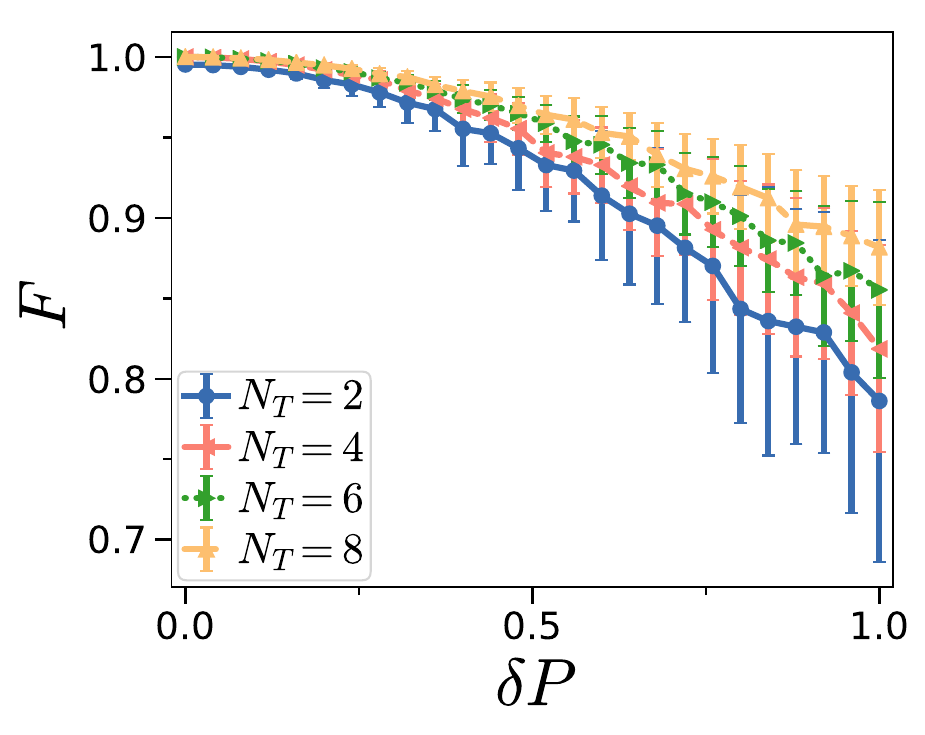}
	\subfigimg[width=0.24\textwidth]{c}{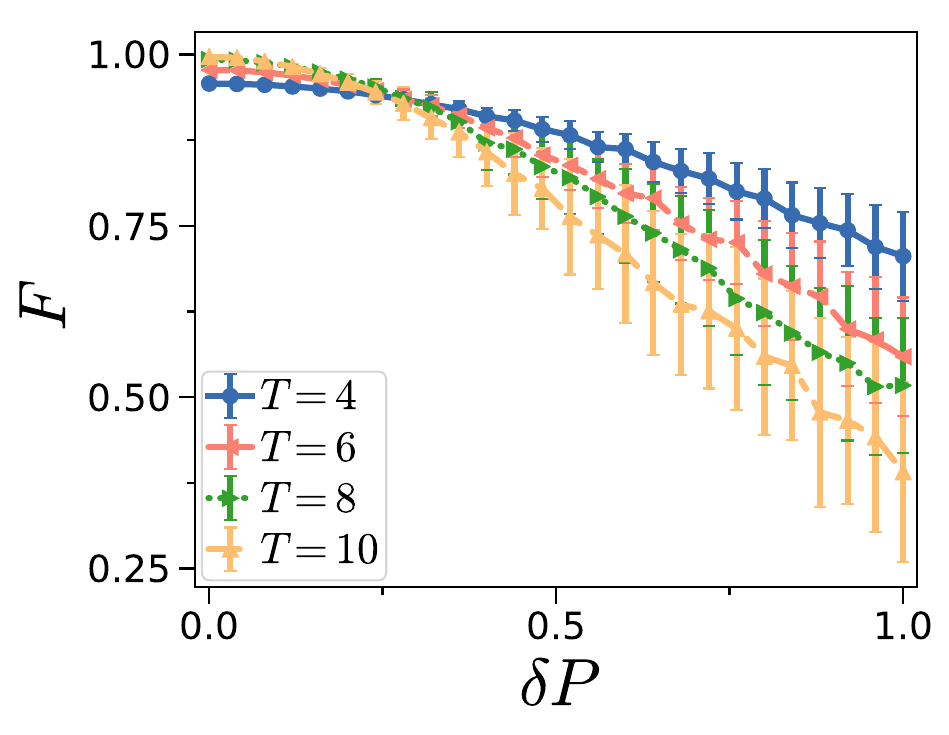}\hfill
	\subfigimg[width=0.24\textwidth]{d}{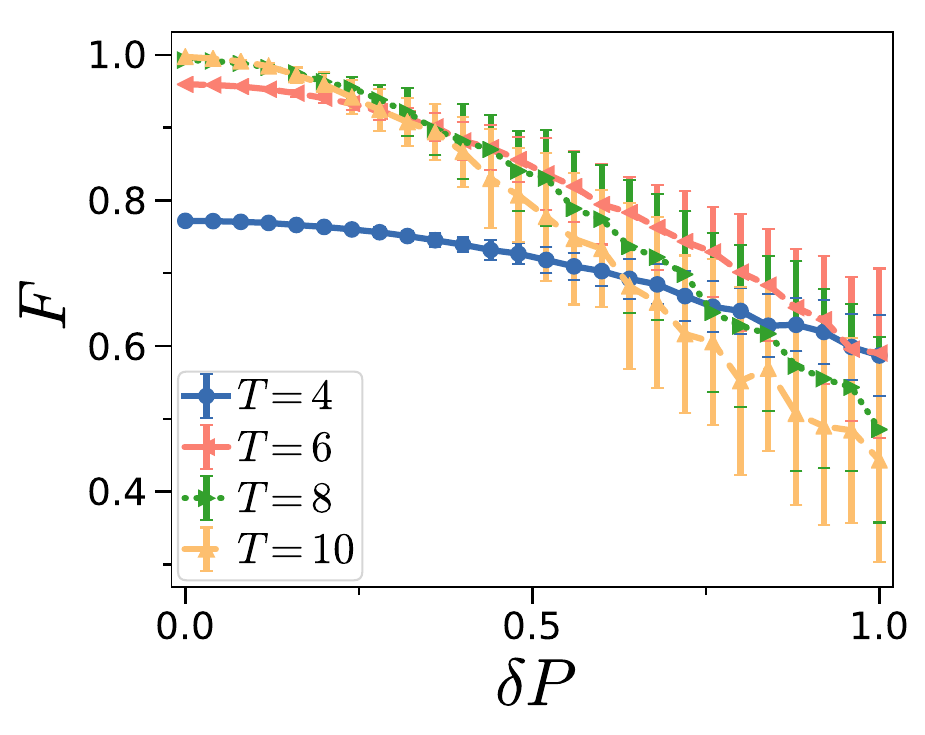}
	\caption{ Full control driving, local potential can take continuous values between $P=-J$ and $P=J$, changing in discrete time-steps. The protocol is optimized with GRAPE. Then, the local potential of the optimal protocol is perturbed with random noise, sampled from uniform distribution between $[-\delta P/2,\delta P/2]$, sampled for every site and timestep. Each point is sampled from 100 noise realizations, dots show mean value and error bars standard deviation of fidelity $F$. \idg{a} Fidelity to create phase winding state $\Omega=1$ for ${N_p=1}$, for varying random perturbation $\delta P$ and protocol time $T$. Robustness against noise decreases with increasing time $T$ \idg{b} Varying protocol timesteps $N_t$ increases with number of timesteps for $\Omega=1$ and ${N_p=1}$. \idg{c} Varying $T$ for $N_\text{p}=2$, $U=J$, $N_\text{T}=6$ and target state $\Omega=1$ \idg{d} Same parameters for entangled superposition state of $\Omega=\{0,1\}$. }
	\label{Robust}
\end{figure}

Experimental realizations are afflicted by experimental uncertainties and noise. In particular, for the full control protocol, where the potential at each site is controlled in time, the actual potential can only be set with finite accuracy. Experimental imperfections may also lead to noise, such that the real value used in the experiment is perturbed from the desired value. We simulate a random fluctuation of the potential in Fig.\ref{Robust}. For each timestep and site, a random potential $P_{i,\text{actual}}(t_n)=P_{i,\text{desired}}(t_n)+\Delta P$ is added, which is sampled from a uniform distribution between $\Delta P\sim[-\delta P,\delta P]$. We observe that for small $\delta P<0.25J$, noise has barely any influence on the resulting dynamics. We observe that increasing time $T$ negatively affects robustness (see Fig.\ref{Robust}a,b,d) and the standard deviation of the fidelity increases. Increasing number of timesteps $N_T$ improves the robustness (see Fig.\ref{Robust}c). We verify that these results persist for both one and two particles (two particles see Fig.\ref{Robust}c,d), as well as for entangled (entangled current see Fig.\ref{Robust}d) and non-entangled winding states.

\section{Supporting numerical results}\label{numerics}
\begin{figure}[htbp]
	\centering
	\subfigimg[width=0.3\textwidth]{a}{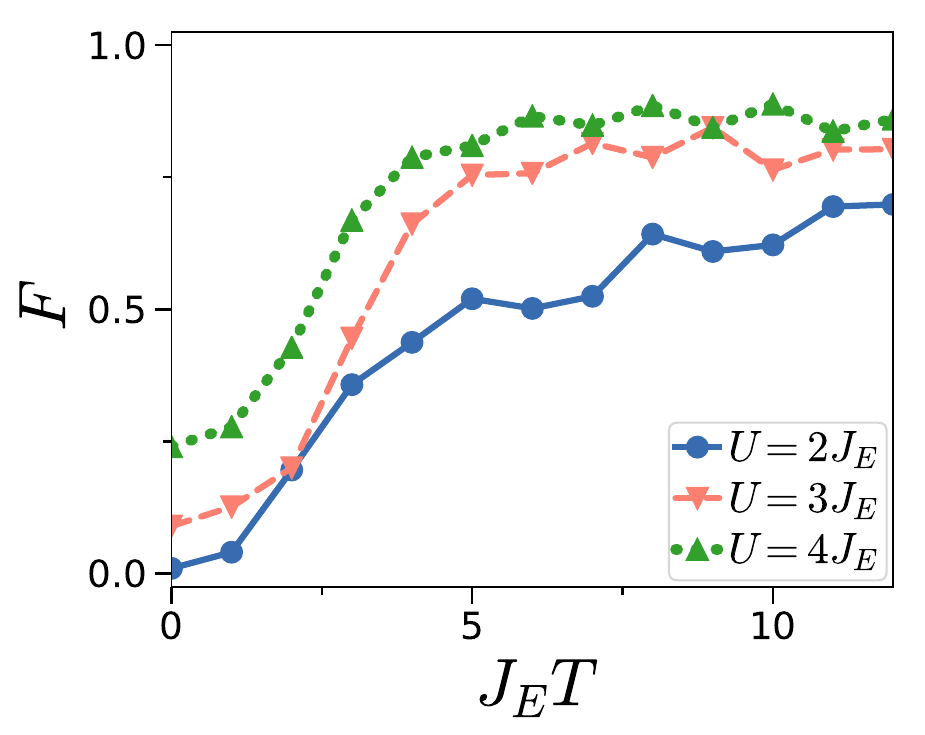}\hfill
	\caption{Fidelity of generating the current state for quantum phase model. Initial state is the ground state without flux, the target state is the ground state of the model for with one flux quantum. Plot shows fidelity of reaching target state $\ket{\Psi_\text{QP}(\Phi_1)}$ for different protocol times $T$.  We restrict the local Hilbert space $\Delta \hat{Q}_m=4$. We choose $N_\text{T}=8$, $L=5$, potential ${\abs{P_\text{max}<J_\text{E}}}$. }
	\label{QP5}
\end{figure}

Here, we present further data to support our findings. To solve the quantum phase model numerically, we  restrict the fluctuations around the mean particle number to $\Delta \hat{Q}_m$. Here, we increase the allowed number of particles to fluctuate In Fig.~\ref{QP5}, we show the fidelity $\Delta \hat{Q}_m=4$ and $L=5$.

\begin{figure}[htbp]
	\centering
	\subfigimg[width=0.24\textwidth]{a}{rewardAll1DComposedPPOInputRAAdtPPON1L12M0E2S0R3i0A0t0_017n6e0_0005v0_5l0_0005U0O1M1m0c6a6u100d2E50000G0_99w1b500P1u1A2I3.pdf}\hfill
	\subfigimg[width=0.24\textwidth]{b}{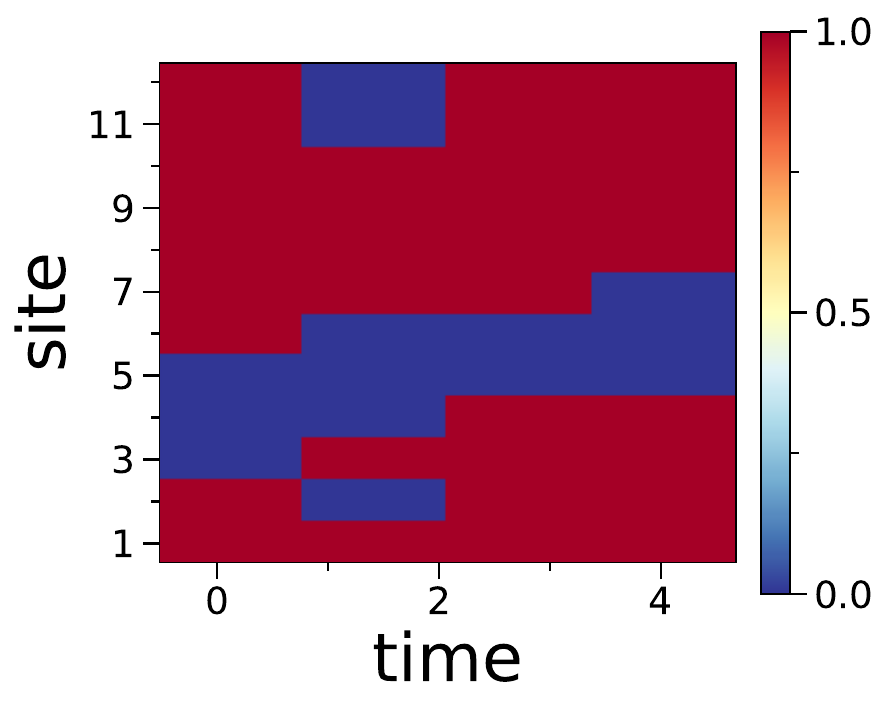}
	\caption{\idg{a} Full control protocol with discretized amplitudes: Local potential can take either the value $P_j(t)=0$ or $P_j(t)=J$.  Fidelity to create phase winding state $\Omega=1$ for ${N_p=1}$, for varying protocol time $T$ and number timesteps $N_\text{T}$.
		\idg{b} Resulting protocol for discretized driving amplitudes. }
	\label{ParamGraphsPot}
\end{figure}

Next, we show the full control protocol with constrained control amplitudes. In Fig.\ref{ParamGraphsPot}, the local potential can either assume $P_i(t)=0$ or $P_i(t)=J$. An example potential is shown in Fig.\ref{ParamGraphsPot}b. The resulting protocol takes a very simplistic form.

\begin{figure}[htbp]
	\centering
	\subfigimg[width=0.24\textwidth]{a}{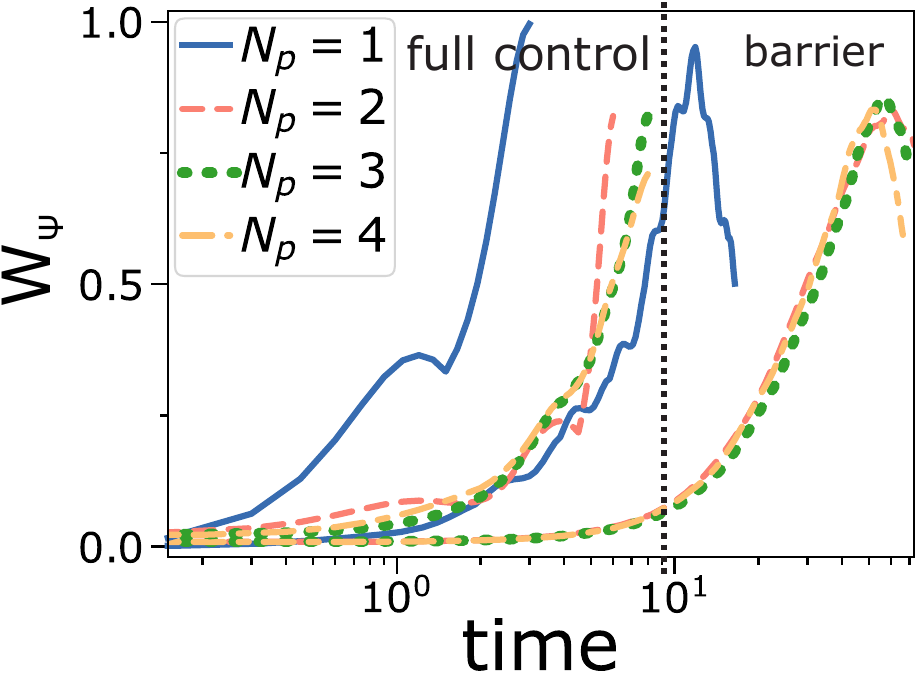}\hfill
	\subfigimg[width=0.24\textwidth]{b}{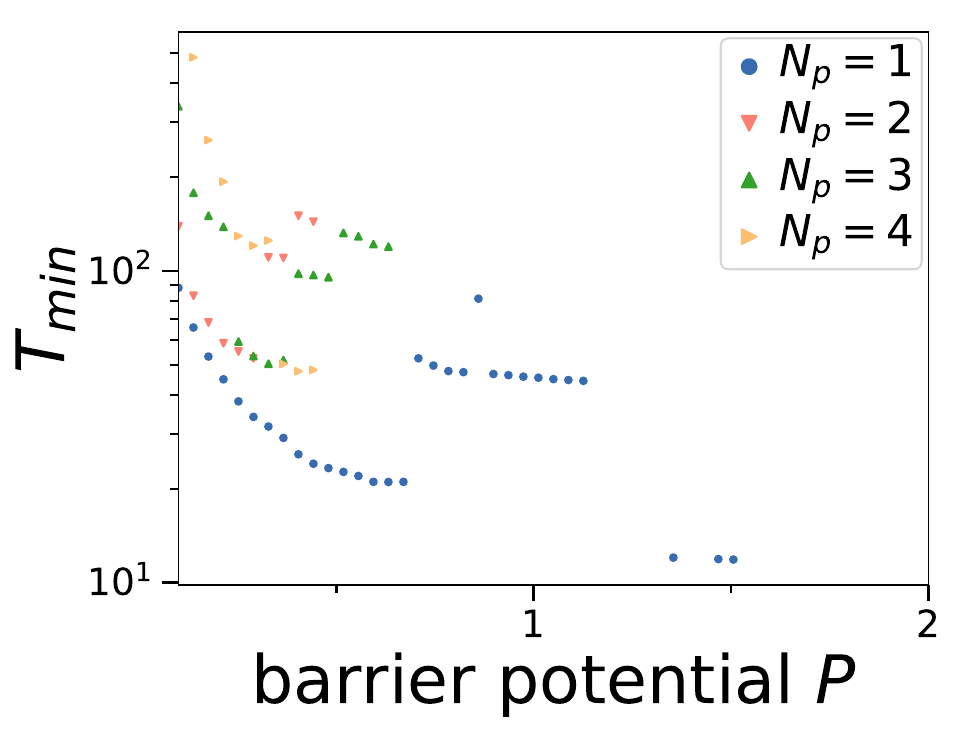}\\
	\subfigimg[width=0.24\textwidth]{c}{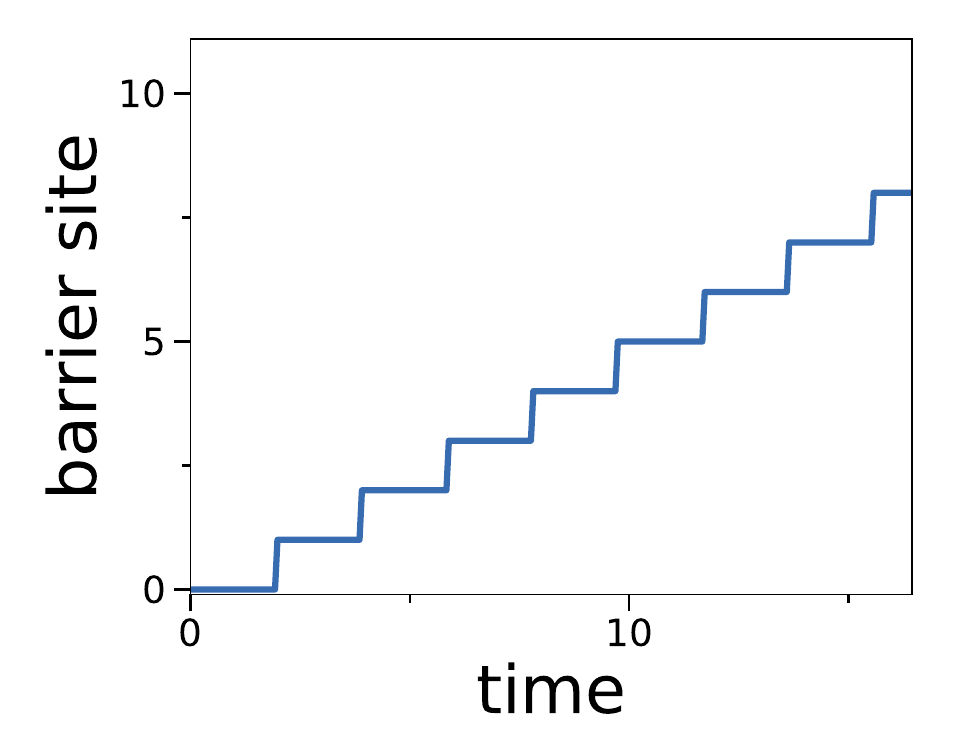}\hfill
	\subfigimg[width=0.24\textwidth]{d}{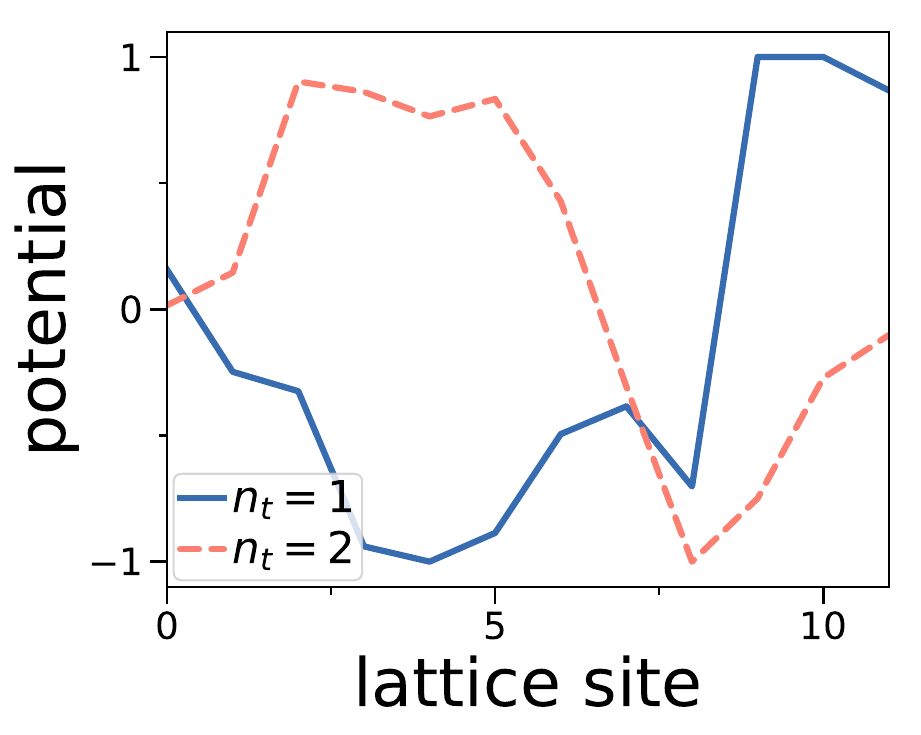}
	\caption{Generation of (entangled) superposition states of zero and one rotational quantum ${\Omega=\{0,1\}}$ in a ring lattice with ${L=12}$ sites. Evolution of certification measure (Eq.\ref{Eq:Fidelity}) during driving. We compare two different protocols: A barrier localized at a single site moving at constant speed (right curves in \idg{a}) or fully controlling the potential (FCP) of every lattice individually  (left curves of \idg{a}).  \idg{c} Minimal time $T_\text{min}$ required to create rotational states above a threshold fidelity ($W_\text{min}=0.95$ for $N_\text{p}=1$, else $W_\text{min}=0.8$) for different values of barrier amplitude. We find best rotation speed of barrier is at ${v\approx0.5J}$.
		\idg{c} best protocol for rotating barrier. Curve shows barrier position over time \idg{d} best protocol for full control over lattice potentials for a protocol of two timesteps $n_\text{t}$. Barrier and full control protocols shown calculated for $N_\text{p}=1$ particles.}
	\label{BarrierMLSup}
\end{figure}

In the main text, we demonstrate the generation of a state with winding number $k=1$ using the barrier protocol. The same protocol can also generate superposition states of $k=0$ and $k=1$.  To generate the entangled state, the barrier is rotated at the same speed as for the $k=1$ case with shorter time $T$. In Fig.~\ref{BarrierMLSup}, we study the dynamics of creating entangled superposition states and compare the FCP against driving a barrier.

\section{Deep reinforcement learning}\label{rl}
\begin{figure*}[htbp]
	\centering
	\subfigimg[width=0.95\textwidth]{}{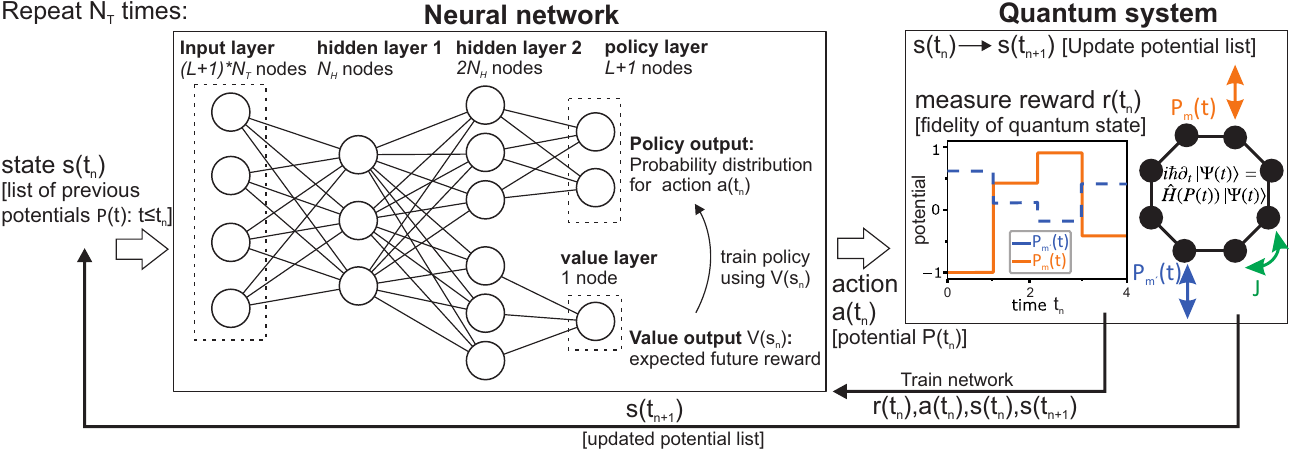}
	\caption{Neural network to optimize protocols to generate quantum states. Deep learning relies on representing a highly complex function (e.g. the quality of the driving protocol) with a neural network, and optimize it using observable data (e.g. measurement outcomes). The quantum system is a lattice ring with $L$ sites where particles can hop between neighboring sites with strength $j$. Each site $m$ has a local potential $P_m(t_n)$ that can be modulated in discrete timesteps $t_n$. The neural network controls the evolution of the quantum system by adjusting $P_m(t_n)$ and optimizes the parameters over many runs. The neural network performs step-wise evolution of the quantum in $N_\text{T}$ discrete time steps $t_\text{n}$ over total runtime $T$. It uses the chosen potentials of previous time steps as an input (state $s(t_n)$), and returns the potentials to be chosen at the next step (action $a(t_n)$) by sampling them from a Gaussian distribution. The training is performed by using a measure for the quantum state (reward $r(t_n)$). }
	\label{NNFull}
\end{figure*}

Here, we describe our machine learning algorithm in more detail. A detailed figure describing the neuronal network structure is shown in Fig.~\ref{NNFull} and pseudo-code  of the algorithm in Table~\ref{PPO}.
We learn the driving protocol via a deep Q-learning network~\cite{mnih2015human}, utilizing the actor-critic method acting on a continuous action space. Our method is using Proximal Policy optimization\cite{schulman2017proximal} and the implementation is based on Tensorflow~\cite{tensorflow2015-whitepaper}.
The quantum system is controlled by an agent, that depending on the state $s_t$ of the system acts with an action $a_t$ using the probabilistic policy $\pi(a_t|s_t)$. The idea of Q-learning is to find the Q-function $Q_\pi(s_t,a_t)$ that estimates the future reward that is paid out at the end the full protocol with this policy. The goal is to learn a policy that can realize long-term rewards over smaller short-term gains. 
The optimal Q-function is determined by the Bellman equation
\begin{align*}
Q(s_t,a_t,\pi)&=\mathbb{E}\left[r_t+\gamma Q(s_{t+1},a_{t+1},\pi)\right]\\
&=\mathbb{E}\left[r_t+\gamma r_{t+1}+\gamma^2 r_{t+2}+\dots\right]\,
\end{align*}
where $\mathbb{E}[.]$ indicates sampling over many instances. $\gamma\le1$ is a discount factor that weighs  future rewards against immediate rewards.
The input to the neural network are the Hamilton parameters at previous timesteps and it outputs the parameters for the policy $\pi(a_t|s_t,\mu,\sigma)$, where the actions are sampled from a normal distribution with mean value $\mu$ and width $\sigma$. $\mu$ is determined by the neural network and  $\sigma$ is optimized as a global variable and decreases during the optimization procedure.  We constrain the possible output values for the potential by mapping values outside of the constraint to the maximally allowed value. 
Proximal policy optimization is based on the actor-critic method.
The idea is to have two neural networks: A policy network and a value network. The policy network (actor) decides on the next action by determining the parameters of the policy. The value based network (critic) evaluates the taken action on how well it solves the task and estimates the future expected reward. It is used as an input to train the policy network. The two networks are trained at the same time using Adam~\cite{kingma2014adam}. 
Better performance can be achieved if the Q-function is split into two parts\cite{wang2015dueling}: $Q(s_t,a_t)=A(s_t,a_t)+V(s_t)$, where $A(s_t,a_t)$ is the advantage function and $V(s_t)$ the value function. $V(s_t)$ gives the expected future reward averaged over the possible actions according to the policy. This is the output of the critic network. $A(s_t,a_t)$ gives the improvement in reward for action $a_t$ compared to the mean of all choices. 
We estimate the Q-function from the value function with $Q(s_t,a_t)= r_t+\gamma V(s_{t+1})$, where $r_t$ is the reward given out under action $a_t$ and $V(s_{t+1})$ is the value function for the next timestep. We then minimize the square of the difference of the value function of the network and the predicted reward in the next timestep $L_\text{V}(\theta)=\mathbb{E}_t\left[(V_\theta(s_t)-y_t)^2\right]$, where $\theta$ are the current network parameters, $y_t=r_t+V(s_{t+1})$ is the calculated reward of the next timestep. 
The advantage function $A(s_t,a_t)=Q(s_t,a_t)-V(s_t)$ tells us how good a certain action $a_t$ is compared to other possible actions. Using above estimation of the Q-function, the advantage function can be approximated.  The advantage function is the input to train the policy network (the actor). Following the idea of proximal policy optimization~\cite{schulman2017proximal}, the goal is to maximize 
\begin{equation}
L_\text{p}(\theta)=\mathbb{E}_t\left[\frac{\pi_\theta(s_t,a_t)}{\pi_{\theta_\text{old}}(s_t,a_t)}A(s_t,a_t)\right]\, ,
\end{equation}
where $\theta$ are the network parameters and $\theta_\text{old}$ are the network parameters of a previous instance. Maximizing $L_\text{p}(\theta)$ for the network parameters $\theta$ over many sampled instances guides the distribution $\pi_\theta(s_t,a_t)$ such that it returns actions $a_t$ with maximal advantage. However, the ratio 
\begin{equation*}
b_t(\theta)=\frac{\pi_\theta(s_t,a_t)}{\pi_{\theta_\text{old}}(s_t,a_t)}
\end{equation*}
can acquire excessive large values, causing too large changes in the policy in every training step and making convergence difficult. It was proposed to use a clipped ratio~\cite{schulman2017proximal}
\begin{equation*}
L_\text{p}(\theta)=\mathbb{E}_t\left[\text{min}\left\{b_t(\theta)A(s_t,a_t),\text{clip}(r_t(\theta),1-\epsilon,1+\epsilon )A(s_t,a_t)\right\}\right]\, ,
\end{equation*}
such that the update at each step stays in reasonable bounds. We use $\epsilon=0.1$.
We optimize the neural network over many epochs $N_\text{E}$. For our results, we show the best protocol that was achieved during the optimization process.
We update the network by randomly sampling $N_\text{train}$ past iterations from a memory (replay Buffer $B$) that stores the last $N_\text{memory}$ epochs.
To reduce premature convergence, we add the entropy of the normal distributions of the policy to the loss function ${L_\text{S}(\theta)=\mathbb{E}_t\left[\frac{1}{2}\sigma\ln(2\pi e)\right]}$. This contribution slows down optimization to avoid convergence to a local minimum. The final loss function to optimize is $L(\theta)=L_\text{p}-c_\text{v}L_\text{V}+c_\text{s}L_\text{S}$, where $c_\text{s}$ and $c_\text{v}$ are hyperparameters. We find $c_\text{s}=0.02$ and $c_\text{V}=0.5$ as good choices.
A sketch of our neural network is shown in Fig.~\ref{NNFull}. The protocol solves the Schr\"odinger equation for a total time $T$ with $N_T$ discrete timesteps of width $\Delta t$, with respective times $t_n$.  The network determines the Hamilton parameters at different sites $m$ of in total $L$ sites. For one epoch, the system runs the network $N_T$ times. Input are the potentials used at previous timesteps $t_n$, and it returns the parameters to be used for the ${n+1}$ timestep. The input vector has length $(L+1)N_T$; it lists the parameters and the corresponding times $t_k$ used up to current timestep $t_k: t_1\dots t_n$. 
The network propagates through two hidden layers of fully connected neurons of size $N_\text{H}$ with ReLu activation functions. The output layer has size ${L+1}$ and uses a linear activation function. For the value function (critic), the output of the last hidden layer is collected to a single node, that represents the value function $V(s_t)$. For the policy (actor), $L$ outputs determine the mean values of the normal distribution that generates the potential at the next timestep $t_{n+1}$ of the protocol.
The neuronal network is trained with the loss function after calculating the full time evolution to time $T$ and measuring all the rewards. 

For the actual implementation, we choose the following parameters: learning rate with Adam $\alpha=0.0002$, $N_\text{H}=200$ neurons in the hidden layer, training over $N_\text{E}=120000$ epochs, training with a randomly sampled batch size $500$, a replay buffer $B$ of $N_\text{train}=500N_\text{T}$ previous results.

\begin{table*}
	\centering
	\fbox{\begin{minipage}{\textwidth}
			\begin{algorithmic}
				\State Randomly initialize critic $V(s|\theta)$ and actor $\mu(s|\theta)$ with weights $\theta$ 
				\State Initialize replay buffer $B$
				\For{epoch=$1,N_\text{E}$}
				\State{Input initial state $s_1$}
				\For{$t=1,N_\text{T}$}
				\State Sample action $a_t=\mu(s_t|\theta)$ from probability distribution
				\State Execute action $a_t$, receive reward $r_t$ and next state $s_{t+1}$
				\State Sample random batch of $N_\text{train}$ transitions ($s_t$,$a_t$,$r_t$,$s_{t+1}$) from $B$
				\State Set $y_t=r_t+\gamma V(s_{t+1})$
				\State Update critic by minimizing loss $L=\frac{1}{M}\sum_i(y_i-V(s_i|\theta))^2$
				\State Calculate advantage function $A(s_t,a_t)=Q(s_t,a_t)-V(s_t)=r_t+\gamma V(s_{t+1})-V(s_t)$
				\State Calculate probability ratio $r_\text{p}=P(a_t|\mu(s_t|\theta))/P(a_t|\mu(s_t|\theta^\text{old}))$ of current policy $\mu(s_t|\theta)$ and previous policy $\mu(s_t|\theta^\text{old})$
				\State function $\text{clip}\left(r_\text{p},c\right)$ clips $r_\text{p}$ between $1-c<r_\text{p}<1+c$
				\State Update actor policy with clipped loss $L=\text{min}\left[r_\text{p}A(s_t,a_t),A(s_t,a_t)\text{clip}\left(r_\text{p},c\right)\right]$
				\EndFor
				\State Store ($s_t$,$a_t$,$r_t$,$s_{t+1}$) $\forall t$ in replay buffer $B$
				
				\EndFor
			\end{algorithmic}
	\end{minipage}}
	\caption{Pseudo-code for our proximal policy optimization algorithm to generate quantum states.}
	\label{PPO}
\end{table*}

\section{Optimization runtime}\label{runtime}
\begin{figure*}[htbp]
	\centering
	\subfigimg[width=0.24\textwidth]{a}{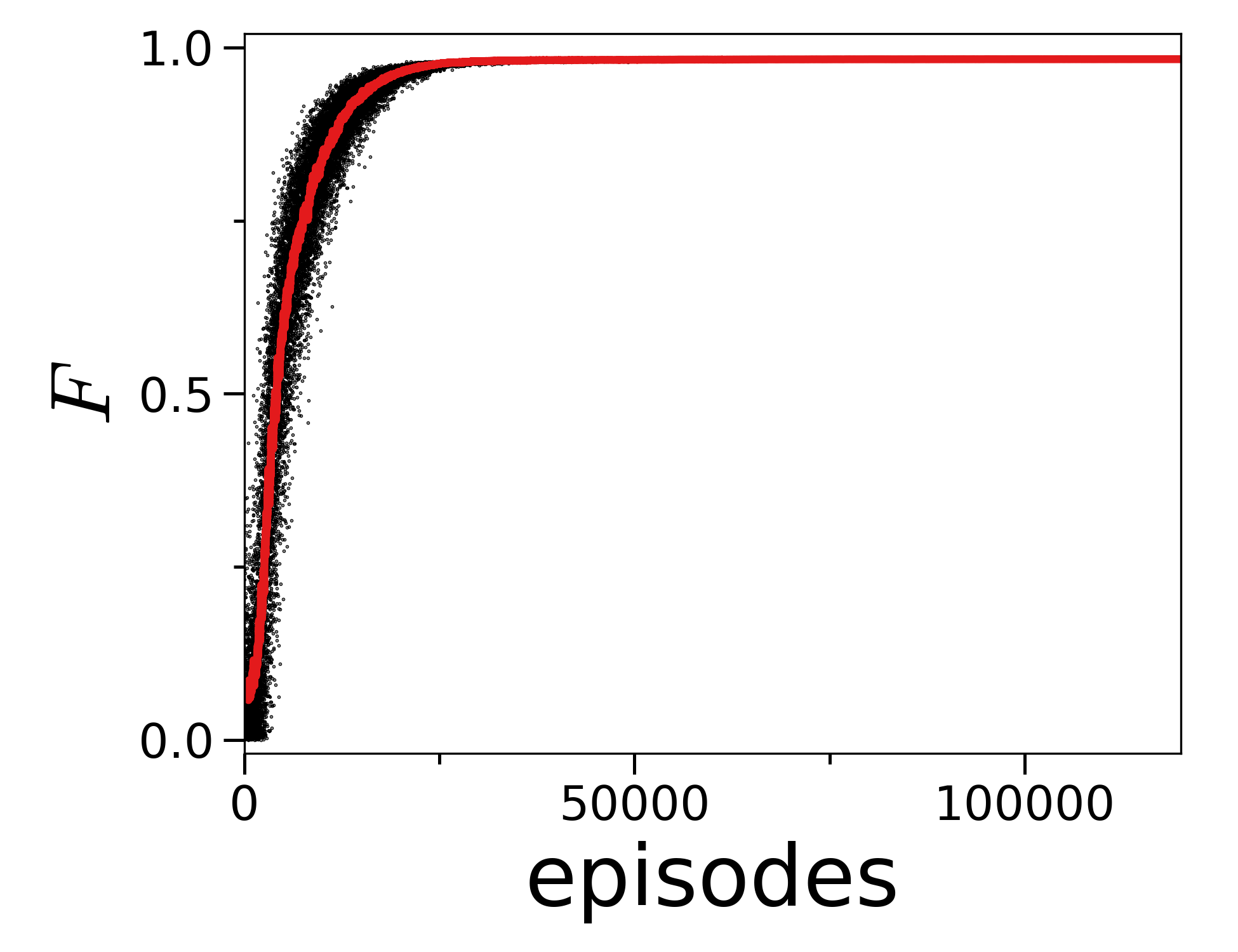}
	\subfigimg[width=0.24\textwidth]{b}{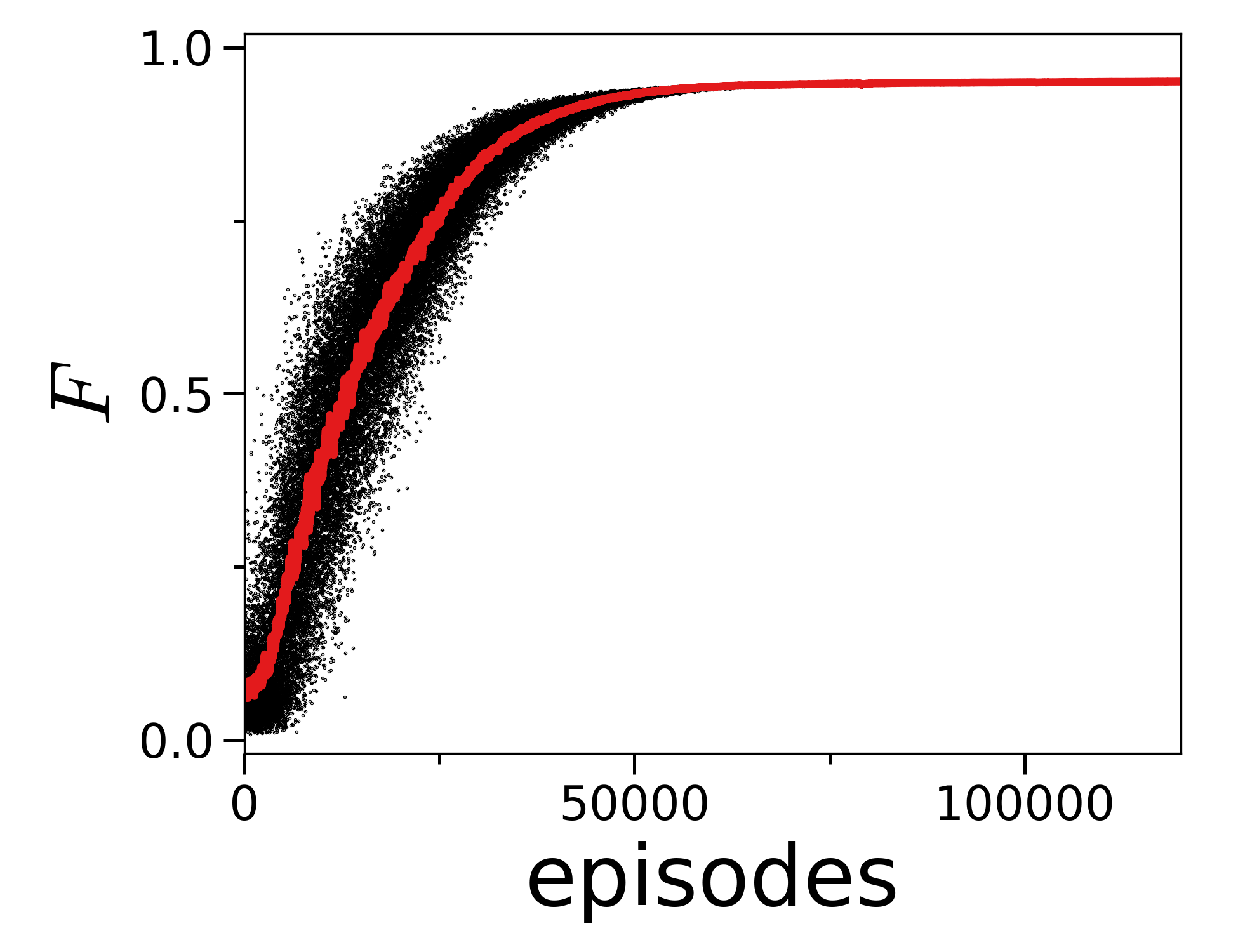}
	\subfigimg[width=0.24\textwidth]{c}{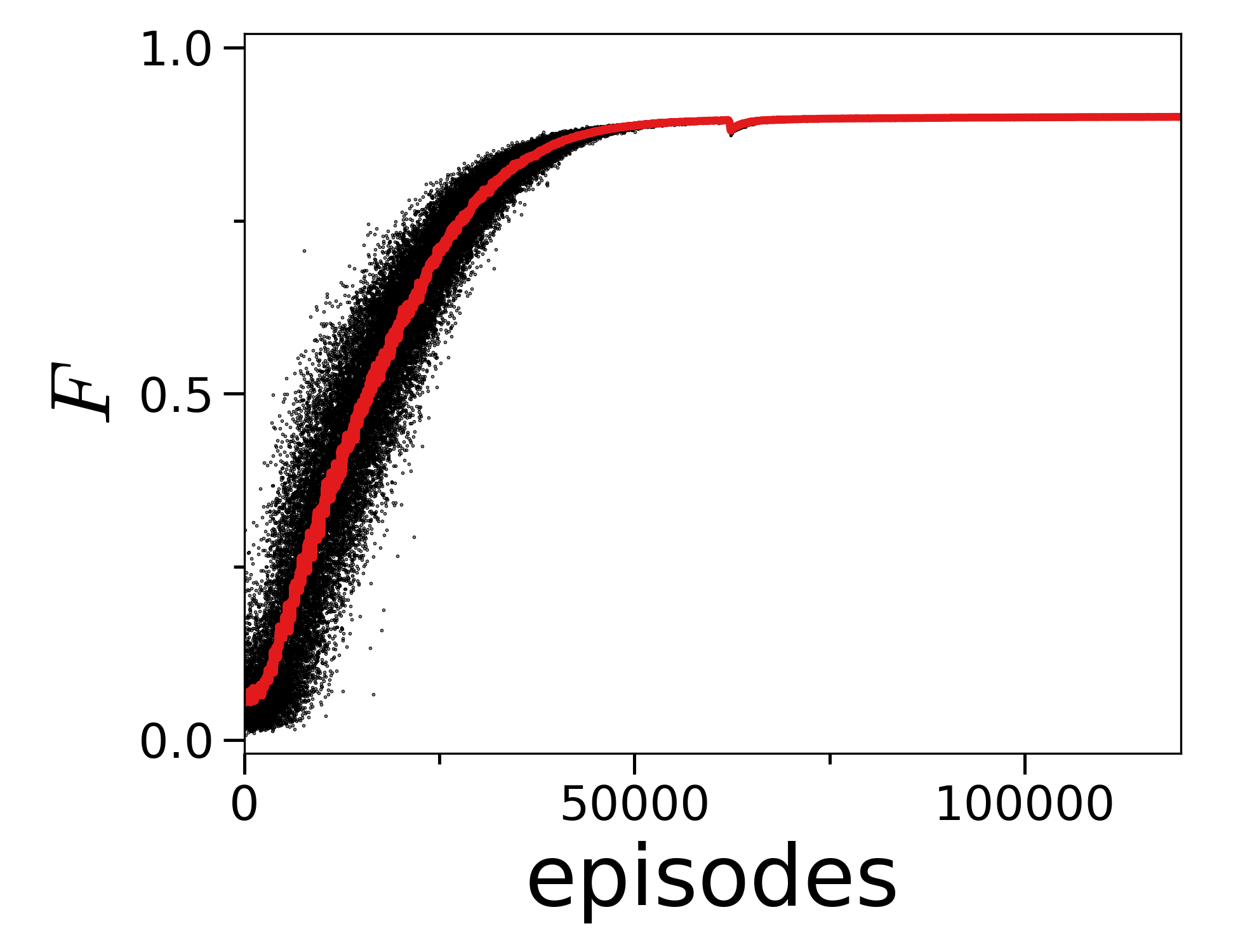}
	\subfigimg[width=0.24\textwidth]{d}{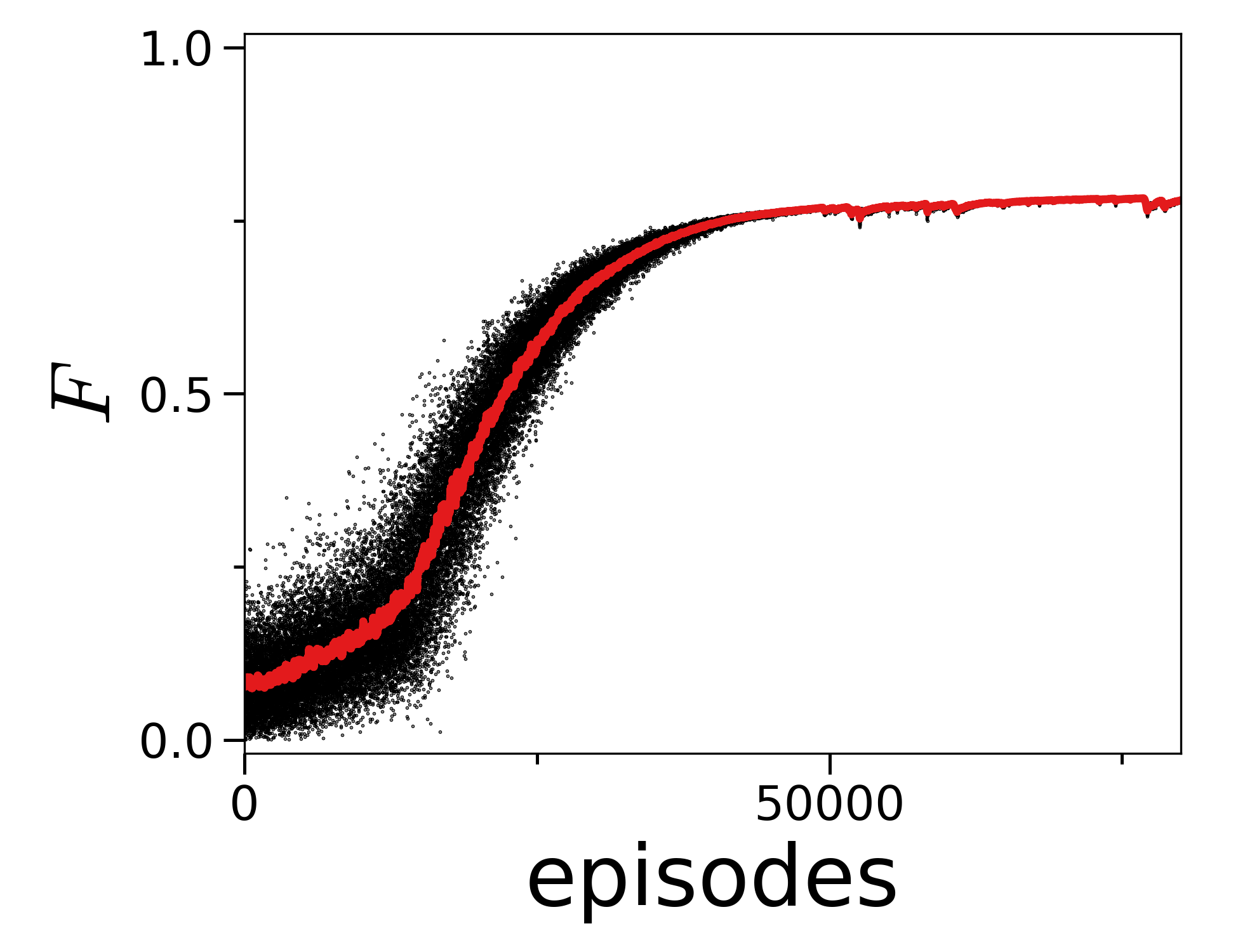}
	\caption{Optimization of the FCP protocol by the neuronal network over the number of epochs (number of protocol runs). We show exemplary data that generated the protocol shown in Fig.~2a of the main text. The dots indicate fidelity achieved during a particular run, while the red line is the moving average over the results.   \idg{a} $N_\text{p}=1$ \idg{b} $N_\text{p}=2$ \idg{c} $N_\text{p}=3$ \idg{d} quantum phase model.
	}
	\label{Training}
\end{figure*}

The machine learning algorithm starts with a randomly initialized neuronal network, that generates the driving sequence. By running repeatedly, the network generates better driving sequences. Finally, we plot the best driving sequence found during the training epochs. Here, we show results on the training procedure. The fidelity achieved increases during the training over many epochs. In Fig.~\ref{Training}, we show the fidelities during the training procedure for the FCP protocols that were used for Fig.~2a of the main text.

\section{Statistics}\label{statistics}
\begin{figure}[htbp]
	\centering
	\subfigimg[width=0.24\textwidth]{a}{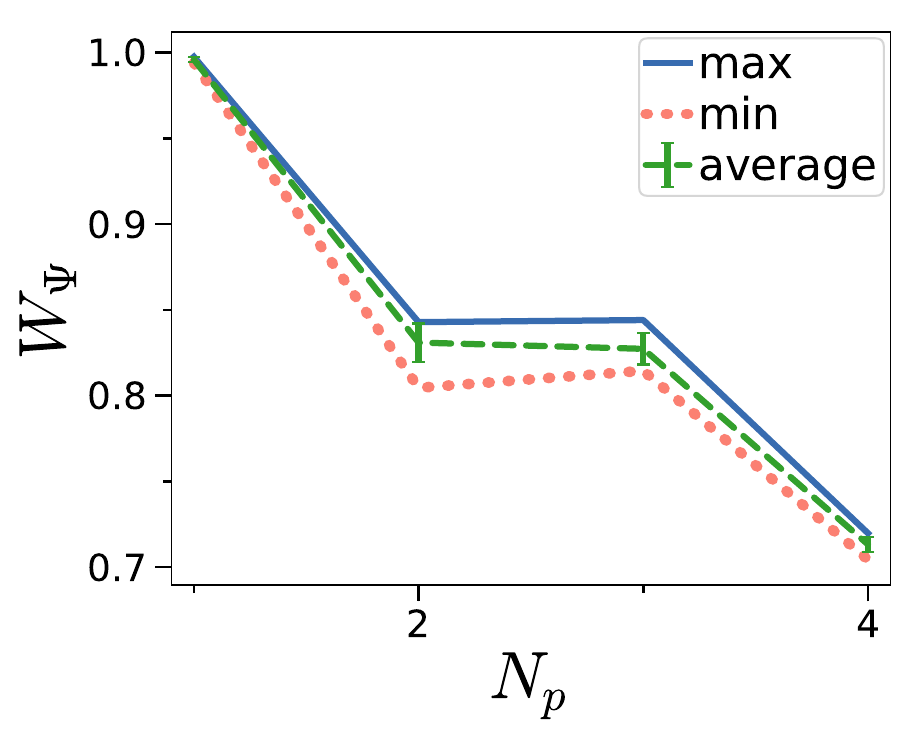}\hfill
	\subfigimg[width=0.24\textwidth]{b}{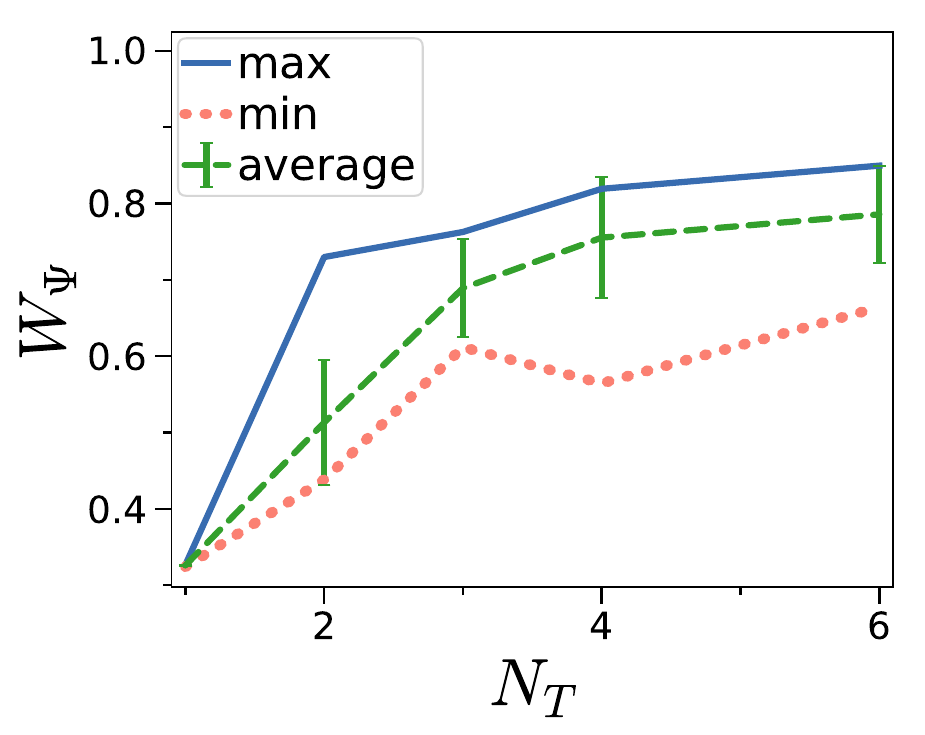}
	\caption{ Statistics (minimum, maximum and average certification measure $W_\Psi$) over 20 repeated runs of the algorithm for different parameter sets \idg{a}  different particle numbers $N_\text{p}$ for $N_\text{T}=4$, $T=8/J$, $U=J$, $\Omega=\{0,1\}$ \idg{b} different protocol steps $N_T$ for a total protocol length of $T=10/J$, $\Omega=\{-1,1\}$, $L=12$ sites for $N_\text{p}=3$ particles and ${U=J}$. Driving with local potential $\abs{P}<P_\text{max}=J$.   }
	\label{Statistics}
\end{figure}

Our goal is to optimize a high-parameter space driving protocol. In general, the optimization landscape is complex, with many local minima. 
We run the machine learning algorithm several times, and look at the convergence of the certification measure. As the algorithm is non-deterministic and not guaranteed to converge to the global minimum, each run can yield different end results. In Fig.~\ref{Statistics}, we show the minimal and maximal certification measure achieved for 20 runs to create entangled states. For reaching $\Omega=\{0,1\}$ we see only a small variation between minimal and maximal achieved certification measure --Fig.~\ref{Statistics}a. Thus, in one run of the algorithm we can be sure that a very good solution is found. 
However, we see a significant spread in certification measure results for higher particle number and more complex entangled states, e.g. $\Omega=\{-1,1\}$ and $N_\text{p}=3$ particles --Fig.~\ref{Statistics}b. Thus, for this parameter set to find the best result, several runs have to performed. 
This implies that the complexity and difficulty of the optimization problem to generate entangles states is highly dependent on the parameters of the problem. We took care to check that the variance of the solutions is within reasonable bounds for our results.

\section{Experimental considerations}\label{exp}
For a cold atom implementation, the driving of the ring lattice can create excitations. Within the Bose-Hubbard approximation, only the first Bloch band is considered. It is assumed that higher Bloch bands are far-detuned in energy and thus do not contribute. In most experiments, the energy gap between the Bloch bands within harmonic approximation of the lattice sites is given by $E_\text{lattice}=2\sqrt{V_0 E_\text{R}}$, where $V_0$ is the potential energy of a sinusoidal confinement and $E_\text{R}$ is the recoil energy \cite{jaksch1998cold}. For typical $V_0=10 E_\text{R}$, we find $E_\text{lattice}=6.3 E_\text{R}$. The nearest-neighbor coupling $J$ can be approximated as $J/E_\text{R}=\frac{4}{\sqrt{\pi}} \left(\frac{V_0}{E_\text{R}}\right)^{3/4}\exp\left(-2 \sqrt{\frac{V_0}{E_\text{R}}} \right)\approx 0.02$ \cite{bloch2008many}. Thus, the energy separation between first and second Bloch band is $\Delta E=E_\text{lattice}/J\approx 315$. From first order perturbation theory, we know that the overlap with higher-order states scales as $P/\Delta E$, where $P_\text{max}$ is the strength of the perturbation. The perturbation of the potential that we apply is on the order of $P_\text{max}=J$, which is much smaller than the energy gap. Thus, we can safely ignore excitations to higher Bloch bands.  
For the quantum phase model, the nearest-neighbor coupling strength scales as $J_\text{E}=JN_\text{p}/L$. To justify the one Bloch band approximation, $J_\text{E}$ has to much less than the energy gap between the Bloch bands. Thus, $J_\text{E}=\frac{JN_\text{p}}{L}\ll E_\text{lattice}$.

For cold atoms, the control parameter to change the local potential $P_j$ is  $V_0$. In our protocol, we change $P_j(t)$ and thus $V_0$ in time on the order of the nearest-neighbor coupling $J$. However, $J$ is actually function of $V_0$ and thus may change due to the driving. However, as $V_0\gg J$, changing $V_0$ on the order of $J$ has a negligible effect on $J$.

We assume a step-wise control of the potential, with sharp changes in the potential. The steps that change the potential have an amplitude on the order of $J$, at timescales of $J$. We detail how  this can be realized in experiment. In the case of cold atoms, the potential is generated by laser pulses. Light-shaping techniques can modify the potential with a frequency of about 20kHz~\cite{gauthier2016direct}. The relevant timescale of the experiment are on the order of $J$, which is far smaller.
For superconducting circuits, the potential is controlled by microwave pulses that modify the circuit potential. The circuit potential can be modulated on the order of 35MHz, while the nearest-neighbor couplings are far slower with $J\approx 4\text{MHz}$~\cite{roushan2017chiral}. Thus, we conclude that the driving parameters can be feasible modulated on the time scales we consider.

\end{document}